\documentclass[fleqn,usenatbib]{mnras}

\usepackage[T1]{fontenc}
\usepackage{ae,aecompl}

\usepackage{graphicx}	
\usepackage{subfig}
\usepackage{amsmath}	
\usepackage{amsfonts} 
\usepackage{amssymb}	
\usepackage{xcolor}
\usepackage[export]{adjustbox} 

\title[Being KLEVER at cosmic noon: ionised gas outflows]{Being KLEVER at cosmic noon: ionised gas outflows are {inconspicuous} in low-mass star-forming galaxies but {prominent} in massive AGN hosts}

\author[Concas et al.]{
Alice Concas,$^{1,2,3,4,5}$\thanks{E-mail: alice.concas@eso.org (AC)}, Roberto Maiolino$^{1,2,6}$,Mirko Curti$^{1,2}$,Connor Hayden-Pawson$^{1,2}$, \newauthor Michele Cirasuolo$^{5}$, Gareth C. Jones$^{1,2}$, Amata Mercurio$^{7}$, Francesco Belfiore$^{4}$,\newauthor   Giovanni Cresci$^{4}$, Fergus Cullen$^{8}$, Filippo Mannucci$^{4}$, Alessandro Marconi$^{3,4}$,\newauthor  Michele Cappellari$^{9}$, Claudia Cicone$^{10}$,  Yingjie Peng$^{11}$  and Paulina Troncoso$^{12}$
\\
$^{1}$Cavendish Laboratory, University of Cambridge, 19 J. J. Thomson Ave., Cambridge CB3 0HE, UK \\
$^{2}$Kavli Institute for Cosmology, University of Cambridge, Madingley Road, Cambridge CB3 0HA, UK \\
$^{3}$Dipartimento di Fisica e Astronomia, Universit\'a di Firenze, Via G. Sansone 1, 50019, Sesto Fiorentino (Firenze), Italy\\
$^{4}$INAF - Osservatorio Astrofisico di Arcetri, Largo E. Fermi 5, 50125, Firenze, Italy\\
$^{5}$European Southern Observatory, Karl-Schwarzschild-Strasse 2, D-85748 Garching bei Muenchen, Germany\\
$^{6}$Department of Physics and Astronomy, University College London, Gower Street, London WC1E 6BT, UK\\
$^{7}$INF - Osservatorio Astronomico di Capodimonte, Via Moiariello 16, I-80131 Napoli, Italy\\
$^{8}$Institute for Astronomy, University of Edinburgh, Royal Observatory, Edinburgh EH9 3HJ, UK\\
$^{9}$Sub-Department of Astrophysics, Department of Physics, University of Oxford, Denys Wilkinson Building, Keble Road, Oxford OX1 3RH, UK\\
$^{10}$Institute of Theoretical Astrophysics, University of Oslo, PO Box 1029, Blindern 0315, Oslo, Norway\\
$^{11}$Kavli Institute for Astronomy and Astrophysics, Peking University, 5 Yiheyuan Road, Beĳing 100871, China\\
$^{12}$Escuela de Ingeniería, Universidad Central de Chile, Avenida Francisco de Aguirre 0405, 171-0614, La Serena, Coquimbo, Chile}

\date{Accepted XXX. Received YYY; in original form ZZZ}

\pubyear{2022}
\usepackage{newtxtext,newtxmath}
\begin{document}
\label{firstpage}
\pagerange{\pageref{firstpage}--\pageref{lastpage}}
\maketitle

\begin{abstract}
We investigate the presence of ionised gas outflows in a sample of 141 main-sequence star-forming galaxies at $1.2<z<2.6$ from the KLEVER {(KMOS Lensed Emission Lines and VElocity Review)} survey. Our sample covers an exceptionally wide range of stellar masses, $8.1<\log(M_\star/M_{\odot})<11.3$, pushing outflow studies into the dwarf regime thanks to gravitationally lensed objects. We stack optical rest-frame emission lines (H$\beta$, [OIII], H$\alpha$ and [NII]) in different mass bins and seek for tracers of gas outflows by using a novel, physically motivated method that improves over the widely used, simplistic double Gaussian fitting. We compare the observed emission lines with the expectations from a {rotating disc (disc+bulge for the most massive galaxies) model}, whereby significant deviations are interpreted as a signature of outflows. We find clear evidence for outflows in the most massive, $\log(M_\star/M_{\odot}) > 10.8$, AGN-dominated galaxies, suggesting that AGNs may be the primary drivers of these gas flows. Surprisingly, at $\log(M_\star/M_{\odot})\leq 9.6$, the observed line profiles are fully consistent with a rotating disc model, indicating that ionised gas outflows in dwarf galaxies might play a negligible role even during the peak of cosmic star-formation activity. Finally, we find that the observed mass loading factor scales with stellar mass as expected from the TNG50 cosmological simulation, but the ionised gas mass accounts for only 2$\%$ of the predicted value. This suggests that either the bulk of the outflowing mass is in other gaseous phases or the current feedback models implemented in cosmological simulations need to be revised.

\end{abstract}

\begin{keywords}
galaxies: evolution -- galaxies: kinematics and dynamics -- galaxies: high-redshift -- galaxies: ISM
\end{keywords}



\section{Introduction}

Numerical simulations based on the Lambda-Cold Dark Matter ($\Lambda$CDM) paradigm have been very successful in reproducing the observed large-scale structure of the Universe \citep{springel_large-scale_2006}. However, under the assumptions that light traces mass and the initial conditions are provided by 
the cosmic microwave background, the same simulations fail to reproduce the observed low star formation efficiency of dwarf and massive galaxies (e.g. \citealp{silk_feedback_2010}). To reconcile theory with observations, state-of-the-art simulations and models of galaxy evolution 
\citep{springel_cosmological_2005,Vogelsberger_2014,schaye_eagle_2015} invoke feedback to suppress further star formation and excessive growth of galaxies. 

{Star-formation (SF) activity is supposed to play a key role at low stellar masses, {below $\log(M_\star/M_{\odot})\sim$10.5 (or halo mass $\log(M_{h}/M_{\odot}) \leq 12$)} by injecting energy and momentum into the interstellar medium (ISM) via stellar outflows and supernovae (SF feedback, e.g. \citealp{Chevalier77, Murray+05}; \citealp{Hopkins+14}). Even more dramatic effects are expected at high stellar masses, on and above $\log(M_\star/M_{\odot})\sim$10.8 (or $\log(M_{h}/M_{\odot}) > 12$), where accreting central supermassive black holes (SMBHs), shining as active galactic nuclei (AGN), and release large amounts of energy and momentum (AGN feedback, see \citealp{Fabian2012,King2015}). Both feedback mechanisms are thought to generate massive gas outflows, capable of reducing the fuel available for star formation by 1) expelling the gas content from the galaxy and/or 2) preventing the accretion of new fresh gas from the circumgalactic and interstellar medium. More recent zoom-in simulations have investigated the role of AGN-driven outflows also in the low-mass regime \citep{Koudmani2019}. The main finding is that, at least in local dwarf galaxies, AGN-driven outflows are not expected to contribute significantly to the direct (ejective) quenching of star formation, but they can make outflows faster and hotter, and therefore contribute to galaxy quenching by preventing cold accretion. Similar simulations have however found that the role of AGN-driven outflows might be more prominent in distant galaxies (z$\sim$1-2) and more important for their quenching \citep{Koudmani2021}.} 

{Although these outflows are thought to be ubiquitous in simulated galaxies, it is now established that massive galactic outflows are not always present in local galaxies. They are detected in peculiar, very active star-forming galaxies undergoing starburst events often connected with galaxy interactions (e.g. \citealp{heckman_nature_1990, Lehnert1996, Rupke2002, Rupke2005a, Rupke2005, Martin2005, Martin2006, Soto2012, Westmoquette2012, RupkeVeilleux2013, Bellocchi2013, Hill_Zakamska2014, Arribas2014, heckman_systematic_2015, Cazzoli2016, Chisholm2017, mcquinn_galactic_2019, Fluetsch2021}) and/or AGN dominated systems (\citealp{morganti_ic_2007,morganti_fast_2015, Feruglio2010,VillarMartin2011,RupkeVeilleux2011, Greene2012, mullaney_narrow-line_2013,  RodriguezZaurin2013, Harrison2014, Cicone2014, Cresci2015, Oosterloo2017,Venturi2018, perna_multi-phase_2019, marasco_galaxy-scale_2020}) but, {as demonstrated by statistical studies, they are rarely detected (both in emission and in absorption) in normal star-forming galaxies which represent the bulk of local galaxy population (\citealp{concas_light_2017, concas_two-faces:_2019,roberts-borsani_prevalence_2019}, but see also \citealp{cicone_outflows_2016}). }}

Gas outflows, however, are expected to be more important at higher
redshift (z$\sim 1-2$), during the peak of the cosmic SF and SMBHs accretion (\citealp{madau_cosmic_2014}), also known as “Cosmic Noon”, where the ejective feedback may be maximised. 
During the past ten years, enormous progress on the detection of gas outflows at Cosmic Noon has been made through the advent of optical and near-IR multi-object spectrographs (MOS, e.g. MOSFIRE on the W. M. Keck telescope, \citealp{MOSFIRE}) and integral field units (IFUs) such as KMOS \citep{2013MsngrKMOS} and SINFONI \citep{2003SPIE_SINFONI,2004MsngrSINFONI} at the Very Large Telescope (VLT). 

{At those redshifts, neutral and ionised gas outflows are mostly probed using rest-frame UV and optical interstellar blue-shifted absorption (e.g. \citealp{Pettini2002, Shapley2003, Weiner2009, steidel_structure_2010,Kornei2012, Martin2012, cimatti_active_2013, Erb2012, Rubin2014, talia_agn-enhanced_2017} ) 
and broad high-velocity nebular emission (e.g. \citealp{genzel_sins_2011, genzel_evidence_2014, newman_sinszc-sinf_2012, Schreiber2014,schreiber_kmos_2019, carniani_ionised_2015, Brusa2015, Brusa2016, harrison_kmos_2016, davies_kiloparsec_2019, freeman_mosdef_2019, swinbank_energetics_2019, Kakkad2020}, see also \citealp{forster_schreiber_star-forming_2020} and \citealp{veilleux_cool_2020} for comprehensive overviews).} 
Absorption lines are sensitive to the entire gas located along the line of sight, able to trace even low gas densities, and therefore could conceivably probe the material ejected over long timescales (\citealp{forster_schreiber_star-forming_2020}). However, estimates of the outflowing mass and/or mass outflow rate based on the absorption features are complex due to their strong dependencies on 1) the chemical enrichment and metal depletion into dust grains of the outflowing material, which are fundamental to translate the observed column densities of metals into hydrogen column density (e.g. \citealp{chisholm_mass_2017}), 2) the geometry and distribution of the absorbing outflowing clouds relative to the stellar continuum light in the background, 3) the radiative effects and the resonant emission filling on the line profiles (\citealp{Prochaska2011, scarlata_semi-analytical_2015}), and 4) the stellar and static ISM absorption contribution (see \citealp{concas_two-faces:_2019}). %
{Moreover, these absorption lines can be contaminated by the gas of faint satellites around the galaxy or residual gas left over from galaxy interactions.} 

Optical rest-frame emission lines are able to trace denser outflowing gas providing an instantaneous snapshot of the ongoing ejective feedback, {therefore, in principle, they are less contaminated by tenuous gas around galaxies}. In the last years they have been used to detect outflows in statistical samples of `normal' massive star-forming galaxies (KROSS by \citealp{swinbank_energetics_2019}, KMOS$^{\text{3D}}$ by \citealp{genzel_evidence_2014, schreiber_kmos_2019}, SINS by \citealp{Schreiber2014}) as well as AGN-dominated systems (KASH by \citealp{harrison_kmos_2016}, SUPER by \citealp{Kakkad2020}).

Despite the success of the aforementioned studies in detecting gas outflows in large samples of galaxies at Cosmic Noon, they focused on 
fairly massive galaxies, with $\log(M_\star/M_{\odot})$> 9-10, leaving the region of dwarf galaxies, where a strong ejective feedback is expected, mostly unexplored. Moreover, previous work was based almost exclusively on a single emission line tracer, mostly the brigh H$\alpha$, or [OIII]$\lambda5007$ emission in case of strong AGNs. Therefore, they have been subject to uncertainties and biases associated with the specific feature used. Indeed, the use of the H$\alpha$ line as outflow tracer is made more complicated by the proximity of the two nitrogen lines, [NII]$\lambda6548$ and [NII]$\lambda6584$, and by the fact that the high-velocity emission near the H$\alpha$ line could be confused with emission from the broad line region, even in galaxies with low-level AGN activity. {On the other hand, the [OIII] emission is 1) more affected by dust extinction, and 2) its flux depends on metallicity and ionisation conditions.} It follows that the simultaneous analysis of multiple emission lines is crucial to reduce all these systematic uncertainties. Moreover, the relative intensities of the rest-frame optical emission lines are essential to put strong constrains on the nature of the ionisation source (SF or AGN activity), gas density and dust content of the outflowing gas (see \citealp{Fluetsch2021} for an application on local galaxies). 

A major difficulty in the study of galactic outflows through emission line profiles is the separation between the emission associated with putative outflows from the emission coming from the gas rotating in the galactic disc. The most common, still controversial, method relies on the decomposition of the observed line into a narrow Gaussian component, which is supposed to trace the virial motions, and a broad component which instead is supposed to trace the outflowing gas. Although this technique has been extensively used, it does not necessarily provide meaningful information about the presence of outflows since the large scale rotational velocity and several observational effects (e.g. the spectral response of the instrument, beam smearing, inclination, etc) may result in to a shape of the line that is not Gaussian (and actually there is not physical reason why the profile should be Gaussian), potentially with broad wings. 

In the past decade, it has been demonstrated that the use of IFU observations might help to minimise the effect of the large velocity gradients by shifting the spectrum of each spaxel according with the observed velocity field (e.g. \citealp{genzel_sins_2011, genzel_evidence_2014, schreiber_kmos_2019, swinbank_energetics_2019, davies_kiloparsec_2019, avery_incidence_2021}). However, it is known that this technique has some weaknesses in the case of unresolved and/or undetermined velocity gradients, mostly due to the `infamous' beam smearing effect, with the appearance of an artificial broad component even without outflow (see \citealp{genzel_evidence_2014}).

In this paper, we aim to overcome all these problems by investigating the incidence and properties of ionised galactic outflows in a sample of 141 star-forming high-redshift ($1.2 <z< 2.6$) galaxies drawn from the KMOS (K-band Multi-Object Spectrograph) LEnsed galaxies Velocity and Emission line Review survey (KLEVER; \citealp{curti_klever_2020, HaydenPawson2021}). 
KLEVER is an ESO Large Programme, (PI: Cirasuolo), aimed at spatially
mapping and resolving the key rest-frame optical nebular diagnostics (from [OII]$\lambda$3727 to [SIII]$\lambda$9530) combining the high multiplexing and integral field units (IFU) capabilities of KMOS (\citealp{kmos2004,2013MsngrKMOS}) on the VLT together with observations in multiple bands (YJ, H, and K). 

We investigate the evidence for ionised galactic outflows through several emission line tracers (H$\beta$, [OIII], H$\alpha$ and [NII] emission) therefore minimising the uncertainty associated with using a single line, and by exploiting line ratios (e.g. [NII]/H$\alpha$ and [OIII]/H$\beta$) to put strong constrains on the main ionisation mechanism  {of} the outflow. Most importantly, we adopt a new physically-motivated method to identify the ionised galactic outflows based on the comparison between the observed emission lines and the prediction of a simple rotating disc model convolved with all the observational effects. This sophisticated technique will allow us to disentangle the emission of the outflowing material from the emission coming from the galactic disc taking into account the large-scale velocity rotations as well as the instrumental resolution and beam smearing effects.

In Section 2, we describe the KLEVER survey, the galaxy sample and observations; in Section 3 we discuss the main limitations of current techniques. In Section 4, we present a new method to identify non-circular motions in the high-velocity tails of the most intense optical rest-frame lines: [OIII]$\lambda$5007, H$\alpha$ and [NII]. The main results and discussion are presented in Section 5. Finally, we summarise our findings in Section 6 and highlight the importance of 1) extending this type of multi-tracer spectroscopic analysis to a much larger sample and 2) providing deep multi-band followups (e.g with ALMA), in order to build a comprehensive picture of the baryon cycle
during the peak of the cosmic star formation history.

Throughout this paper, where required we have assumed a $\Lambda$CDM cosmology with $H_{0}=70$ km s$^{-1}$ Mpc$^{-1}$, $\Omega_{M}=0.3,$ and $\Omega_{\Lambda}=0.7$ and a \cite{Chabrier2003} initial mass function (IMF). 1 arcsec corresponds to $8.4$ kpc at $z =1.4$ and $8.3$ kpc at $z =2.2$. 

\section{The KLEVER program}
KLEVER (KMOS Lensed Emission Lines and VElocity Review) is an ESO large program (197.A-0717, PI: Michele Cirasuolo) aimed at determining the spatially-resolved properties, kinematics and dynamics of the ionised gas in a sample of $\sim 200$ typical star-forming galaxies in the redshift range $1.2-2.6$. 
The goal of our KMOS observations is to provide a full coverage of the near-infrared region of the spectrum by observing each galaxy in the YJ, H, and K bands, hence providing information on most of the brightest rest-frame optical emission lines: [OII]$\lambda\lambda3727,29$, H$\beta$, [OIII$\lambda4959$, [OIII]$\lambda5007$, H$\alpha$, [NII]$\lambda6584$ and [SII]$\lambda\lambda6717,31$. 

In this paper, we present the results related to the galaxy-integrated emission-line profiles of star-forming main sequence galaxies at $1.2<z<2.6$. The spatially resolved properties of the KLEVER survey will be discussed in future papers.

\subsection{Observations and data reduction}
   \begin{figure*}
   \centering
   \includegraphics[angle=0,width=\hsize]{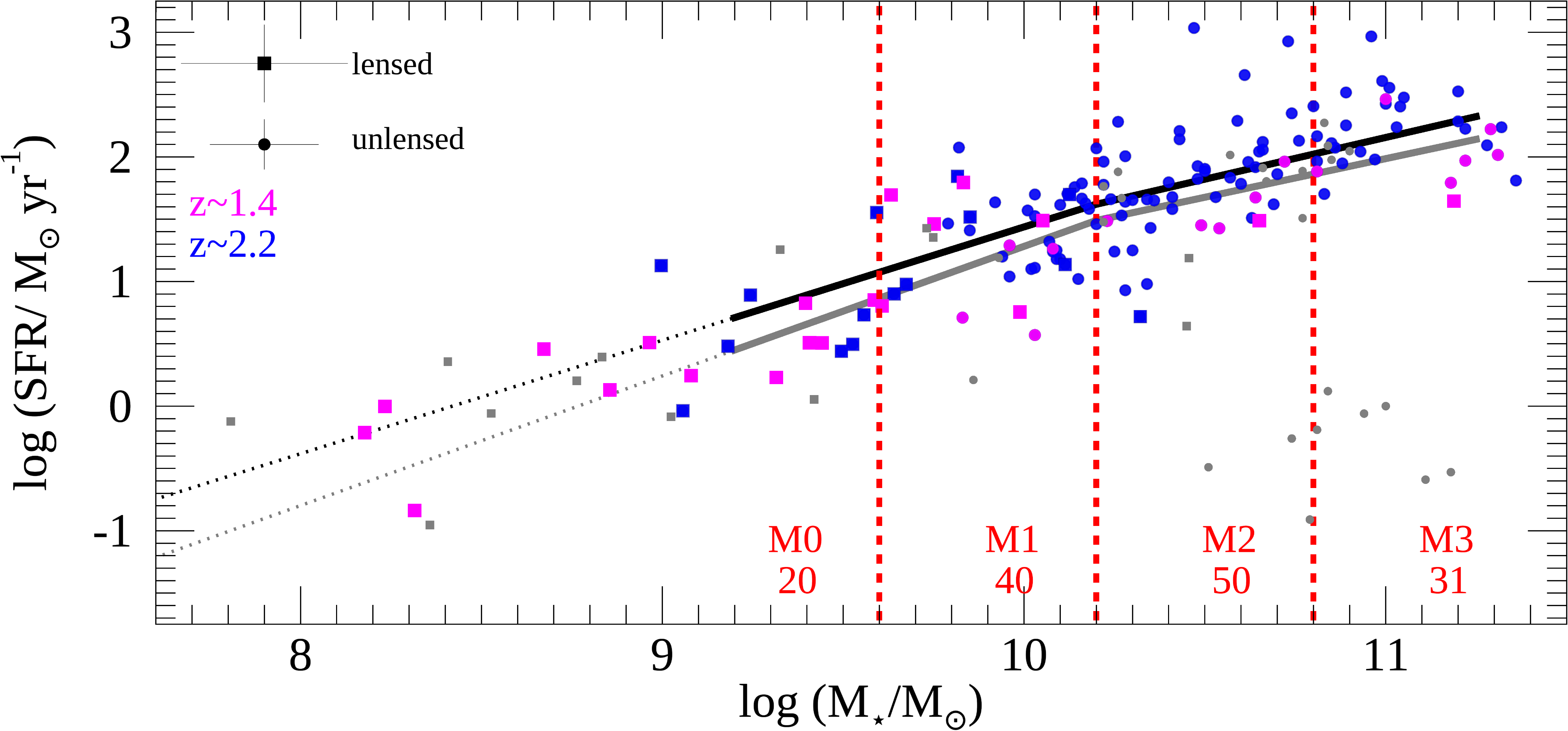}
   \caption[caption.]
  {{Distribution of the full KLEVER sample in the SFR-M$\star$ plane. 141 galaxies with high-quality spectra used in this work are indicated with magenta and blue colours, according with their redshift (respectively for $z\sim 1.4$ and $z\sim 2.2$). For reference, grey symbols mark the remaining KLEVER galaxies. Squares and circles represent lensed and unlensed objects. Average error bars are shown in the top left corner. The solid grey and black lines indicate the main sequence of star-forming galaxies at $z\sim 1.4$ and $z\sim 2.2$ from \cite{whitaker_constraining_2014}. The dotted lines are the extrapolated MS below M$\star = 10^{9.2}$ M$\sun$. The red dashed lines represent the $4$ stellar mass bins used for the stacking.  
  }}
              \label{SFR_Mstar}%
   \end{figure*}

KLEVER observations were performed with the K-band Multi Object integral field spectrograph at VLT (KMOS, \citealp{2013MsngrKMOS}), for a total of $216$ h. 
KMOS has $24$ integral-field units (IFUs) that operate simultaneously and can be independently positioned inside a circular field of $7.2$ arcmin in diameter. Each IFU has a square field of view of $2.8 \times 2.8$ arcsec with a uniform spatial sampling of $0.2\times 0.2$ arcsec. 
Our observations are obtained in the YJ, H and K gratings, yielding a spectral resolution $R=3582, 4045$ and $4227$ (median FWHM=$80$kms$^{-1}$) over the spectral ranges of $1.016-1.353$, $1.445-1.859$ and $1.934-2.473 \upmu$m, respectively.
The KMOS observations were taken during {ESO Periods $97-102$}.
Complete descriptions of the KLEVER survey, sample selection, observations and the data reduction are provided in Hayden-Pawson et al. (in preparation). In this section we report a brief summary of observations and data reduction.

The KLEVER sample consists of {$60$} gravitationally lensed galaxies in well-studied cluster fields, specifically the HST Frontiers Fields \citep{Lotz2017} and the HST-CLASH sample \citep{Postman2012} programs, as well as $132$ more massive unlensed galaxies in the southern CANDELS fields: Ultra Deep Survey (UDS), Cosmological Evolution Survey (COSMOS), and the Great Observatories Origins Deep Survey (GOODS-S). The combination of bright unlensed targets together with the fainter lensed sample enables us to explore an exceptionally wide range of stellar masses, specifically $\log(M_\star/M_{\odot}) \in [8.1-11.4]$.

The observations of lensed objects were carried out in service mode during Periods 97-102, from  April 2016 to October 2018. The integration times ranged from 3 h in YJ and H band to 4-5 h in K band on source. The targets were selected using the spectroscopic redshifts provided by: the CLASH-VLT survey  (\citealp{MsngrRosati2014}, see also \citealp{Balestra2016}) conducted with VIsible Multi-Object Spectrograph (VIMOS) on the VLT (ESO VLT Large program 186.A-0798. PI: P. Rosati), the Grism Lens-Amplified Survey from Space (GLASS, \citealp{Treu2015}, see also \citealp{Karman2015} and \citealp{Caminha2016}) and the Multi Unit Spectroscopic Explorer (MUSE) observations of MACS J1149.5+2223 at VLT (prog.ID 294.A-5032, PI:C. Grillo, \citealp{Grillo2016}). In particular, we selected targets in two redshifts ranges: 
\begin{enumerate}
    \item 1.2<z<1.65, to observe H$\beta$ + [OIII]$\lambda4959$ + [OIII]$\lambda5007$ in the YJ band, H$\alpha$+[NII]$\lambda6584$+ [SII]$\lambda\lambda6717,31$ in the H band, and [SIII]$\lambda\lambda9068,9530$ in the K band;
    
    \item 2<z<2.6 to have [OII]$\lambda\lambda3727,29$ in the YJ band, H$\beta$+[OIII$\lambda4959$+[OIII]$\lambda5007$ in the H band, and H$\alpha$+[NII]$\lambda6584$+[SII]$\lambda\lambda6717,31$ in the K band.
\end{enumerate} 

The second half of the KLEVER sample consist of $132$ unlensed galaxies selected from the KMOS$^{\text{3D}}$ Survey \citep{Wisnioski2015,wisnioski_kmos^3d_2019}, for which the H$\alpha$+[NII] spectra, obtained with the K-band observation of KMOS, were publicly available at the time of the observations. For these brighter targets we observe the missing bands (YJ and H) to obtain full wavelength coverage as for the fainter lensed sample. The observations of the unlesed objects were carried out in service mode during Periods 97-102. The integration times, in this case, range from 4 to 6 h on source in YJ and H band and $\sim$9 h in K band observations taken from KMOS$^{\text{3D}}$. 

In this paper, we present preliminary results based on $35$ lensed galaxies members of the MACS$1149$, MACS$0416$ and RXJ$2248$ clusters and $106$ more massive unlensed galaxies for a total of $141$ {galaxies with high-quality (SNR>5) H$\alpha$ emission line (see Section \ref{1Dspectra}).}

For all our observations, we adopted an A-B-A nodding with dithering strategy for sky sampling and subtraction. 
The KMOS data was reduced using ESO-KMOS pipeline (version 2.6.6). Sky subtraction within the pipeline was enhanced using the SKYTWEAK technique \citep{Davies2007}, which greatly reduces the sky-line residuals. In each OB, three IFUs, one for each KMOS detector, were dedicated to observing bright stars. The average seeing of the observations was then determined by fitting collapsed images of these stars with an elliptical Gaussian. Typical values for the seeing in our observations range from FWHM$=$0.5 to 0.6 arcseconds. Over the course of each OB, it is typical for the telescope to drift from its acquired position. Therefore, to properly align and combine individual exposures, both within a single OB and across different OBs, the centroid positions of the observed stars were measured in each exposure. Shifts between different exposures were then calculated and applied to the scientific sources observed on the same detector of the corresponding reference star. All exposures for each galaxy were then stacked using a 3-$\sigma$ clipping to produce the final data cubes. Individual exposures with large seeing values (>0.8 arcseconds), or unusually large shifts in centroid position of the observed star were excluded from this final stacking. Finally, we rebinned the cubes on to a 0.1 arcsec pixel scale. Full details of the observations and data reduction will be presented in Hayden-Pawson et al. (in preparation).

\subsection{Galaxy-integrated spectra}\label{1Dspectra}
Thanks to the multi-band coverage offered by KLEVER, we can extract the galaxy-integrated spectrum for each galaxy at different wavelengths. In this study we will focus on the brightest emission lines of the rest-frame optical {spectrum: H$\beta$, [OIII]$\lambda5007$, H$\alpha$, [NII] and [SII].} 
To obtain the galaxy-integrated spectra, we summed the spectra from all of the pixels within a circular aperture of diameter of 1.3 arcseconds (corresponding to a projected diameter of $\sim11$ kpc for sources at z$\sim1.4-2.2 $). {Such an aperture provides a good balance between sampling the majority of the flux within the galaxy and assuring a low noise level in the galaxy-integrated spectra. We test the stability of the results by applying a smaller aperture (i.e., with a diameter of 0.6 arcseconds) on low mass galaxies (stellar mass $\log(M_\star/M_{\odot})<9.6$)
and find no significant differences.} In each band, the aperture was centred at the peak of the continuum emission or to the peak of the strongest emission line ([OIII]$\lambda5007$ or H$\alpha$) if the continuum was not detected. If neither clear emission lines nor continuum emission was seen, then the aperture was placed at the nominal position of the galaxy (based on its optical position) which is located at the spatial centre of the cube.

{Of the full near-IR spectroscopic sample ($60$ lensed and $135$ unlensed galaxies), we select 143 objects (35 lensed and 108 unlensed) that have the high-quality and signal-to-noise ratio, $S/N >5$ in the H$\alpha$ line and do not have strong contamination by atmospheric OH sky emission around the [OIII]$\lambda5007$ and H$\alpha$ emission.} {[OIII]$\lambda5007$ is detected with SNR>3 for $80\%$ of the sample, H$\beta$ in $66\%$, [NII] in $65\%$, and [SII] in $40\%$.} 

{The presence of galaxy interaction and/or mergers affects the distribution of the gas in the galaxies, causing perturbations of emission lines shape which are not connected with the presence of outflowing material driven by AGN or SF activity. As a consequence, galaxies undergoing a mergers event must be carefully avoided. In the rest of the paper we then exclude COS$4\_06963$ and COS$4\_11363$ as they are already known to be part of galaxy interactions (see \citealp{genzel_evidence_2014}).} 

{The final sample of KLEVER galaxies analysed within this paper then consists of 141 galaxies, 35 of which are lensed and 106 unlensed.}

\subsection{Global galaxy properties: M$\star$, SFR and A$_{v}$}\label{global_prop}

Since our targets lie in the most well-studied galaxy clusters and in the deep extragalactic survey field, a variety of archival photometry and derived quantities for the physical properties exist. 
In particular, global properties, such as SFR, M$\star$ and A$_{v}$ were derived through the broad-band spectral fitting technique using spectral templates derived from the \cite{BruzualCharlot2003} evolutionary code assuming a \cite{Chabrier2003} IMF. Details of the derivations for the lensed and unlensed galaxies are given in the references below. We note that the methods and model assumptions were similar for the two sub-samples. All SFR and M$\star$ measurements are converted to a \cite{Chabrier2003} IMF, when necessary, ensuring consistency for the present study.
 The final sample consists of the following two subsets.
\begin{itemize}
    \item $106$ galaxies in CANDELS fields. The SFR, M$\star$ and A$_{v}$ are taken from the KMOS$^{\text{3D}}$ Data Release\footnote{https://www.mpe.mpg.de/ir/KMOS3D/data} (\citealp{wisnioski_kmos^3d_2019}). They are 
    obtained with the FAST \citep{Kriek2009} code assuming: solar metallicity, \cite{calzetti_dust_2000}  reddening law, and either constant or exponentially declining star formation histories (SFHs). Star-formation rates are determined from the same SED fits or, for objects observed in at least one of the mid- to far-IR bands with the Spitzer/MIPS and Herschel/PACS instruments, from rest-UV+IR luminosities through the Herschel-calibrated ladder of SFR indicators of \cite{Wuyts2011}. {As reported by \cite{SWuyts2016}, the typical uncertainty associated with the parameters are: 0.15 dex for M$\star$, 0.25 dex for SFRs inferred from SED fitting and UV + MIPS 24$\mu m$ photometry and 0.1 dex for SFR from UV+Herschel photometry.}
    \item $35$ lensed galaxies. Following the same procedure adopted in \cite{curti_klever_2020}, we estimate the SFR, M$\star$ and A$_{v}$ using the high-z extension of the MAGPHYS program \citep{daCunha2008}. MAGPHYS adopts the two-component model of \cite{charlot_simple_2000} to describe the attenuation of stellar emission at ultraviolet, optical, and near-infrared wavelengths. We used the deep Frontier Field HST images\footnote{https://irsa.ipac.caltech.edu/data/SPITZER/Frontier/} \citep{Lotz2017} in the three optical bands F435W, F606W, F814W, and in the four NIR bands: F105W, F125W, F140W, F160W, plus the IRAC 3.6 and 4.5 data acquired by Spitzer \citep{castellano_astrodeep_2016,di_criscienzo_astrodeep_2017,bradac_hubble_2019} to perform the SED fitting and derive the M$\star$, SFR and A$_{v}$. Finally, the intrinsic SFR, M$\star$ and A$_{v}$ are obtained by correcting the estimates derived through SED-fitting, considering the median magnification factor \footnote{The median magnification factor has been calculated using the distribution of different magnification values obtained from different lensing models provided at https$://$archive.stsci.edu/prepds/frontier/lensmodels/} for each source. The uncertainties on stellar masses and SFRs are derived from the 1$\sigma$ interval of the resulting likelihood distribution and include the contribution from statistical uncertainties on the magnification, but they do not account for systematic uncertainties on the lensing model. {The M$\star$ and SFR uncertainties vary across the sample, being higher for the strongly lensed systems and spanning a range of [0.1,0.94] and [0.11,0.93], respectively. The average uncertainties are: 0.23 dex for M$\star$ and 0.31 dex for SFR.}
\end{itemize}

The combination of bright unlensed targets together with the fainter lensed sample enables us to explore an exceptional wide range of stellar masses $\log(M_\star/M_{\odot})\in[8.1-11.4]$ in two redshift ranges z$\sim 1.4$ and $2.2$.
The distribution of the final sample in SFR versus stellar mass is shown in Figure \ref{SFR_Mstar}.
{The vast majority of our galaxies lies on the main sequence of star-forming galaxies (MS), with  $64\%$ of them within 1$\sigma$ (0.34 dex as reported by \citealp{Whitaker2012}) and $91\%$ of them within 2$\sigma$ of the main sequence as defined by \cite{whitaker_constraining_2014}.}

{\section{Searching for galactic outflows: Previous methods}}\label{outflows_methods}%

   \begin{figure*} 
   \centering
   \includegraphics[angle=0,width=\hsize]{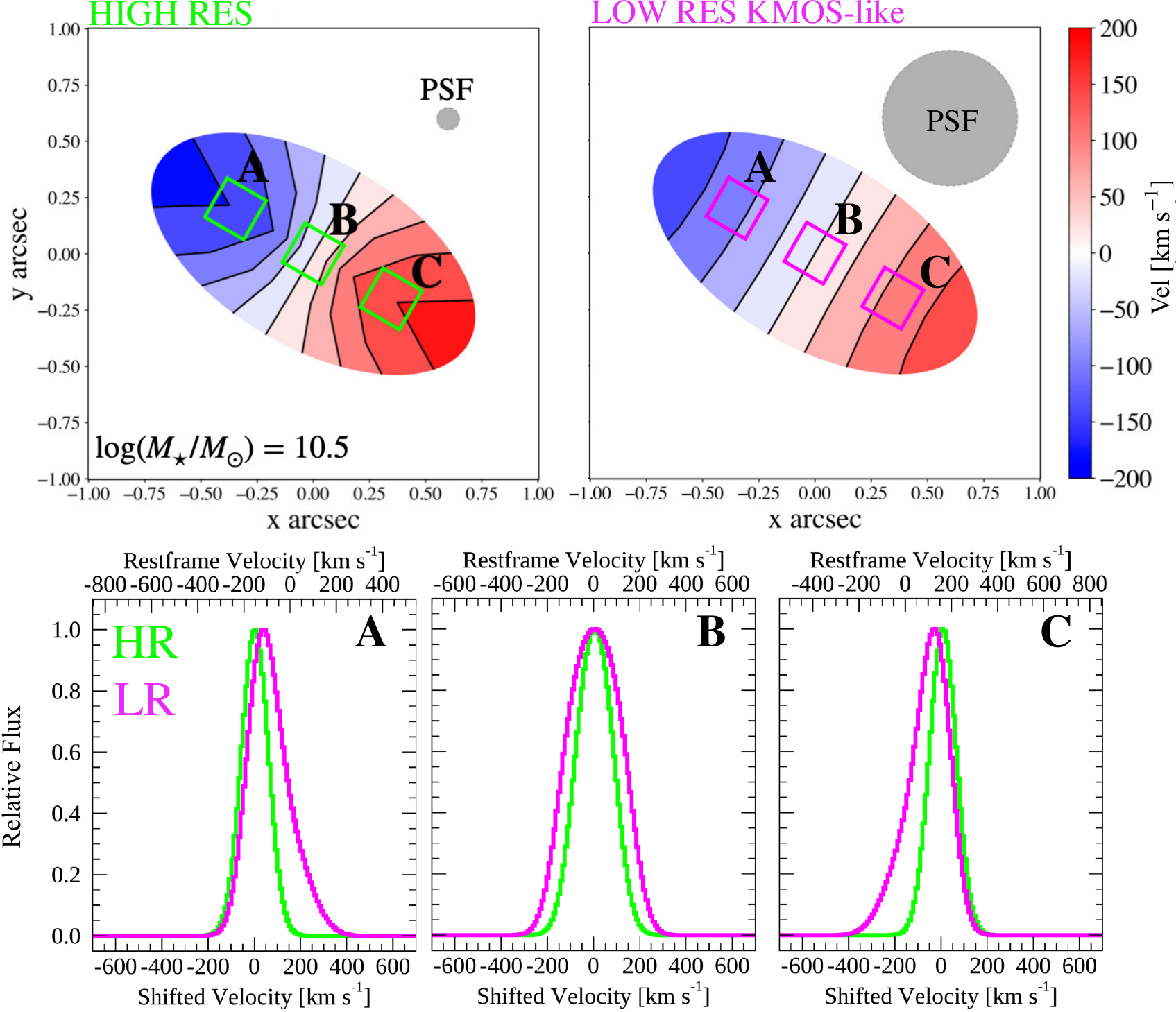}
   \caption[caption.]
  {{Beam smearing effect in the "velocity-subtracted" method} (\citealp{shapiro_sins_2009,genzel_sins_2011,genzel_evidence_2014,schreiber_kmos_2019,davies_kiloparsec_2019,avery_incidence_2021,swinbank_energetics_2019}) applied to a rotating disc model with stellar mass of $\log(M_\star/M_{\odot})=10.5$ (without central bulge) and an effective radius of $R_{e,\star}= 3$ kpc (see Section \ref{vel_sub_method} and \ref{mock} for more details). We show the comparison between moment 1 maps (\textit{top panels}) and single pixel spectra (\textit{middle panels}) for a rotating disc model of ionised gas (e.g. H$\alpha$), observed at at high-z with high and low (KMOS-like) spatial resolution (respectively, HR and LR) with a PSF of FWHM=0.1 and 0.6 arcsec. Note the effect of the beam smearing: 1) the velocity gradients of the HR moment 1 map (top left panel) are smoothed and disappears in the LR map (top right panel); 2) in the single pixels, the symmetric, Gaussian line profiles observed in the LR case (green lines in A, B and C bottom panels) turn into broad and (except for the central pixel, B), asymmetric LR profiles (magenta curves in A and C bottom panels) characterised by strong wings as they are contaminated by the flux belonging to adjacent regions. In the bottom panels, the restframe velocity is shown in the top axis and the shifted velocity, obtained after the application of the velocity-shifted method, in the bottom axis. 
  }
              \label{mock_HR_LR_comparison}%
   \end{figure*}

   \begin{figure*}
   \centering
   \includegraphics[angle=90,width=\hsize]{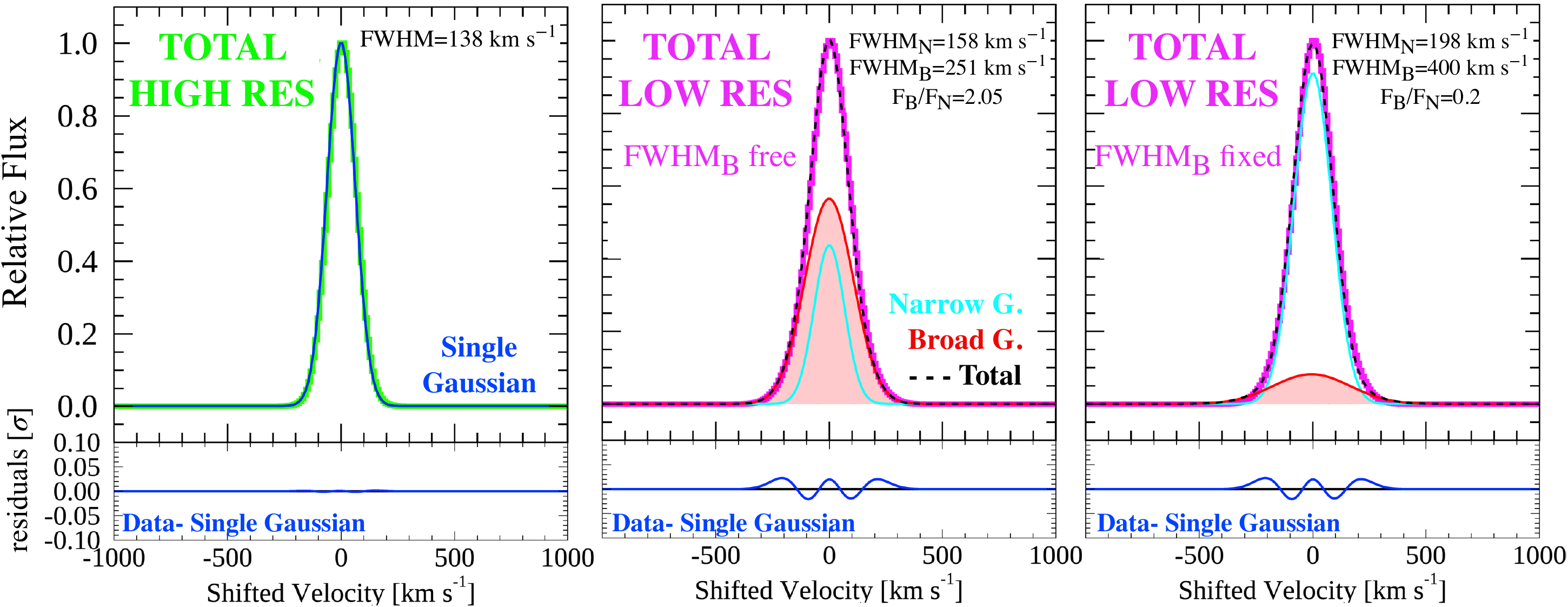}
   \caption[caption.]
  {{Example of artificial outflows detection (red area) due to beam smearing effect in the galaxy-integrated emission line during the application of the velocity-subtracted method applied to the rotating disc models presented in Fig. \ref{mock_HR_LR_comparison}. The 
 high-resolution (HR) velocity-shifted spectrum (green curve in the left panel) well reproduced by a single Gaussian component (blue line) becomes broader and contaminated by high-velocity wings in the low resolution (LR) case (magenta curves in the middle and right panels), as demonstrated by the strong residuals (data- Single Gaussian fit in the bottom panels). An artificial broad component (red area) emerges when the LR case (magenta line) is fitted with a narrow (cyan line) and broad component (red line). The flux enclosed in the broad component (F$_{\text{B}}$) is sensitive to the priors adopted in the fit as can be appreciated by comparing middle with right panel where priors of \cite{swinbank_energetics_2019} (free FWHM$_{\text{B}}$) and \cite{schreiber_kmos_2019} (fixed FWHM$_{\text{B}} \geq 400$ km s$^{-1}$) are respectively adopted.}

   }
              \label{mock_HR_LR_KROSS_KMOS3Dlike}%
   \end{figure*}


As already mentioned, the most common method used to identify galactic outflows in galaxy-integrated spectra is the decomposition of the observed emission-line profiles with a narrow and broad Gaussian component, which are supposed to trace, respectively, virial motions within the galaxy and the outflowing gas. Although this technique is very useful for quantifying {the global flux underling the emission line}, it does not necessarily provide meaningful information about the presence of outflows, as even regular virial motions in the galaxy (i.e. not associated with outflow) can in principle result in a line of sight velocity profile distribution that is not a single Gaussian {(see some examples of line profiles expected for a simple rotating disc model reported in Appendix \ref{1D_not_gaussian}).} 
It is therefore not obvious whether the narrow and broad Gaussian components can be unambiguously associated to rotating disc and non-circular motions, respectively, since the large scale rotational velocity and {several observational effects like the beam smearing effect and/or the spectral response of the instrument may alter the shape of the line.} {While the latter could be taken into account during the study of the line profiles, as done for SINFONI data by \cite{schreiber_sinszc-sinf_2018} (see in particular their Appendix B for an example of non-Gaussian line spread function), the first two effects (i.e. large scale rotational velocity and beam smearing) are more difficult to disentangle as we will discuss more in detail in this section.}
\\
\subsection{The velocity-subtracted method}\label{vel_sub_method}
{A common technique used to minimise the effect of the line broadening due to circular motions is the removal of the large-scale velocity gradient from each data cube using the emission line velocity field (e.g. \citealp{shapiro_sins_2009,genzel_sins_2011,genzel_evidence_2014,schreiber_kmos_2019,davies_kiloparsec_2019,avery_incidence_2021,swinbank_energetics_2019}). Throughout the rest of the paper, we will refer to this approach as "velocity-subtracted" method. It consists of the following steps. The spectrum of each pixel in the observed data cube is fitted, in proximity of the H$\alpha$ line, with a single Gaussian profile to extract the velocity map (or moment 1) of the galaxy. Then, the derived velocity field is applied in reverse to the observed data cubes by shifting the spectrum of each pixel accordingly with the measured velocity (i.e. the peak of the emission line will corresponds to the zero velocity). This procedure will creates the "velocity-subtracted" datacube which is subsequently used to extract the one dimensional spectrum and search for signatures of outflowing gas. In particular, the spectrum is decomposed in two Gaussian components: the narrow component associated with the emission from the gas located in the galactic disc and the broad component which is interpreted as evidence of outflowing material. }


{As shown by \cite{swinbank_energetics_2019} this technique definitely reduces the overall broadening of the galaxy-integrated emission line (see their Fig. 1.) but it is known to have some limitations in compact and/or unresolved sources (\citealp{genzel_evidence_2014}). Using a mock rotating disc model tailored to KMOS observations (i.e. PSF=$0.6$ arcsec, pixel size of $0.2$ arcsec and spectral resolution of $\delta\lambda =33$ km s $^{-1}$) and presented in Section \ref{mock}, we tested the velocity-subtracted method finding that the combination of KMOS-like low spatial resolution and unresolved velocity gradients{, also called beam-smearing effect,} can generate a large amount of flux at very high velocity which can easily be mistaken for outflowing gas.} 

{ To fully appreciate the effect of the beam smearing in the velocity-subtracted method we perform the following test. We simulate 
 mock H$\alpha$ observations of a rotating galactic disc for a galaxy with $\log(M_\star/M_{\odot})=10.5$,
effective radius of $R_{e,\star}= 3$ kpc (note that this is the typical size of a star-forming galaxy observed at z=2 according with \citealp{van_der_wel_3d-hstcandels_2014}), without a bulge component, with inclination of $60$ deg. This model is spatially degraded to simulate the effects of observations at high and low (KMOS-like) spatial resolution (HR and LR respectively), assuming a PSF with FWHM of 0.1 and 0.6 arcsec.} Following the steps of the velocity-subtracted method we first calculate the moment 1 maps for the HR  and LR case, reported respectively, in the top left and right panels of Fig. \ref{mock_HR_LR_comparison}). 
{As expected, the beam smearing reduces the gradients in the velocity fields as it can be fully appreciated by comparing the HR map (top left panel) with the smeared LR (top right panel, see Section 3.3 of \citealp{diteodoro_3d_2015} for several examples of beam smearing effects on velocity gradients observed in nearby galaxies). In the lower panels of Fig. \ref{mock_HR_LR_comparison}, we present three examples of emission line profiles extracted in  three different spaxels, on the approaching side of the disc, at the center and on the receding side (A, B and C in the figure) for the HR (green) and LR (magenta) case. Note that we report the original restframe velocity in the top axis (where we can appreciate the rotation of our disc-model) and the corresponding shifted velocity after the application of the velocity-subtracted method in the bottom axis. The LR line profiles (magenta lines) appear to be broader compared to the HR ones (green lines), they are non-Gaussian and, except for the central spaxel (B), the profiles are not symmetrical, showing clear wings (see A and C middle panels) as they are contaminated with flux belonging to adjacent regions due to the coarse resolution.} 

We complete our test by extracting the galaxy integrated total spectrum from the velocity-shifted HR and LR data cubes. The results of the HR and LR spectrum are shown in the left, middle and right panel of Fig. \ref{mock_HR_LR_KROSS_KMOS3Dlike}.
We find that, in the HR case (green curve reported in the left panel), the velocity-shifted galaxy integrated total spectrum (green line in the bottom left panel) can be reproduced by a single Gaussian component (blue curve) as also indicated by the zero-level residuals (data- single Gaussian in the lower panel). The situation is different for the LR case (magenta curve in the middle and right panels), where a single Gaussian fit is not adequate to reproduce the expected emission line as demonstrated by the strong residuals (data- single Gaussian, blue line) shown in the lower middle and right panels. The LR velocity-shifted line emission is better reproduced by a combination (black dashed line) of a narrow Gaussian component (cyan) and a broad component (red area). As demonstrated by this simple test, the beam smearing effect can affect the velocity-shifted method with a resulting artificial broad component even in a simulated rotating disc model without any outflowing flux.  

{The demarcation between narrow and broad Gaussian component differs in different works in a rather arbitrary fashion. Some authors restrict the line width of the broad and narrow component assuming FWHM$_B \geq 380$ km s $^{-1}$ and FWHM$_N < 380$ km s $^{-1}$ \citep{genzel_evidence_2014} or FWHM$_B \geq 400$ km s $^{-1}$ ({{in some cases fixed to FWHM$_B = 400$ km s $^{-1}$}}) and FWHM$_N < 400$ km s $^{-1}$ \citep{schreiber_kmos_2019} or FWHM$_B\geq 353$ km s $^{-1}$ and FWHM$_N < 353$ km s $^{-1}$ \citep{davies_kiloparsec_2019}. \cite{avery_incidence_2021} require that $\sigma_B>\sigma_N +50$km/s$^{-1}$.  In other cases, the fit is allow to vary without any restriction (see \citealp{swinbank_energetics_2019}). {As an example, in the middle and right panel of Fig. \ref{mock_HR_LR_KROSS_KMOS3Dlike} we show the different fit and associated quantities (especially FWHM and relative fluxes) obtained assuming two different prescription: 1) no restriction, as adopted by \cite{swinbank_energetics_2019}, in the middle panel and 2) FWHM$_B \geq 400$ km s $^{-1}$ and FWHM$_N < 400$ km s $^{-1}$ as proposed by \citep{schreiber_kmos_2019}.} If we let the fit free to vary without any restriction the flux of broad component strongly increases reaching values that could be even higher than the narrow component. It is clear that different assumptions in the fit will have an impact in the obtained line parameters (FWHM and flux of the narrow and broad components) and in particular in the flux associated to the broad or artificial "outflow" component.}


{Finally, we find that the contribution of the broad component to the total line profile could be even larger if a bulge is included in our simulation. We will show an example of such an effect in Appendix \ref{vel_sub_residuals} and explore the variation of the artificial broad flux as a function of galaxy properties, observational effects (inclination, spatial and spectral resolution) and priors adopted in the Gaussian fit in a forthcoming paper (Concas et al. in prep). }
{We urge the reader to take into account this "spurious contamination" when the properties of the outflowing gas are estimated using the "velocity-subtracted" method as they could provide an overestimation of the detection rates and mass of the outflowing gas.}

{\section{SEARCHING FOR GALACTIC OUTFLOWS: 
The new method}}\label{newmethod}

{In this paper we propose a new physically motivated method to disentangle the emission of the disc (or disc+bulge for the most massive galaxies) from the outflow. We adopt a 
novel strategy that relies on the comparison between the kinematics of the ionised gas and the prediction of a rotating disc (and/or disc+bulge) model.} Specifically, we compare our galaxy-integrated spectra 
with the expectations of a 3D rotating disc (or disc+bulge for the most massive galaxies) model tailored to the KLEVER sample. 
This has three major advantages: (1) we can directly compare observations with models without modifying the original data-cubes as in the "velocity-subtracted" method, 
(2) our disc (disc+bulge) (narrow) component is certainly tracing the flux associated with the virial motions and is not merely the results of a pure mathematical Gaussian fit, and (3) the large scale rotational velocity and the spatial and spectral resolution of our observations, which may alter the intrinsic shape of the emission line, are properly taken into account.
\\
{The fundamental steps of our new method are the following:
\begin{itemize}
\item We build galaxy-integrated emission line (i.e. H$\beta$, [OIII], H$\alpha$, [NII] and [SII]) templates of mock rotating disc and disc+bulge observations  
    as described in Section \ref{mock};
\item We fit each observed spectrum with its templates to find a best-fit rotating disc (disc+bulge) model for each galaxy in our sample (Section \ref{bestfit});
\item We stack the observed spectra and the corresponding best-fit templates in bins of stellar mass (using the procedure that will be presented in Section \ref{stacking}), obtaining four observed and four mock averaged stacked spectra;
\item Finally, we compare each observed averaged spectrum with the corresponding simulated stacked spectrum to search for possible deviations attributable to the presence of non-circular motions, like gaseous outflows (Section \ref{searching_outflows}).
\end{itemize}}

{Details of each step of our disc-decomposition method are provided in the following subsections. As the bulge component is only assumed for the most massive galaxies (more details are provided in the next section), in the following part of the paper we will simply refer to the method as disc-decomposition.}
\\

{The main assumption in this approach is that non-circular motions are only due to the overflowing gas, but we should keep in mind that they could also originate from other phenomena, such as the presence of bars, spiral arms, tidal disturbances and/or undetected galaxy interactions. For this reason, it is reasonable to think that our outflow fluxes, and consequently masses and mass outflow rates (discussed in Section \ref{results}), are probably overestimated.}



\subsection{Rotating disc and disc+bulge models}\label{mock}
We use the publicly available \large{KIN}\small{EMATIC} \large{M}\small{OLECULAR} \large{S}\small{IMULATION} routine (\large{K}\small{IN}\large{MS} \normalsize Python version \footnote{https://github.com/TimothyADavis/KinMSpy}; \citealp{davis_atlas3d_2013}) to produce mock observations of rotating galactic disc and assess the contribution of  circular motions on the shape of the observed emission lines ( H$\beta$, [OIII], H$\alpha$, [NII] and [SII]). This tool produces a simulated data cube that can be compared to the observed data taking into account the effects of beam-smearing, spectral and spatial resolution. It has been applied in various works to 1) simulate observations of molecular and atomic gas distributions \citep{davis_evolution_2019}, 2) investigate the kinematics of molecular gas in local early-type galaxies \citep{davis_atlas3d_2013}, low excitation radio galaxies \citep{ruffa_agn_2019} 3) atomic gas in high-redshift galaxies \citep{DeBreuck2014}, and 4) to estimate black hole masses (e.g. \citealp{NorthDavis2019}, \citealp{davis_revealing_2020}).

In this work, we take advantage of the great flexibility of the code to simulate the ionised gas distribution observed in the infrared spectrum, generating mock cubes for all of our 141 galaxies. {In particular, we 1) generate a 3D rotating disk model for each galaxy according to the physical parameters described below; 2) for each galaxy we simulate the integrated spectrum assuming 41 different inclinations in the range of i=[10-90] deg;  3) we fit each observed spectrum with its 41 templates to recover the overall shape of the line profile and find the best-fit rotating model. We end up with a best-fit rotating disc model associated to each of our galaxy spectra.}

We set up \large{K}\small{IN}\large{MS} \normalsize to simulate our KMOS observations, imposing an angular resolution with a point spread function PSF of FWHM=0.6 arcsec and pixel size of 0.1 arcsec. The cubes are created with very high spectral resolution, with  FWHM of $\delta \lambda =10$km/s, and then convolved with a Gaussian filter to reproduce the observed median KMOS spectral resolution of $\delta \lambda =33$km/s.

To create our mock observations {we assume a surface brightness profile and kinematic distribution using} various archival photometry and physical properties such as stellar masses, global structural parameters, and ionised gas kinematics already available for our galaxy sample or derived using empirical relations obtained for similar datasets {and local well studied galaxies. For each galaxy in our sample we define a 3D rotating disk model by using the observational parameters ($M_\star$, $R_{e,\star}$, n and z) and gas distribution and kinematics as described in the following section.}

{\subsubsection{Gas distribution}}
For our modelling, we assume that the ionised gas is distributed accordingly to a \cite{sersic1963} profile:
\begin{equation}
 I(r) = I_{0} \cdot \exp \left\{ -b_{n} \left[ \left( \frac{r}{R_{e, gas}}\right)^{(1/n)} -1 \right]    \right\} ,
\end{equation}
where n is the S\'{e}rsic index, and  R$_{e, gas}$ is the effective radius of the ionised gas. {Following the recent results of \cite{wilman_regulation_2020} obtained with the full KMOS$^{\text{3D}}$ sample, we translate R$_{e, gas}$ as a function of the effective radius of the stars, $R_{e,\star}$, assuming $R_{e, gas}=1.26 R_{e,\star}$ for massive galaxies, i.e. $\log(M_\star/M_{\odot}) > 9.6$, and $R_{e, gas}=R_{e,\star}$ for less massive systems (see also \citealp{nelson_where_2016, ubler_evolution_2019}). The surface brightness profile is then described with two observable parameters: $R_{e,\star}$ and $n$. For all the unlensed galaxies, we used the  n, and $R_{e,\star}$ values measured from H-band imaging by \cite{van_der_wel_3d-hstcandels_2014}.} For the remaining 35 lensed systems,
we assign an average values of $n=[1; 1.3; 2.; 2.5]$ for galaxies in the following stellar mass ranges $\log(M_\star/M_{\odot})=[\leq 10.; 10.-10.8; 10.8-11.; \geq11.]$ following the median n values presented by \cite{lang_bulge_2014}. The value of $R_{e,\star}$ is assigned using empirical $M_\star - R_{e,\star}$ relation provided by equation 3 of \cite{van_der_wel_3d-hstcandels_2014}. Note that for low-mass galaxies we adopt $n=1$ so the surface-brightness profile corresponds to an exponential disc.
\\


\begin{figure} 
\centering
\subfloat{\includegraphics[width=\hsize]{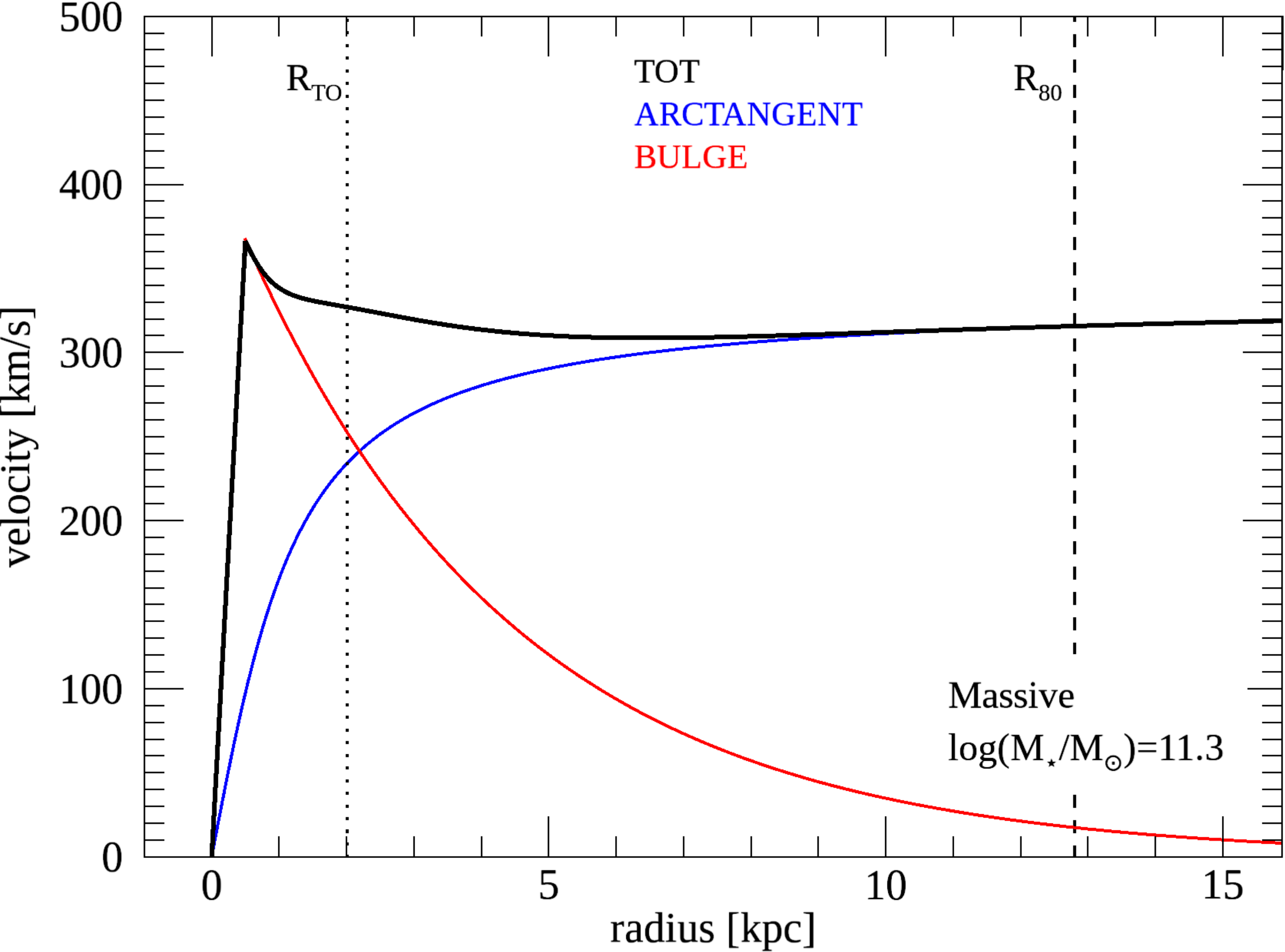}}
\vfill
\subfloat{\includegraphics[width=\hsize]{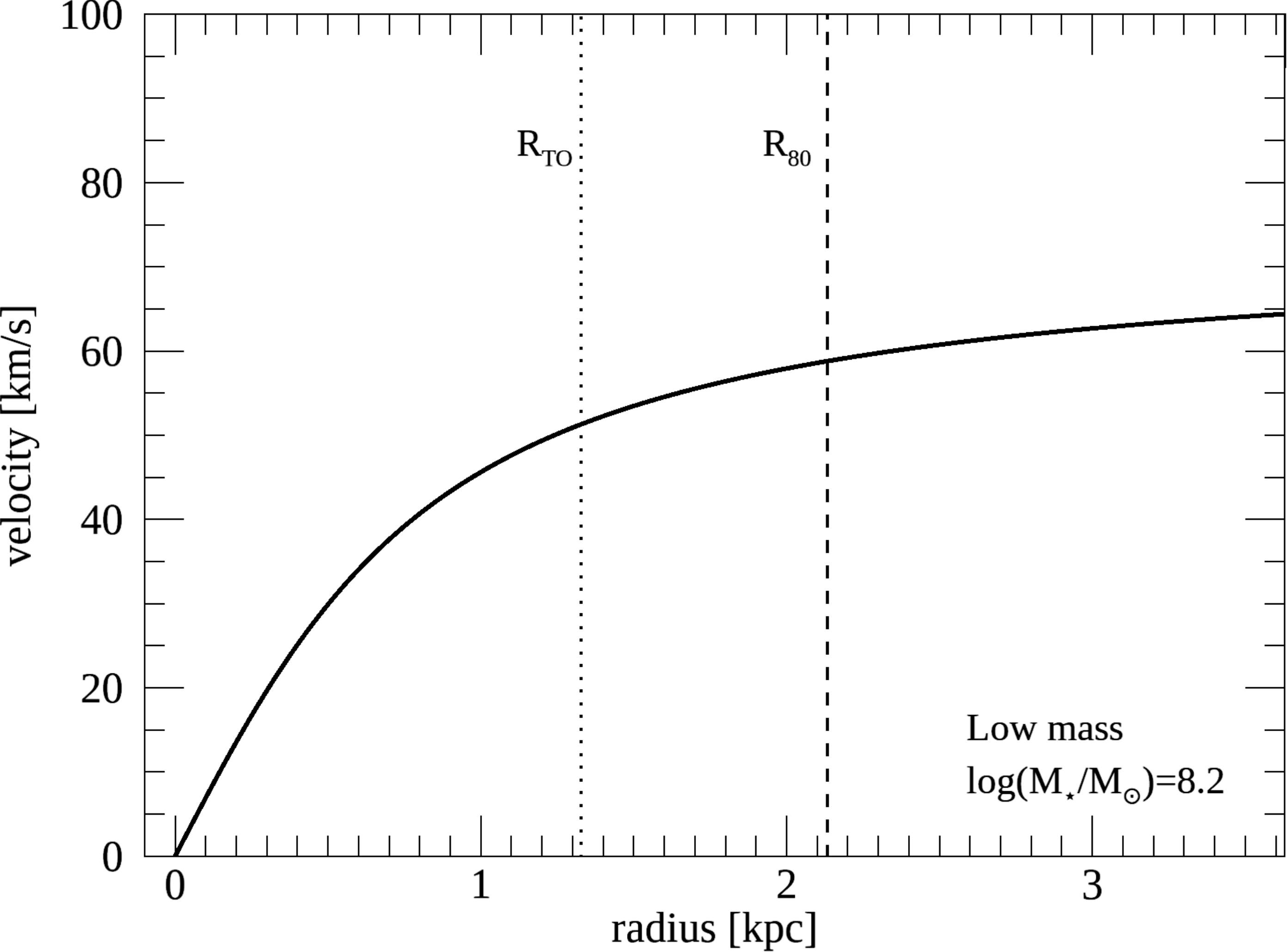}}
   \caption[caption.]{Example of rotation curves models adopted for massive ($\log(M_\star/M_{\odot})=11.3$) and dwarf ($\log(M_\star/M_{\odot})=8.3$) galaxies, top and bottom panel, respectively. As explained in the text, for low mass galaxies, below {$\log(M_\star/M_{\odot})<10.2$}, we assume a simple arc-tangent model which smoothly rises, reaching a maximum velocity asymptotically. For galaxies above {$\log(M_\star/M_{\odot})>10.2$}, where the bulge contribution is expected to be important (see \citealp{lang_bulge_2014}), we add a central component to the arc-tangent model to simulate the steep rotation curve in the centre with a central peak followed by a decline of $\sim 20\%$ typically observed on galaxies with concentrated light distributions and luminous bulges \citep{noordermeer_mass_2007}. In both cases the curves are normalised at $V_{80}$ using the empirical Tully-Fisher relation derived by \cite{teodoro_flat_2016}.}
                \label{vel_curves_profiles}%
\end{figure}

{\subsubsection{Gas kinematics}}
We assume that the ionised gas rotation curves follows a simple arc-tangent model \citep{Courteau1997} of the form:
\begin{equation}
V_{rot}(r)=\frac{2}{\pi} V_{MAX} \arctan{\left ( \frac{r} {R_{TO}} \right)} \quad ,
\end{equation}
which smoothly rises, reaching a maximum velocity $V_{MAX}$ asymptotically at an infinite radius. The turn-over radius $R_{TO}$ is the radius at which the rotation curve starts to flatten out. 
Using the values of $R_{TO}$ and $R_{e,\star}$ reported by \cite{reyes_calibrated_2011} for a sample of $189$ well studied local disc galaxies, we find that the $R_{TO}$ can be expressed as a function of the effective radius by the following relation: $R_{TO}=R_{e,\star}\times\left [(-0.232 \times \log(M_\star/M_{\odot}))+3. \right]$. The rotation velocities are commonly evaluated at some suitable optical radius. 
A common choice is the radius containing $80$ per cent of the total optical light, $R_{80}$. 
We convert our half-light radius $R_{e,\star}$ to $R_{80}$ using the $R_{80}/R_{50}-n$ relations of \cite{miller_new_2019}: $R_{80}/R_{e,\star} (n)= 0.0012n^{3}-0.0123n^{2} +0.5092n+1.2646$ obtained for galaxies in the 3D$/$CANDELS survey. We can then express asymptotic velocity, $V_{MAX}$ in terms of observable parameters: 
\begin{equation}
V_{MAX}= \frac{\pi \;V_{80} }{2 \arctan{ \left (R_{80}/R_{TO} \right)}}\quad ,
\end{equation}
where $V_{80}$ is the velocity measured at $R_{80}$ and is gauged from the Tully-Fisher relation \citep{TullyFisher1977} derived by \cite{teodoro_flat_2016} for a sample of $z\sim1$ galaxies:
 \begin{equation}
\log{(M\star/M_{\odot})}= 1.88 + 3.80 \log{(V)}\quad ,
\end{equation}
where we assume that $V_{80}=V$. {Specifically, the velocity, $V$, is derived by \cite{teodoro_flat_2016} by using the average circular velocity along the flat part of the rotation curve, $V_{\text{flat}}$. Our assumption of $V_{80}=V_{\text{flat}}$ is fully justified by studies on local galaxies where it has been shown that the two velocity definitions are comparable (\citealp{Hammer2007}), within a 20$\%$ variation (\citealp{lelli_baryonic_2019}), which is negligible for the purpose of this paper.}

As reported by \cite{noordermeer_mass_2007}, the shape of the central rotations curves depends on the concentration of the stellar light distribution and the bulge-to-disc ratio (B/T). Galaxies with high B/T reach the maximum of their rotation curves at smaller radii than galaxies with small bulges. After the maximum the curves decline with the asymptotic velocity, which is typically 10-20\% lower than the maximum \citep{noordermeer_mass_2007}. This effect could be particularly important in the most massive galaxies in KLEVER where the bulge component is {expected to be} prominent. Indeed, as reported by \cite{lang_bulge_2014} for a sample of similar galaxies at z$=1-2$, the median B/T, {measured by deep high-resolution HST imaging}, is around $25\% $ for intermediate masses,$10.< \log(M_\star/M_{\odot})< 11.$, and increases with stellar mass reaching a maximum of 40-50\% above $\log(M_\star/M_{\odot})=11$. { Dynamical evidences of central bulges have been reported even at higher z thanks to sub-kpc spatial resolution observations obtained with ALMA (e.g., \citealp{Lelli2021, Rizzo2021}).}

To take into account the effect of the bulge on the rotation curves {of our high mass ($\log(M_\star/M_{\odot})>10.2$) galaxies,} we simulate the effect observed in \cite{noordermeer_mass_2007} by adding, a central exponential function with a peak of $V_{\text{peak}}=1.25\times V_{80}$ (as observed on local galaxies by \citealp{noordermeer_mass_2007}) at R$_{\text{peak}}$=0.5 kpc from the galaxy center (consistently with the mean effective radius of bulges observed at high z by \citealp{lang_bulge_2014}) to our arc-tangent model. {The scaling radius (R$_{\text{peak}}$) of the bulge component is assumed to be fixed for all massive galaxies irrespective of their mass, as small variations in the real R$_{\text{peak}}$ cannot provide any difference in our templates due to the low spatial resolution of our KMOS data (PSF=0.6 arcsec that corresponds to $\sim$5 kpc at z=2). We also test the possible degeneracy between outflow detection and the presence of a central bulge at high stellar masses by using mock templates without considering a bulge at high masses,$\log(M_\star/M_{\odot})>10.2$. We find that the results obtained considering or excluding a bulge, in terms of outflow detection, fluxes and outflowing mass, are consistent within 1$\sigma$. As bulges are statistically observed in similar massive galaxies at these redshifts ( \citealp{lang_bulge_2014,nelson_where_2016}), the disc+bulge model is used as fiducial one at high stellar masses, $\log(M_\star/M_{\odot})>10.2$, while no bulge component is included in less massive systems ($\log(M_\star/M_{\odot})<10.2$).}

{In conclusion, our velocity curve profile is then described by the following observational parameters: $M_\star$ (used to define $R_{TO}$, $V_{MAX}$, and the presence of a bulge component), $R_{e,\star}$ ( to define $R_{TO}$ and  $V_{MAX}$ by $R_{80}$) and n (to define $V_{MAX}$ by $R_{80}$).} An example final velocity curve profile for a low and high mass galaxy is shown in Fig. \ref{vel_curves_profiles}.

The internal velocity dispersion of the gas, $\sigma_{gas}$, is assumed to be spatially constant and is determined using the empirical relation provide by \cite{ubler_evolution_2019}: $\sigma_{gas}= 23.3 + 9.8 \cdot z $.
\\

{\subsubsection{From cubes to mock galaxy integrated spectra}}
{Having determined a physical model for each galaxy in our sample, we project it on four different inclinations, $i=10, 40, 60$ and $90$ degrees. We produce 8 mock emission line data cubes for each galaxy, 4 for the H$\alpha$ and 4 for the [OIII]$\lambda5007$ emission. The other lines, specifically, H$\beta$, [OIII$\lambda4959$, [NII] and [SII] are added in a second steps as described below. Each emission line follows the exact same distribution and velocity field. 
We extract the galaxy-integrated H$\alpha$ and [OIII] spectra following the prescription used for our KMOS observations (see Sect \ref{1Dspectra}).} {Some examples of galaxy-integrated spectra obtained with our modelling for galaxies with $\log(M_\star/M_{\odot})= 9.0, 10.0, 10.5, 11.0$ and  $i=10, 40, 60$ and $90$ deg are showed in Appendix \ref{1D_not_gaussian}.} 

{To take into account variations of the line profiles due to different inclinations the galaxy integrated H$\alpha$ and [OIII] mock emission are linearly interpolated in a fine grid of $\delta i =2$ degree between $i=10-90$ degree, providing 41 templates. 
For each galaxy, the corresponding 41 H$\alpha$ and 41 [OIII] mocks are normalised to the peak of the observed emission lines. The H$\beta$, [OIII$\lambda4959$, [NII] and [SII] emission lines are added, respectively, to the [OIII] and H$\alpha$ mock spectrum by assuming the same line shape of the [OIII] or H$\alpha$ emission. As for the strongest lines they are normalised to the observed line peaks.}
\\

\subsection{Best-fit templates}\label{bestfit}
{Each observed galaxy integrated spectrum in the H$\alpha$ and [OIII] region is fitted with the corresponding 41 templates.} The best fit rotating disc model is obtained by minimising the chi squared. {Note that the best-fit models are used only to recover the overall shape of the observed emission lines and not to derive a precise measure of the galactic disc inclination as the inclination effect on the integrated spectrum is degenerate with other physical properties, i.e. the gas velocity dispersion.} {Similar effects, e.g. broadening of the emission line, can be
obtained with a small inclination and high gas velocity dispersion or with a high inclination and low velocity dispersion. For this reason, we do not attempt to associate a physical meaning to the inclination derived with the galaxy-integrated fit.}

Some examples of observed galaxy integrated spectrum, in the [OIII] and H$\alpha$ region compared with the best fit mock rotating disc model are reported in Appendix \ref{obs_mock_single_EX}. 

{
Furthermore, as reported { in Section \ref{perturbations} and Appendix \ref{figurePerturbations}, we verified that our final results are stable under random perturbations on the model parameters (i.e., n, R$_{e,\star}$, R$_{e,\text{gas}}$, V and $\sigma_{\text{gas}}$).} 
}
\\

\begin{table}
\small
\centering
\setlength{\tabcolsep}{1.pt}
\begin{tabular}{l|c|c|c|c}

\hline
\hline
Stack   & N$_{\text{gal}}$       &   <z>    &        <M$\star$>      & <SFR>               \\
                    &            &             &  $\log(M_{\star}/M_{\odot})$      &    $\log(M_{\odot}$ yr$^{-1} )$               \\
\hline
\hline

$8.0<\log(M_{\star}/M_{\odot})< 9.6 $    &   $20$        & $1.7 \pm 0.4$    & 		$9.1 \pm 0.5$      &   $0.5 \pm 0.57$   \\  
$9.6<\log(M_{\star}/M_{\odot})< 10.2 $      &   $40$        & $2.1 \pm 0.4$    & 		$10.0 \pm 0.2$    &   $1.3 \pm 0.55$  \\  
$10.2<\log(M_{\star}/M_{\odot})< 10.8 $       &   $50$        & $2.2 \pm 0.4$    & 		$10.4 \pm 0.2$    &   $1.66 \pm 0.55$  \\  
$\log(M_{\star}/M_{\odot})>10.8 $      &   $33$        & $2.1 \pm 0.3$    & 		$11.0 \pm 0.2$    &   $2.1 \pm 0.51$ \\  
\hline
TOT     &  $141$     & $2.1 \pm 0.4$    &   $10.3 \pm 0.6$     &     $1.6 \pm 0.7$       \\
\hline
\hline
\end{tabular}
\caption{{Global properties of the galaxy sample. See Section \ref{global_prop}}}
\label{tab:InfoSample}
\end{table}
\normalsize

\subsection{From galaxy-integrated to stacked spectra}\label{stacking}

The galaxy-integrated spectra are fundamental to determine the spectroscopic redshift of our sources and find the best fit rotating disc model. However, the detection of modest flux originating from the outflowing material is very challenging in most of the spectra. This is especially true in the case of the [OIII] emission line which lies, for most of our targets, in the H-band, where the sky emission is dominated by strong and crammed lines. In most of our spectra the residual sky lines contaminate the oxygen emission making the study of lines in individual objects very challenging. To overcome this problem, as well as to increase the sensitivity of the spectra ( to detect modest flux associated with the high velocities), we performed a stacking technique on the galaxy integrated spectra {as well as in the corresponding best-fit rotation disc mocks derived in Sect. \ref{bestfit}}.

Before stacking, the observed galaxy-integrated spectra are shifted to the rest-frame velocity by using the redshift provided by the [OIII]$\lambda5007$ and H$\alpha$ lines. In the few cases where the stellar continuum is detected its median value is calculated and removed from the original spectrum. 
To accommodate the variation in fluxes from galaxy to galaxy, due to small errors on flux calibration, variation in SFR and dust reddening, spectra are normalised at the peak of the flux density of the [OIII]$\lambda5007$ and H$\alpha$. The spectra are then re-binned to a common grid of wavelength. As mentioned before, in the integrated spectra, some residual sky lines are still present. By comparing the galaxy-integrated noise spectrum with the \cite{rousselot_night-sky_nodate} catalogue we note that the presence of strongest sky lines correspond to a peak in the noise. For this reason, we mask all the wavelengths where the noise flux is above 1.5 the median value. Such a mask provides the optimal balance between accuracy of the final stacked spectrum and the number of single spectral pixels used to calculate the mean at each channel. 

Finally, the spectra are averaged together according the standard variance-weighted procedure (see \citealp{bischetti_widespread_2019} for a similar analysis). The stacked flux $ F_{\lambda}$ at the wavelength (or spectral channel) $\lambda$ is calculated as follows: 
\begin{equation}\label{stackingEq}
F_{\lambda} = \frac{1.}{W_{\lambda}} \sum^{n}_{j=1} \left (f_{j,\lambda} \cdot w_{j,\lambda} \right)\quad ,
\end{equation} 
where $n$ is the number of galaxies used in each bin and $f_{j,\lambda}$ and $w_{j,\lambda}$ are, respectively,  the flux and the weight factor of galaxy $j$ at wavelength $\lambda$. 
The weight factor is defined as $w_{j,\lambda}= \frac{1}{\sigma^{2}_{j,\lambda}}$, where $\sigma_{j,\lambda}$ is the noise level of source $j$ estimated at $\lambda$. The relative weight $W_{\lambda}$ is then defined as:
 \begin{equation}\label{weights}
     W_{\lambda} = \sum^{n}_{j=1} w_{j,\lambda}=  \sum^{n}_{j=1}            \frac{1}{\sigma^{2}_{j,\lambda}}\quad .
\end{equation}

To verify that each final spectrum is not dominated by the presence of a few luminous outliers but instead represents the general trend of the bin we use the bootstrapping technique. 
For each stacked spectrum presented in this paper we build 1000 realisations 
by randomly re-sampling with replacements the $10\%$ of galaxies in the bin. Specifically, for a general sample with N galaxies we randomly select $10\%$ of the objects in the bin and replace them with other randomly selected $10\%$. In the new stacks repetitions are allowed, therefore the same object can occur more than once or never. {Each quantity in this manuscript is calculated for each of the $1000$ realisations. The error associated with a generic measure, hereinafter referred as "statistical-error" ($\sigma _{\text{stat}}$), is the standard deviation of the distribution obtained from the new realisations.}

   \begin{figure}
   \centering
   \includegraphics[angle=-90,width=\hsize]{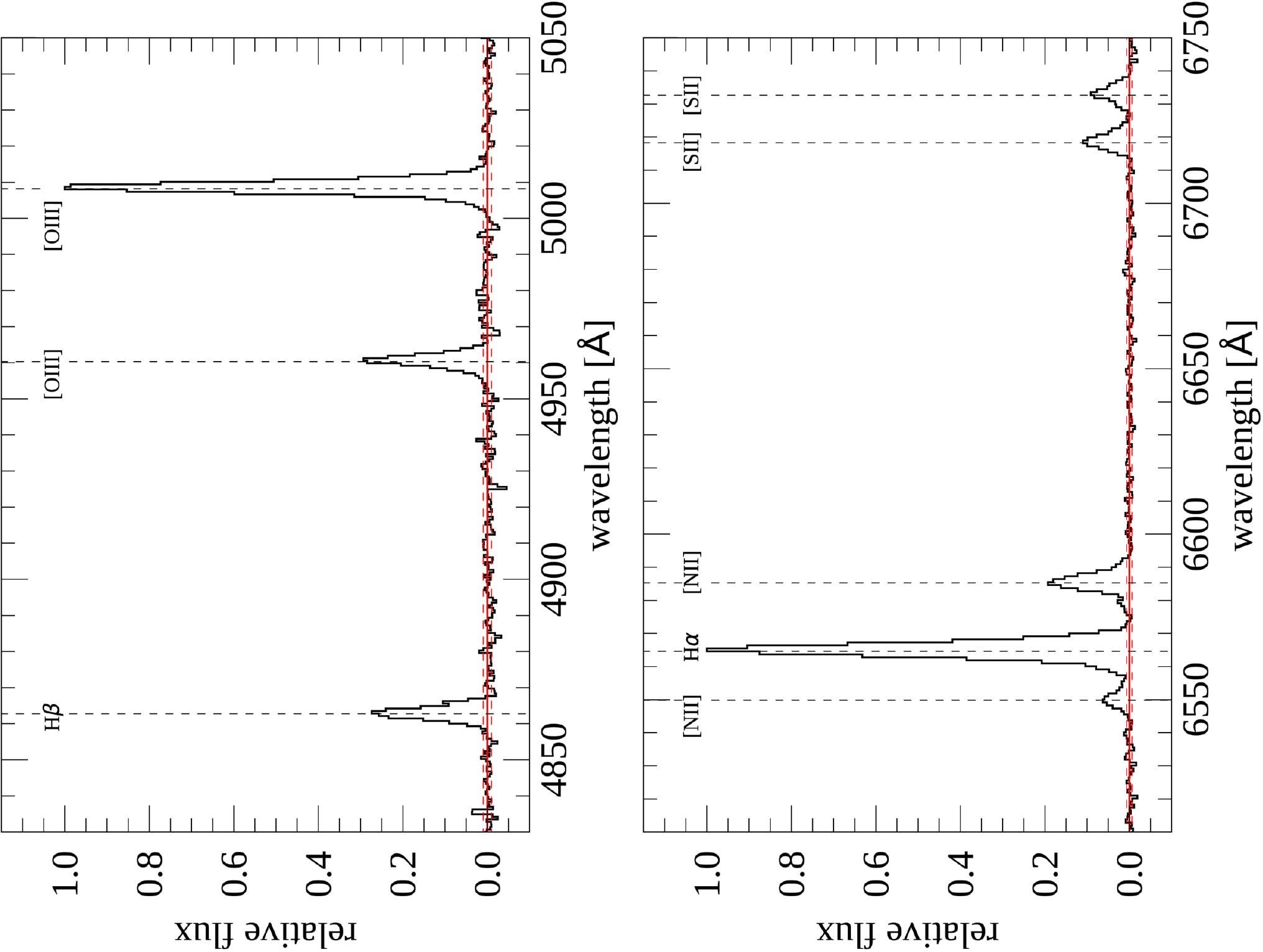}
   \caption[caption.]
  {Example of our final stacked spectrum of galaxies at $10.2 < \log(M_\star/M_{\odot}) < 10.8 $, in the [OIII]$\lambda5007$ and H$\alpha$ regions (\textit{top} and \textit{bottom panel}, respectively).
  The solid red lines shows the flux at the zero level. The red dashed lines represent the root mean square (RMS) calculated in the spectral region free of emission lines and highlights the high-quality achieved from our method. The ratio between the the oxygen lines and nitrogen lines are perfectly consistent with the theoretical prediction, [OIII]$\lambda4959$/[OIII]$\lambda5007=0.33$ and [NII]$\lambda6548$/[NII]$\lambda6584= 0.34$ \citep{OsterbrockFerland2006}}
              \label{M1_stacked_spectra}%
   \end{figure}

   \begin{figure*}
   \centering
   \includegraphics[angle=90,width=\hsize]{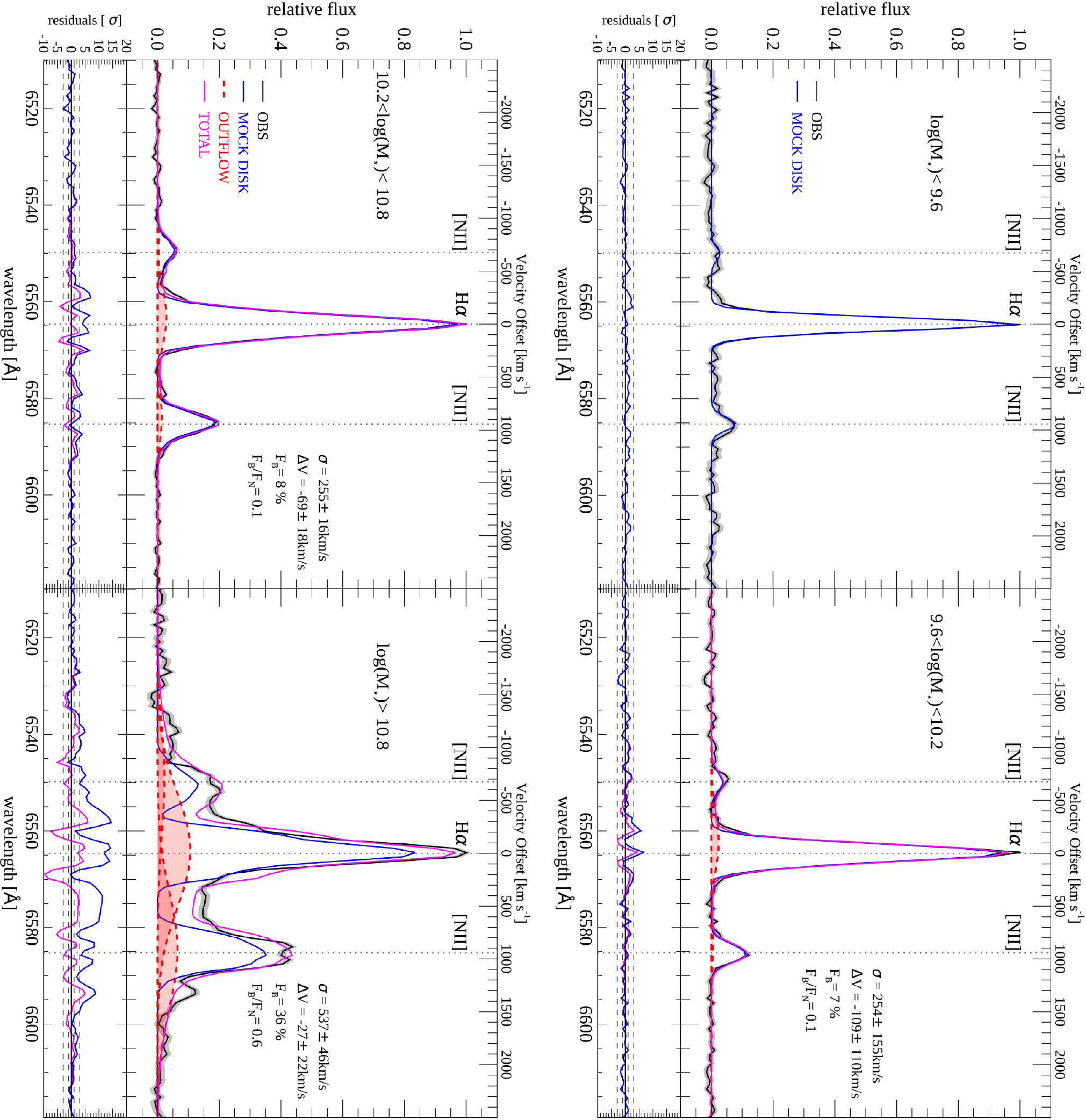}
   \caption[caption.]
  {Stacked spectra in the H$\alpha$ region for the four mass bins presented in Fig. 1. The stellar mass increases from top to bottom. Observed stacked spectra and $\sigma_{RMS}$ (black lines and grey shaded area) are compared to: 1) the expectation of a simple rotating disc model (blue curves) and, 2) the combination of the disc model plus a broad Gaussian component which represents the non-circular motions or galactic outflow (magenta lines). Residuals from the best fits (data-disc model) are reported in lower panels. The line profiles of dwarf galaxies, $\log(M_\star/M_{\odot})< 9.6$ (top left panel) are perfectly reproduced by a simple rotating disc model, no flux excess is observed above 3$\sigma$. A significant, flux excess is detected (above 3$\sigma$) at higher stellar masses (top right and bottom panels). In these cases, the inclusion into the disc model of a broad and blue shifted Gaussian component (red lines) significantly improves the fit as can be appreciated by the new residuals.  The vertical dotted lines in the spectra indicate the expected location of nebular emission lines, [NII]$\lambda6548$, H$\alpha$ and [NII]$\lambda6584$. The reported errors are computed taking into account the RMS of each averaged spectrum. 
  }
              \label{ha_stacked_obs_mock}%
   \end{figure*}

   \begin{figure*}
   \centering
   \includegraphics[angle=90,width=\hsize]{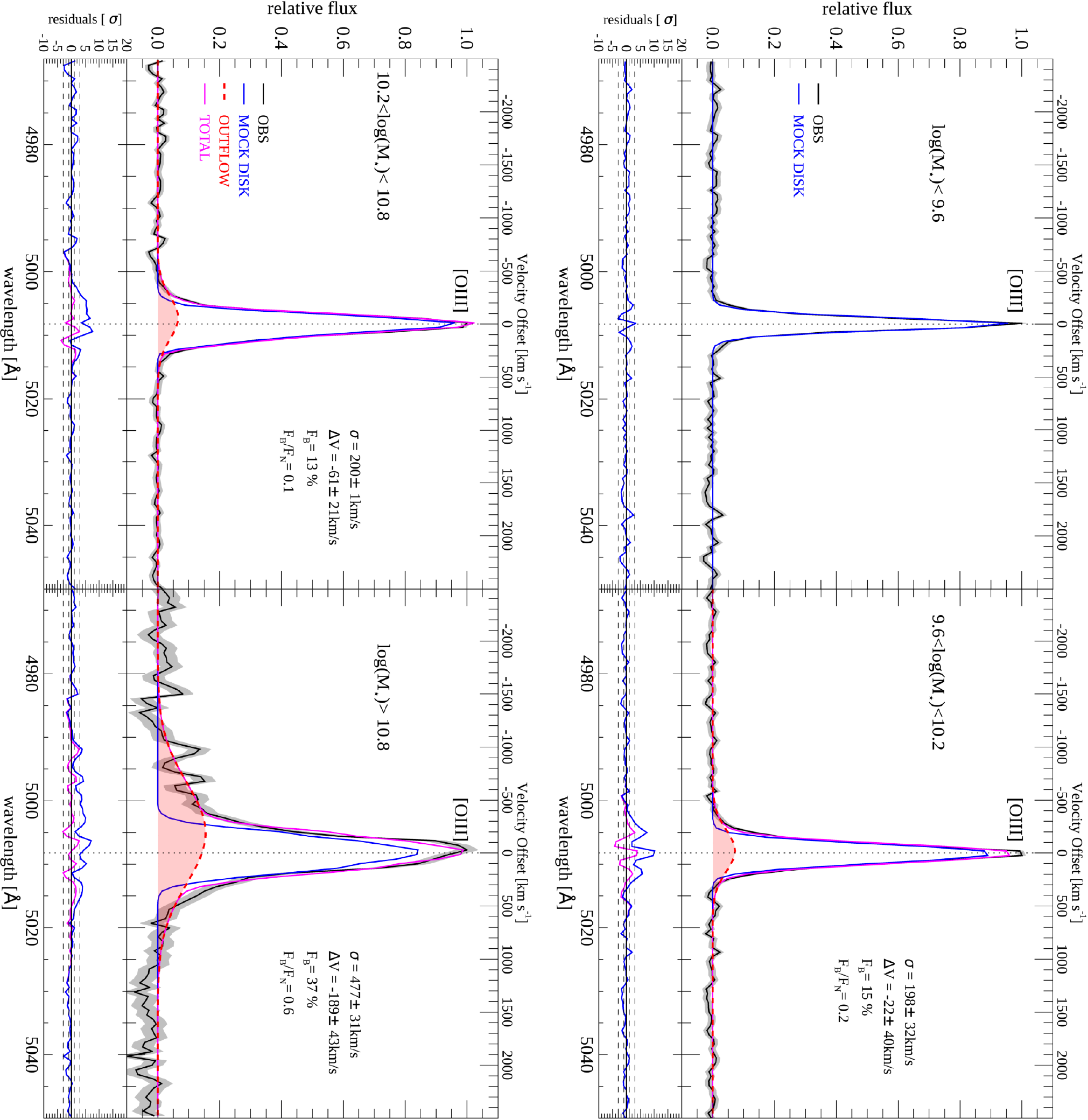}
   \caption[caption.]
  {Stacked spectra in the [OIII] region for the four mass bins presented in Fig. 1. Same description as in Fig. \ref{ha_stacked_obs_mock}. As observed in the H$\alpha$ and [NII] lines, the residuals from the rotating disc model are consistent with the noise in the low mass galaxies (top right panel). Significant flux excess, above 3$\sigma$ level is detected for high M$\star$ galaxies. For these most massive bins, $\log(M_\star/M_{\odot})> 9.6$, the line profiles are better reproduced by a disc model plus a broad Gaussian component as indicated by the residuals below the fit.}
              \label{o3_stacked_obs_mock}%
   \end{figure*}

To study the variation of the gas kinematics and presence of outflows as a function of stellar mass we applied the stacking technique in four bins of stellar mass with a width of $\Delta\log(M_\star/M_{\odot}) = 0.6$ for galaxies with $\log(M_\star/M_{\odot})> 9.6$ and larger for the less massive objects, where the statistics are reduced, see Fig. \ref{SFR_Mstar}. 
The boundaries of such grid allow us to take into account the uncertainties in the stellar mass measurements which are of the order of 0.4 dex in the case of field galaxies and even larger for the lensed objects. 
Global information about each mass bin is given in Table \ref{tab:InfoSample}. An example of the high-quality spectrum produced by our method is shown in Fig. \ref{M1_stacked_spectra}. 
Note that the RMS, seen as dashed red lines in the figure, is very low ($\sim 1.\%$ of the H$\alpha$ or [OIII] peak in all bins except for the [OIII] region of the most massive bin, $\log(M_\star/M_{\odot})> 10.8$, where it is $\sim  3\%$). The ratio between the two oxygen lines and nitrogen lines are in all cases consistent with the theoretical predictions, [OIII]$\lambda4959$/[OIII]$\lambda5007=0.33$ ( \citealp{StoreyZeippen2000}) [NII]$\lambda6548$/[NII]$\lambda6584= 0.34$ \citep{OsterbrockFerland2006}, and represents a further confirmation of the validity of our stacking procedure. 

We also explore the possible connection between the outflow detection and the star-formation activity by binning our galaxies in bins of SFRs at fixed stellar mass. For each mass bin, we select all galaxies located above and below 0.3 dex from the star-forming main sequence (MS, $\Delta MS=0.3$) using the MS relation of \cite{whitaker_constraining_2014}. As already shown in Fig. \ref{SFR_Mstar}, KLEVER galaxies are mostly located in the proximity of the MS so the resulting number of objects in each new SFR bin is quite low, $\sim7$ galaxies per bin on average. As expected, the stacked spectra binned by SFR at fixed stellar mass are noisier than those binned by mass alone, due to the smaller number of objects. The RMS increases of a factor $2-7$ depending on the case, making the outflows detection more challenging. 
The connection between the SF activity and the properties of the outflow will be explored in a forthcoming paper where we will exploit  much larger statistics. Throughout the remainder of the paper, we will refer only to the stellar mass bins.

The same stacking technique with the same weight factors is applied to the best fit mock models obtained in Sect. \ref{bestfit}. {We end up with four observed stacked spectra and four corresponding rotating disc stacked spectra.}The comparison between the observed and mock averaged spectra will be presented in the following section.

\subsection{Searching for outflows}\label{searching_outflows}
{Finally, the stacked rotating disc models described in the previous sub-sections are used to isolate the contribution of the virial motions from our observed emission lines and investigate the contribution of possible non-circular motions, like gaseous outflows.}

Specifically, we fit each emission line in the stacked spectra with two different models: the rotating \textit{Disc-model} and the \textit{Disc-Gaussian model}.

In the \textit{Disc-model}, the amplitude of the mock lines is allowed to vary in the fit to take into account small variations due to the noise. {The relative peak of the different emission lines is fixed.} The amplitude, $A$, is the only free parameter in the fit: $A \times$ F$_\text{Disc}$, where F$_\text{Disc}$ is the flux of the mock rotating disc.

The \textit{Disc-Gaussian} fit is performed by adding to the \textit{Disc-model} a Gaussian component accounting for the presence of non-circular motions. 
The fit is performed separately for lines in the H$\alpha$ and in the [OIII] region to take into account possible variations on the spectral resolution as the two regions are observed with different bands of KMOS. In particular, we require that Gaussian line shifts and widths are the same for [NII]$\lambda6548$, H$\alpha$, [NII]$\lambda6584$ and [SII] lines and, likewise for the H$\beta$, [OIII]$\lambda4959$ and [OIII]$\lambda5007$ emission. The ratio between the two nitrogen line fluxes is fixed to the theoretical value ([NII]$\lambda6548$/[NII]$\lambda6584= 0.34$; \citealp{OsterbrockFerland2006}). 
Therefore, the fit in the H$\alpha$ region has a total of 7 free parameters: the line width, $\sigma_{B}$, the velocity shift between the peak of the Gaussian and the systemic velocity, $\Delta v$, the H$\alpha$, [NII], [SII] and [SII] line flux, and the amplitude of the disc component, $A$.
The fit in the [OIII] region will have 5 free parameters: the line width, $\sigma_{B}$, the velocity shift, $\Delta v$, the H$\beta$, [OIII] line flux and the amplitude of the disc component, $A$.

{The errors on all measured quantities are obtained following two different criteria. First, using the bootstrapping realisations presented in Section \ref{stacking}, we quantify the "statistical-error", $\sigma_{\text{stat}}$ which takes into account the variations of the spectrum within the galaxy bin. Then, we quantify the error due to the residual noise in the final stacked spectrum. In this case, we randomly perturb the stacked spectrum within the RMS (calculated nearby the emission of interests) and we generate 1000 new realisations of the averaged spectrum. Also in this case, the error associated with each parameter is estimated by repeating the fit in all new realisations and taking the variance of the distribution, $\sigma_{\text{RMS}}$. Therefore, for each general parameter $P$ in the paper we have $\sigma_{\text{stat}}$ and $\sigma_{\text{RMS}}$. 
}

To establish the necessity of a broad Gaussian component on top of the rotating disc model, we evaluate the statistical improvement of the fit using the variation of the reduced chi squared, $\chi^{2}_{R}$ and the Bayesian Information Criterion (BIC \footnote{The BIC is defined as: BIC $=\chi^2 + p \times \ln(n)$, where $\chi^2$ is the chi squared of the fit, $p$ is the number of free parameters and $n$ is the number of flux points used in the fit (\citealp{Schwarz1978}, see \citealp{Liddle2007} as a review on information criteria).}, \citealp{Schwarz1978}).  Specifically, the more complex \textit{Disc-Gaussian} model is chosen as best fit only if the $\chi^{2}_{R}$ improves more than $1\sigma$ ($\chi^{2}_{R,\text{Disc}}- \chi^{2}_{R,\text{Disc-Gauss}} > 1 \sigma_{\text{RMS}}$) and if the BIC variation, $\Delta BIC= BIC_{\text{Disc}}-BIC_{\text{Disc-Gauss}}$ is bigger than 10 \footnote{{According with \cite{fabo14}, $\Delta BIC>10$ represents  a very strong evidence against "a candidate model" or, in this case, against the simplest Disc model.}}.
{We check that assuming a different value for the $\Delta BIC$ limit, e.g. $\Delta BIC=$0 or 20, does not affect our conclusions.}

\section{Results}\label{results}
\subsection{Detection of outflows}
In this section we search for evidence of ionised outflows traced by the brightest emission lines in the optical rest frame averaged spectrum of 141 star forming galaxies at $1.2 < z<2.6$ by following the method described in the previous Section. Here we report the results of the comparison between the observed stacked spectra and the averaged mock rotating disc models.

Figures \ref{ha_stacked_obs_mock} and \ref{o3_stacked_obs_mock} show the resulting best-fit obtained, respectively, for the H$\alpha$+[NII] and [OIII] emission lines in each mass bin. 

At low stellar masses, $\log(M_\star/M_{\odot})< 9.6$, the observed emission lines are well reproduced by the \textit{Disc-model} (blue curve) as can be fully appreciated from the top-left panels of Fig. \ref{ha_stacked_obs_mock} and \ref{o3_stacked_obs_mock}, respectively for the H$\alpha$+[NII] and [OIII] region.
The residuals of the \textit{Disc-model} fit (data-model) are below the $3\sigma$ noise level, even in the high-velocity tails in all the emission lines. The goodness of the fit is statistically confirmed by the $\chi^{2}_{R}$ value which is perfectly consistent with unity:$\chi^{2}_{R}$=0.8$\pm 0.2$ and 1.1$\pm 0.2$ ($\sigma_{\text{RMS}}$), respectively for the H$\alpha$+[NII] and [OIII] region.  
The addition of a Gaussian broad component to the fit does not lead to a substantial improvement of the fit as indicated by the fact that the $\chi^{2}_{R}$ obtained with the \textit{Disc} fit is statistically consistent (within 1$\sigma$) with that derived with the more complex \textit{Disc$-$Gaussian} model. This is also confirmed by the very low values of the $\Delta BIC$, $\Delta$ BIC $< 10$. The \textit{Disc$-$Gaussian} fit is therefore rejected in the case of low mass objects, $\log(M_\star/M_{\odot})<9.6$ and the \textit{Disc} model is chosen as the more statistically appropriate. 
This result is very surprising considering that, in the dwarf regime, the ionised gas traced by H$\alpha$, [NII] and [OIII], is expected to be strongly affected by turbulence and non-circular motions driven by intense stellar feedback in galaxies located at cosmic noon. 

At higher stellar masses, above $\log(M_\star/M_{\odot})> 9.6$, the \textit{Disc} model (blue curves) is not able to fully reproduce the observed emission lines especially in proximity of the line wings. We start to detect some residuals above the 3$\sigma$ level in the middle mass bins $9.6< \log(M_\star/M_{\odot})< 10.2$ and $10.2< \log(M_\star/M_{\odot})< 10.8$, top-right and bottom-left panels, respectively. 
The fit slightly improves with the \textit{Disc-Gaussian} model (magenta curves) thanks to the addition of a broad Gaussian component (red areas). This effect become particularly strong in the most massive bin, $\log(M_\star/M_{\odot})> 10.8$ (bottom right panels). In this case, strong residuals from the \textit{Disc} model are detected (above 6 and 10$\sigma$ in the H$\alpha$+[NII] and [OIII], respectively), the $\chi^{2}_{R}$ strongly decreases (a factor 3.4 and 2.3 in the  H$\alpha$+[NII] and [OIII] case) and the BIC variation is conspicuous, $\Delta BIC > 100$. In particular, we note that the $\chi^{2}_{R}$ obtained for the H$\alpha$+[NII] complex, is yet larger than unity, $\chi^{2}_{R}\sim 4$, suggesting that the line profile might be more even complex in this case. 

We explore the possible presence of Type 1 AGNs and resulting very broad emission originating from the broad line regions (BLR) of the central AGN in the hydrogen lines, H$\beta$ and H$\alpha$. We visually inspected the integrated spectra of all the galaxies in KLEVER finding that only one system presents a very broad H$\beta$ emission (larger than the broad [OIII] line) which is clearly contaminated by the gas in the BLR. We find that including or excluding the suspected Type 1 AGN does not change significantly the results obtained for the H$\alpha$ region (we obtain similar best-fit parameters) but it strongly reduces the signal to noise in the [OIII] region making the oxygen decomposition really challenging. Since the [OIII] line cannot be contaminated from the dense BLR and its best-fit parameters are perfectly consistent with the H$\alpha$ values, we conclude that the contamination due to the presence of Type 1 AGNs in the most massive bin is not significant and does not affect our final results. 

Finally, we find that the second Gaussian component, when detected (only in stellar mass bins with $\log(M_\star/M_{\odot})> 9.6$), appears to be broad with $\sigma> 200$ km s$^{-1}$ and blue-shifted with respect to the systemic velocity (or the disc component with $\Delta V\sim 50-170$ km s$^{-1}$ depending on the case) that could indicate the presence of  massive galactic outflows. 

In summary, we do not detect any statistical evidence (above 3$\sigma$) of perturbed ionised flux associated with non-circular motions (outflows) in galaxies with stellar mass below  $\log(M_\star/M_{\odot})<9.6$. For more massive systems ($\log(M_\star/M_{\odot})>9.6$), instead, it is clear that the kinematics of the ionised gas is more complex, probably due to the presence of non-circular motions like massive gaseous outflows. 
Our derived parameters and associated uncertainties of the best fit \textit{Disc} and \textit{Disc-Gaussian} models obtained for all our bins are reported in Table \ref{tab:InfoSample2}. 

\begin{figure} 
\centering
\includegraphics[width=\hsize]{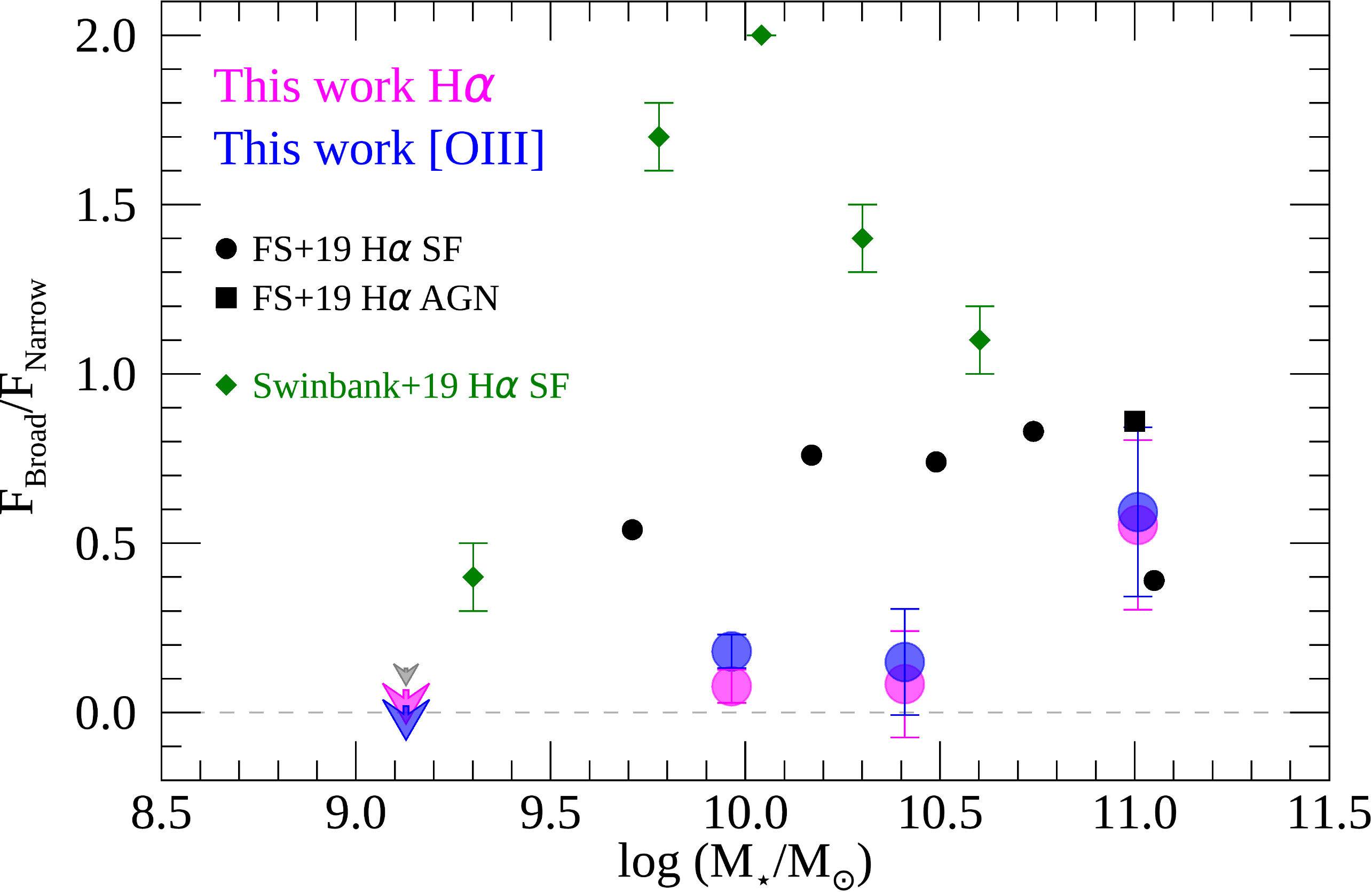}
   \caption[caption.]{ 
   Broad to narrow/disc flux ratio of the [OIII] and H$\alpha$ line as a function of stellar mass, for H$\alpha$ and [OIII] line, respectively, magenta and blue symbols. Black symbols are  H$\alpha$ stacks from \cite{schreiber_kmos_2019} obtained for star-forming galaxies (circles) and AGNs (considering the AGN best spectra values reported in  \citealp{schreiber_kmos_2019}, square). Green diamonds represent the stacked of H$\alpha$ in z$\sim1$ star-forming galaxies from the KROSS team \citep{swinbank_energetics_2019}. Our massive bin is fully consistent with the results of KMOS$^{\text{3D}}$, however, we find lower fluxes at low stellar masses. The difference is even stronger with KROSS values. 
   The broad flux, $F_{\text{B}}$, is not detected (above 3$\sigma$) in the dwarf galaxy bin, for which we measure only an upper limit indicated by the arrow. {The small grey arrow is the upper limit obtained in the unlikely extreme case of a broad Gaussian component with FWHM$\sim 700-1000$ km s$^{-1}$ and $\Delta v = 0$ km s$^{-1}$.} The errors are the statistical errors computed with the bootstrapping realisations presented in Section \ref{stacking}.}
                \label{flux_broad}%
\end{figure}

\begin{table*}
\small
\centering
\begin{tabular}{l|c|c|c|c|c|c|c|c|c|c}

\hline
\hline
Stack   & F$_{\text{Broad}}$ &  F$_{\text{Broad}}$/F$_{\text{Narrow}}$  & $\sigma_{\text{Broad}}$  &  $\Delta_v$  &   v$_{\text{out}}$ &  $\dot M_{\text{out}}$       & $\log (\eta)$ \\
        &    $\%$        &                                 &    [km s$^{-1}$]    &  [km s$^{-1}$] &  [km s$^{-1}$]    &  [M$_{\odot}$yr$^{-1}$] &        \\
\hline
\hline
\multicolumn{8}{|c|}{H$\alpha$ line} \\
\hline                                      
$8.0<\log(M_{\star}/M_{\odot})< 9.6 $    & $\leq 6.2 $  & $\leq 0.07$  & $197\pm 22$  & $-66$ &  $460\pm 44$ & $0.06$ & $-1.58$  \\  
$9.6<\log(M_{\star}/M_{\odot})< 10.2 $   & $7.2 \pm 2.7 $  & $0.08 \pm 0.03$ & $254 \pm 155$  & $-109 \pm 110$ & $616\pm 331$ & $0.36$  & $-1.48$ \\  
$10.2<\log(M_{\star}/M_{\odot})< 10.8 $  & $8.0 \pm 0.9 $  & $0.1 \pm 0.01$ & $255 \pm 16$  & $-69 \pm 18$ & $578\pm 39$ & $0.7$  & $-1.54$      \\  
$\log(M_{\star}/M_{\odot})>10.8 $        & $36.0\pm 1.1 $  & $0.56 \pm 0.03$ & $537 \pm 46$  & $-27 \pm 22$ & $1101\pm 82$ & $2.3$  & $-1.09$      \\  
\hline
\multicolumn{8}{|c|}{[OIII]$\lambda5007$ line} \\
\hline                                      
$8.0<\log(M_{\star}/M_{\odot})< 9.6 $    &   $\leq 1.8 $  & $ \leq 0.02$ & $196 \pm 73$  & $-100\pm 59 $ &  $492\pm 111 $ & $0.02$ & $-2.02$     \\  
$9.6<\log(M_{\star}/M_{\odot})< 10.2 $   &   $15.3 \pm 1.2 $  & $0.18 \pm 0.02$ & $198 \pm 32$  & $-22 \pm 40$ & $418\pm 48$ & $0.36$  & $-1.48$      \\ 
$10.2<\log(M_{\star}/M_{\odot})< 10.8 $  &   $13.0 \pm 1.4 $  & $0.15 \pm 0.02$ & $200 \pm 1$  & $-61 \pm 21$ & $461\pm 21$ & $0.8$  & $-1.47$      \\  
$\log(M_{\star}/M_{\odot})>10.8 $        &   $37.2\pm 2.2 $  & $0.59 \pm 0.06$ & $461 \pm 31$  & $-189 \pm 43$ & $1142\pm 91$ & $2.1$  & $-1.12$      \\  
\hline
\hline
\end{tabular}
\caption{{Properties of the non-circular motions and outflows obtained for each mass bin presented in the paper. The parameters errors are calculated taking into account the RMS on the final averaged spectra (see Section \ref{results}).}}
\label{tab:InfoSample2}
\end{table*}
\normalsize

\subsection{Outflow prominence as a function of galaxy stellar mass}
We investigate the variation of the the flux enclosed into the broad Gaussian component, F$_{\text{Broad}}$, as a function of stellar mass. We find similar results for the H$\alpha$ and [OIII] lines.

For the less massive systems, where the broad component is not detected, we use the values obtained in the \textit{Disc-Gaussian} fit as an upper limit of the flux associated with non-circular motions,  F$_{\text{Broad}}< 6\%$ in H$\alpha$ and [NII] and  F$_{\text{Broad}}< 2\%$ in the [OIII] line. 

{For galaxies at stellar mass below $\log(M_\star/M_{\odot})=10.8$, we find that} the maximum flux enclosed in the broad Gaussian component is less than the $10-16$ per cent of the total flux in both the [OIII] and H$\alpha$ emission line. 
In the most massive galaxies the flux encapsulated in the broad component is more than the $36-37\%$ of that of the total line for the H$\alpha$, [NII] and [OIII] line. 

Similar prominent broad-line emissions have been seen in the stacked H$\alpha$ spectra of z$=1-2$ galaxies (e.g. \citealp{genzel_sins_2011,genzel_evidence_2014,schreiber_kmos_2019,swinbank_energetics_2019}). To better compare our results with previous findings, the prominence of the component is expressed in terms of the ratio between the flux of the broad Gaussian and the narrow-disc component, F$_{\text{Broad}}$/F$_{\text{Narrow}}$. 
Figure \ref{flux_broad} shows the variation of F$_{\text{Broad}}$/F$_{\text{Narrow}}$ as a function of stellar mass. In the highest mass bin, the KLEVER values {( F$_{\text{Broad}}$/F$_{\text{Narrow}}= 0.56\pm 0.25$, for the H$\alpha$; errors are computed with the bootstrapping realisations)} are fully consistent with the F$_{\text{Broad}}$/F$_{\text{Narrow}}$ found by \cite{schreiber_kmos_2019} using the H$\alpha$ spectra of the KMOS$^{\text{3D}}$ galaxy sample. Note that our H$\alpha$ ( and [OIII]) flux ratio resides between the F$_{\text{Broad}}$/F$_{\text{Narrow}}$ of the star-forming galaxies and the best value obtained for AGNs in the KMOS$^{\text{3D}}$ sample. This is perfectly explained by the fact that our massive bin in KLEVER is a combination of "normal" SF galaxies and AGNs. This agreement found at high masses is reassuring given the fact that our method is substantially different from that adopted from the KMOS$^{\text{3D}}$ team.

At low stellar masses, below $\log(M{\star}/M_{\odot})=10.8$, our flux ratios are 
lower compared to the previous studies based on similar galaxies observed with KMOS (e.g. \citealp{schreiber_kmos_2019} and \citealp{swinbank_energetics_2019}). As shown in Fig. \ref{flux_broad}, our F$_{\text{Broad}}$/F$_{\text{Narrow}}$ are $\sim 7-8$ times lower than the values reported by the KMOS$^{\text{3D}}$ team \cite{schreiber_kmos_2019} and even lower (up to $25$ times, green diamonds) compared to the values found by the KROSS team \citep{swinbank_energetics_2019}. {To better understand what factors drive such a discrepancy, we explore differences in the methods used to derive F$_{\text{Broad}}$/F$_{\text{Narrow}}$ and differences in the samples studied.}

\subsubsection{Detailed comparison with previous observations}\label{KLEVER_vs_KMOS3D_KROSS}
{As already mentioned in Section \ref{vel_sub_method}, both KMOS$^{\text{3D}}$ (\citealp{schreiber_kmos_2019}) and KROSS (\citealp{swinbank_energetics_2019}) team used the "velocity-subtracted" technique to isolate the virial motions in galaxy spectra.}

{Performing the same "velocity-subtracted" method on our KLEVER galaxies and following all the fundamental steps described by \cite{schreiber_kmos_2019} for the KMOS$^{\text{3D}}$ data-set and \cite{swinbank_energetics_2019} for the KROSS galaxy sample, {we find that the discrepancy observed in Fig. \ref{flux_broad} is relieved in the case of KMOS$^{\text{3D}}$ and disappears for KROSS:} the F$_{\text{Broad}}$/F$_{\text{Narrow}}$ derived for the KLEVER galaxies using the KMOS$^{\text{3D}}$ and KROSS approach are consistent with the results previously determine by \cite{schreiber_kmos_2019} and \cite{swinbank_energetics_2019} {with the only exception for the lowest mass bin presented in \cite{schreiber_kmos_2019} (see Appendix \ref{comparisonKMOS3D_KROSS} for more details).} 
 This suggests that the discrepancy presented in Fig. \ref{flux_broad} is only apparent and primary driven by the adopted method. At this point, one might wonder, 
"Why does the F$_{\text{Broad}}$/F$_{\text{Narrow}}$ ratio obtained for the KLEVER galaxies following the velocity-subtracted method presented in \cite{schreiber_kmos_2019} and \cite{swinbank_energetics_2019} feature a higher value than the ratio obtained with the rotating disc technique presented in this paper?" Can the limitations of the velocity-subtracted method discussed in Section \ref{vel_sub_method} for a single galaxy be responsible for this large amount of F$_{\text{Broad}}$ observed in the stacked spectra?}


{To test this possibility, we use our mock rotating discs. As reported in Appendix \ref{comparisonKMOS3D_KROSS}, we repeat the  KMOS$^{\text{3D}}$ and KROSS analysis directly on the mock rotating discs created for the galaxies in the dwarf ($\log(M_\star/M_{\odot})<9.6$) and medium stellar mass bin ($10.2< \log(M_\star/M_{\odot}) <10.8$). We find that the artificial F$_{\text{Broad}}$/F$_{\text{Narrow}}$ obtained in the mock rotating disc, where no outflow is present, could explain the large flux previously determined in the observed data analysed with the KMOS$^{\text{3D}}$ and KROSS technique (see the comparison between mock and observations in Fig. \ref{mock_stacked_KMOS3D} and \ref{mock_stacked_KROSS} in Appendix \ref{comparisonKMOS3D_KROSS}).} {This simple exercise, therefore, confirms that the broad flux could be easily overestimated in the case of the velocity-subtracted method providing higher values of F$_{\text{Broad}}$/F$_{\text{Narrow}}$ than in our rotating disc method, and it demonstrates that the differences between our method and previous observations are only apparent
and {primarily attributable} to the beam smearing effect on the velocity-subtracted method.}



For a more in-depth, analysis and discussion on the limitations of the velocity-subtracted method, we refer the reader to a forthcoming paper (Concas et al. in preparation).

{\subsection{Detectability test on low mass galaxies}}
{We check the upper limit obtained for low mass galaxies by using the mock stacked spectrum obtained for this mass bin. In particular, we simulate the presence of outflowing material by 1) adding a Gaussian component to the galaxy-integrated disc model H$\alpha$+[NII] spectrum, 2) perturbing the global model (disc+outflow) according with the observed noise and, 3) fitting the mock observation with the \textit{Disc} and the \textit{Disc-Gaussian} model. Using the criteria presented in Sec. 4.4, if the broad Gaussian component is assumed to have a FWHM$\sim 400$ km s$^{-1}$ (as reported by \citealp{schreiber_kmos_2019} for less massive galaxies) and a line shift $\Delta v = -100$ km s$^{-1}$, we start to detect the broad Gaussian component with  F$_{\text{Broad}}= 10\%$, F$_{\text{Broad}}$/F$_{\text{Narrow}}=0.1-0.12$ 
Similar values are obtained assuming a very broad and centred line, $\Delta v = 0$ km s$^{-1}$, and higher line width, reaching a maximum value of F$_{\text{Broad}}= 11-12\%$ and F$_{\text{Broad}}$/F$_{\text{Narrow}}=0.13-0.14$ (grey circle in Fig. \ref{flux_broad}) for the very unlikely and extreme case of FWHM$\sim 700-1000$ km s$^{-1}$ .}
\\
\\


   \begin{figure}
   \centering
   {\includegraphics[angle=0,width=\hsize]{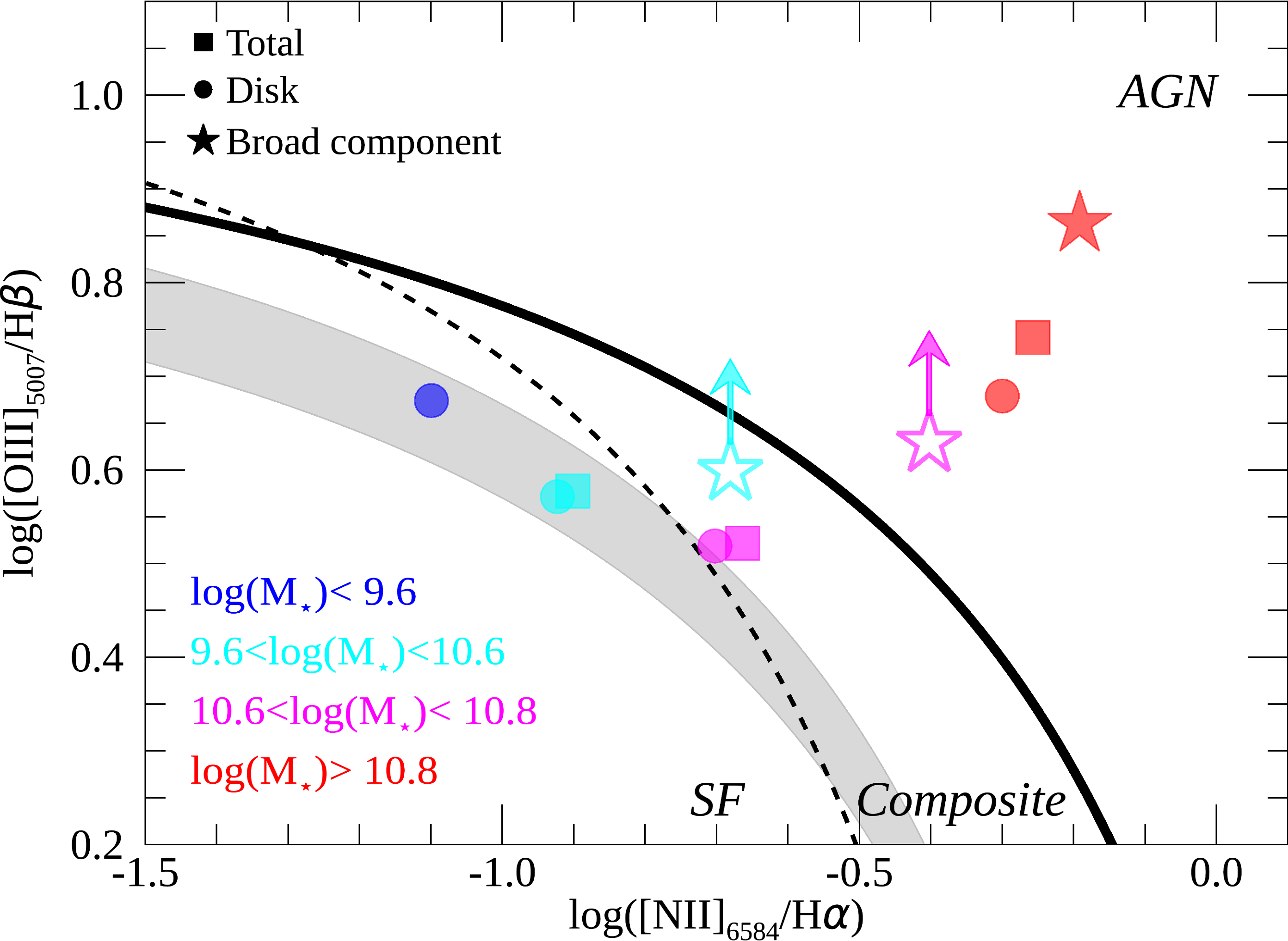}}
   \caption[caption.]{Distribution of the KLEVER stacked spectra in the BPT line-ratio diagram. The four mass bins are represented with different colours according to the label. Squares, circles and stars represent the ratios obtained for the total emission, the disc (or narrow) and the broad Gaussian component, respectively. Empty stars represent the vertical lower limits for the broad component in the medium mass bins ($9.6<\log(M{\star}/M_{\odot})<10.2$ and $10.2<\log(M{\star}/M_{\odot})<10.8$) for which the broad H$\beta$ emission is not detected. 
   The solid and dashed curves are the theoretical and empirical demarcation between the star-forming locus and AGN galaxies from of \cite{Kewley2001} and \cite{Kauffmann2003}, respectively. The emission of the most massive bin is clearly dominated by the AGN activity. At stellar masses below $\log(M{\star}/M_{\odot})<10.8$ the disc component lies on top of the locus of the high-z galaxies (grey shaded area) derived by \cite{strom_nebular_2017} using the KBSS survey. 
  The flux associated with the non-circular motions, or broad component (stars) appears to be shifted to the right part of the diagram towards the AGN region. 
 }
                \label{BPT_stacked}%
   \end{figure}

\subsection{Gas excitation mechanism and AGN-driven outflows}
We investigate the nature of the excitation of the ionised gas in each of our stacked spectra, by using the so-called BPT diagnostic diagrams (\citealt*{BPT1981}). In particular we calculate the [OIII]$/$H$\beta$ and [NII]$/$H$\alpha$ ratios for the global emission line, the disc, and the broad Gaussian component. Figure \ref{BPT_stacked} shows the position of our galaxies on the [O III]$\lambda5007$/H$\beta$ versus [N II]$\lambda6584$/H$\alpha$ diagram with line of demarcation between the different excitation mechanisms identified by \cite{Kauffmann2003} and  \cite{Kewley2001} for local galaxies. The grey shaded area show the average position of high-z galaxies as inferred for the KBSS survey (\citealp{strom_nebular_2017}, $z \sim 2.3$ ). 
The total emission line ratios (filled squares in Fig. \ref{BPT_stacked}) of low and intermediate mass bins ($\log(M{\star}/M_{\odot})<10.8$) occupy the area of the diagram expected for stellar photoionization (or SF activity) and/or from a combination of SF and AGN activity. The most massive bin, instead, is clearly dominated by the AGN excitation. 

The same result is found for the disc components (circles) with the only difference that the disc ratios appear to be slightly shifted towards the SF region compared to the global values for the $\log(M{\star}/M_{\odot})<9.6$,  $10.2<\log(M{\star}/M_{\odot})<10.8$ and $\log(M{\star}/M_{\odot})>10.8$ bin. We also observed that, emission line ratios of the disc components of galaxies below $\log(M{\star}/M_{\odot})<10.8$ tend to lie on top of the locus of $z\sim2.3 $ galaxies observed by the KBSS survey (\citealp{strom_nebular_2017}). Note that given the errors our disc ratios are also consistent with the locations observed in the FMOS \citep{kashino_fmos-cosmos_2019} and MOSDEF \citep{shapley_mosdef_2015} surveys. 

The broad Gaussian component ratio obtained for the most massive bin ($\log(M{\star}/M_{\odot})>10.8$, red star) is clearly dominated by AGN activity. For the medium mass bins, $9.6<\log(M{\star}/M_{\odot})<10.2$ and $10.2<\log(M{\star}/M_{\odot})<10.8$, we do not detect a broad Gaussian component in the H$\beta$ line so we provide a lower limit in the vertical position (open stars in the figure).
We observe that, in these medium mass bins, the flux associated with non-circular motions are always shifted to the right part of the panel, towards the AGN region (see the open stars) compared to the global and/or disc components. This result could suggest a possible connection between the non-circular motions and the AGN activity or shocks. Note that for the less massive bin ($\log(M{\star}/M_{\odot})<9.6$) the broad component is not detected in any emission line 
hence, for these perturbed components, we cannot infer any information from the BPT diagram.

\begin{figure}
   \centering
   {\includegraphics[angle=0,width=\hsize]{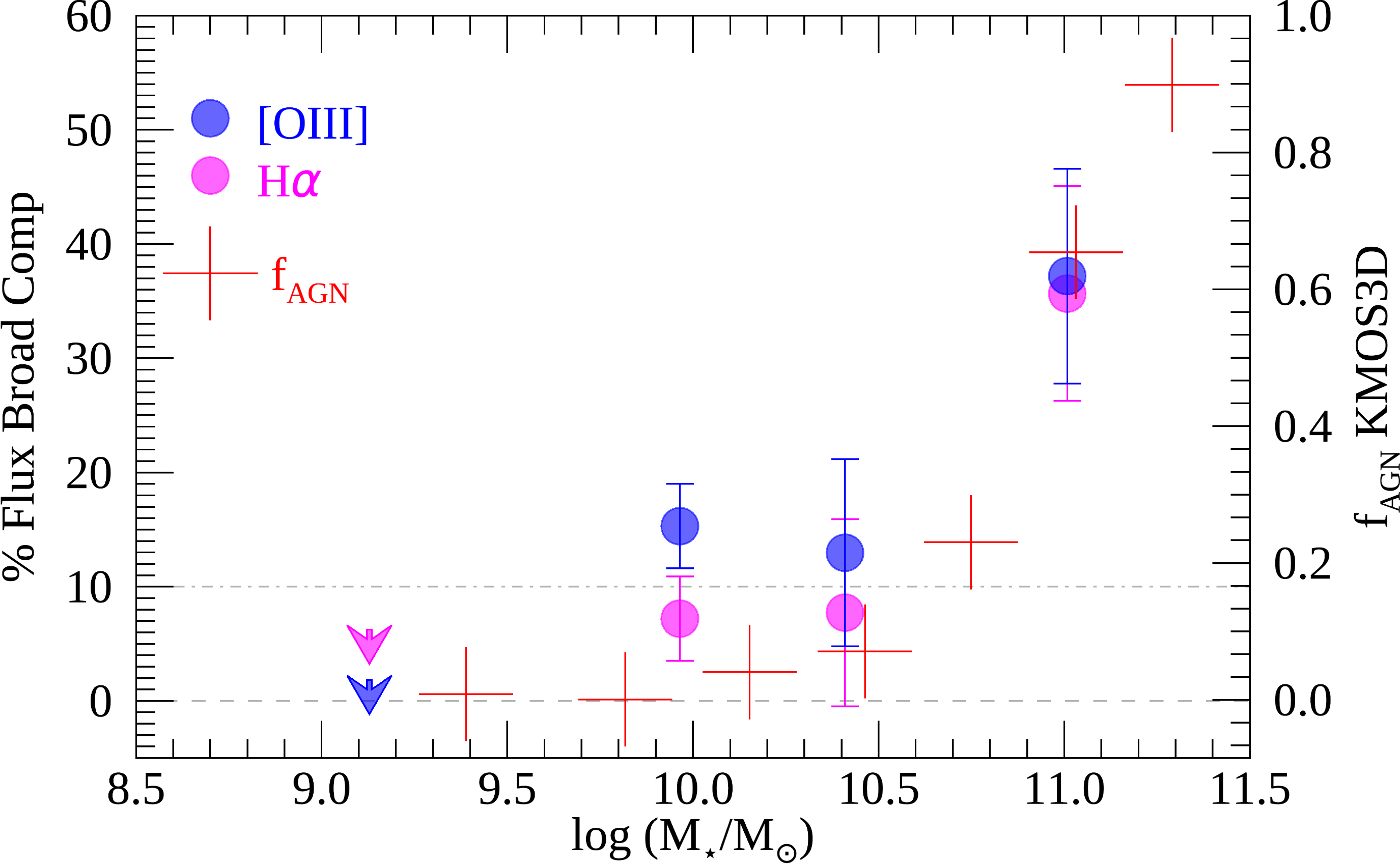}}
   \caption[caption.]{Variation with stellar mass of the flux percentage associated with non circular motions (F$_{\text{Broad}}$, possible outflows) detected on KLEVER stacked spectra (magenta and blue symbols, for H$\alpha$ and [OIII] line respectively) and the incidence of AGN activity (f$_{\text{AGN}}$, red cross) reported by the KMOS$^{\text{3D}}$ team on Fig. 13 of \cite{schreiber_kmos_2019}. The f$_{\text{AGN}}$ is scaled according to the vertical axis on the right side panel. The arrows denote the upper limits and the errors are the statistical errors computed with the bootstrapping realisations presented in Section \ref{stacking}.}
                \label{Fout_FAGN_comparison}%
   \end{figure}

To further explore the possible presence of AGNs and their connection with the non-circular motions in our stacked spectra we use the incidence of AGN activity (f$_{AGN}$) reported by \cite{schreiber_kmos_2019} for the KMOS$^{\text{3D}}$ survey. 
Fig. \ref{Fout_FAGN_comparison} compares f$_{AGN}$ (red crosses) with the flux associated with non-circular motions (F$_B$, same value reported in Fig. \ref{flux_broad}) as a function of the stellar mass. This figure shows that both f$_{AGN}$ and F$_B$ correlates with $\log(M{\star})$ starting from negligible values below $\log(M{\star}/M_{\odot})<9.6$ and reaching a maximum at > 11. The correlation is very strong in both cases, showing a Pearson rank correlation factor of $\rho=0.86$ and $0.81$ for AGN and broad flux. Even more interesting is the fact that the number of AGNs expected at intermediate masses, $10.0< \log(M{\star}/M_{\odot})<10.8$, is not zero, going from $\sim 10\%$ to $\sim 25\%$ suggesting that this medium mass bins may be in principle host some AGN with possible presence of AGN-driven outflows. Although this correlation alone may not be sufficient to establish a causal link between the detection of non-circular motions and AGN activity it corroborates the indication already suggested by the line ratios shown before. Since the KLEVER sample is a subsample of the KMOS$^{\text{3D}}$ survey, the comparison with the f$_{AGN}$ is only qualitative. For this reason other physical mechanisms able to generate the observed non-circular motions, such as SF-driven outflows, presence of shocks, spiral arms, bars etc. cannot be fully excluded.

\subsection{Outflow physical properties: mass, velocity and density}\label{outflowsProp}
We now determine the physical properties of the outflowing ionised gas in each mass bin focusing on the broad line flux observed in the brightest emission line, H$\alpha$. 
In Appendix \ref{massLoadingo3}, we also report the outflows properties obtained using the [OIII] line. Briefly, if using [OIII] we find similar results in terms of outflow velocities but lower masses and corresponding lower $\dot M_{out}$ and $\eta$, consistent with previous studies (see \citealp{ carniani_ionised_2015,marasco_galaxy-scale_2020}). Since the mass values obtained with the [OIII] line depend on the chemical enrichment of the outflowing gas and likely do not properly account for the lower ionisation phases (see Appendix \ref{massLoadingo3}), we will focus our analysis on the more reliable H$\alpha$ emission.

The individual H$\alpha$ luminosity of each galaxy has been corrected for the dust attenuation using the visual extinction of the stellar light estimated from the best-fitting SED modelling, A$_{V\star}$. 
Following \cite{schreiber_kmos_2019}, we adopt the Calzetti reddening law \citep{calzetti_dust_2000} considering an extra dust attenuation on the nebular gas:  A$_{\text{gas}}=$A$_{V\star}\times (1.9-1.5 \times$ A$_{V\star})$ (see also \citealp{Wuyts2013}). 
The luminosities of the lensed objects have been corrected for the magnification factor as already done for the M$\star$ and SFR. Note that the resulting mass loading factor defined as $\eta= \dot M_{out}/SFR$ will not depend on the adopted magnification as the M$_{out}$ and the SFR share the same dependence on  magnification.

To estimate the luminosity of the H$\alpha$ broad Gaussian component, we computed the total weighted L$_{\text{stack}}$ by applying Equation \ref{stackingEq} and \ref{weights} to the individual H$\alpha$ luminosities of our targets. We therefore decompose the global weighted L$_{\text{stack}}$ into the disc and broad Gaussian component using the flux percentages (F$_{\text{Narrow}}$ or F$_{\text{Disc}}$ and F$_{\text{Broad}}$) defined in the previous section as follows: $L_{\text{Disc}} = L_{\text{stack}} \times \text{F}_{\text{Narrow}}$  and L$_{B} = \text{L}_{\text{stack}} \times \text{F}_{\text{Broad}}$. The H$\alpha$ luminosity associated with the disc component, L$_{\text{Disc}}$, is then converted to SFR assuming the \cite{kennicutt_star_2012} relation and applying a scaling factor of 1.06 to convert from \cite{Kroupa1993} to \cite{Chabrier2003} IMF. 

Assuming that the outflowing material can be described by a collection of ionised clouds sharing the same electron density, $n_e$, the mass of the outflowing gas can be inferred from the extinction corrected, H$\alpha$ luminosity, L$^{\text{H}\alpha}_{\text{B}}$ as follows (see \citealp{marasco_galaxy-scale_2020}, \citealp{cresci_muse_2017}):

\begin{equation}
    \text{M}^{\text{H}\alpha}_{\text{out}}=3.2\times 10^{5} \left ( \frac{ \text{L}^{\text{H}\alpha}_{\text{B}} }{10^{40} \text{ erg s}^{-1}} \right)  \left ( \frac{100 \text{ cm}^{-3}}{n_e}\right) \text{M}_{\odot}
\end{equation}
The same masses can be obtained using equation 2 of \cite{schreiber_kmos_2019}. We estimate the electron density, $n_e$ directly from our data, using the [SII]$\lambda6717$/[SII]$\lambda6731$ ratio (see \citealp{Osterbrock1989}). For the most massive bin, $\log(M{\star}/M_{\odot})>10.8$, we find $n_e=1480$ cm$^{-3}$, following \cite{sanders_mosdef_2016}:
\begin{equation}
    n_e= \frac{cR-ab}{a-R}
\end{equation}
where, a=0.4315, b=2107, c=627.1 and $R=$[SII]$\lambda6717$/[SII]$\lambda6731$. Comparably high electron densities have been detected for the AGN systems in the KMOS$^{\text{3D}}$ sample \citep{schreiber_kmos_2019} and in local galaxies (e.g. \citealp{Perna2017}, \citealp{Mingozzi2019}, \citealp{Fluetsch2021}). Unfortunately, the direct measure of $n_e$ is possible only for the most massive bin, where the [SII] broad component is detected. For the rest of our sample, the [SII] lines are too weak to estimate a robust broad component, in this case we assume {an average value of $n_e= 380 ~ \text{cm}^{-3}$ (as found by \citealp{schreiber_kmos_2019} for a subsample of 33 KMOS$^{\text{3D}}$ galaxies with SF-driven outflow detection) and a range of variability of $n_e = [200-600]~ \text{cm}^{-3}$ (consistently with the electron densities of outflowing gas reported in the literature for nearby well studied and high-z galaxies, e.g. \citealp{heckman_nature_1990, Arribas2014, Mingozzi2019, fluetsch_properties_2020-1, Davies_electronDensity2020}). As reported in the next Section, this $n_e$ variation is taken into account by assuming 0.3 dex uncertainty in the measurements of the mass loading factor.} 

{As it is well-known, and recently fully discussed by \citealp{Davies_electronDensity2020}, the [SII] method used to determine the electron density has 3 main disadvantages: 1) it cannot probe high densities (i.e., $n_e>10^4$ cm$^{-3}$), where the [SII] ratio saturates, 2) the [SII] emission could be contaminated by the stellar absorption at 6716 \AA , and 3) it could underestimate the real $n_e$ value in the case of AGNs, as most of the [SII] is emitted from a partially ionised zone, where the gas is mostly neutral. 
We note that the first two effects do not affect our results. We do not observe an extremely high $n_e$ in the broad component of massive systems (where we measure $n_e=1480$ cm$^{-3}$) and, we do not expect to have such high values in the outflow at lower masses (according to estimates of $n_e$ in local and high-z outflows, e.g. \citealp{heckman_nature_1990, Arribas2014, Mingozzi2019,  schreiber_kmos_2019,fluetsch_properties_2020-1, Davies_electronDensity2020}). Regarding point 2) the stellar continuum is not detected in the majority of our KMOS data, so the stellar absorption contamination is expected to be negligible in our case. The only effect that may affect our $n_e$ value is the region traced by the [SII] in case of AGN ionisation. As discussed by \cite{Davies_electronDensity2020}, the electron density determined from the [SII] ratio could be underestimated compared to other methods (i.e. based on auroral and trans-auroral lines). We stress here that a higher value of $n_e$ would have the effect of reducing the outflowing gas masses and mass loading factors, hence exacerbating the difference with the current cosmological simulations presented in the next Section.}

In the case of a multi-conical or spherical outflow and a constant outflow velocity ( v$_{\text{out}}$), the mass outflow rate ($\dot M_{\text{out}}$) is defined as \citep{lutz_molecular_2020}:
\begin{equation}\begin{split}\label{Mdot}
\dot M_{\text{out}} &= \text{\textit{C}} \; \frac{ \text{M}_{\text{out}} \; v_{\text{out}}}{ \text{R}_{\text{out}}} = \\ &= 
1.02 \times 10^{-9} 
\left ( \frac{v_{\text{out}}}{\text{km s}^{-1}}  \right)
\left ( \frac{\text{M}_{\text{out}}}{\text{M}\odot}  \right)
\left ( \frac{\text{kpc}}{\text{R}_{\text{out}}}  \right)
\textit{C} \; \text{M}_{\odot} \; yr^{-1}
\end{split}\end{equation}
where the multiplicative factor C depends on the assumed outflow history and R$_{\text{out}}$ is the radius of the outflow. 
Similar to \cite{genzel_sins_2011,genzel_evidence_2014,schreiber_kmos_2019} we adopt a constant outflow rate started at $-t=$-R$_{\text{out}}/v_{\text{out}}$ which gives C$=1$. In this model, the outflowing gas density radially decreases with $\rho \propto R^{-2}$. 

uncertaintyTo be consistent with previous works, the outflow speed, v$_{\text{out}}$, is calculated as v$_{\text{out}}=\Delta v -2 \times \sigma_{\text{B}}$ (see \citealp{Veilleux2005,genzel_sins_2011,genzel_evidence_2014,freeman_mosdef_2019}), and the outer radius of the outflow is assumed to be R$_{\text{out}}=$R$_{e}$ (see \citealp{Schreiber2014,schreiber_kmos_2019}). As proposed by \cite{Schreiber2014} this assumption is justified by the typical sizes of ionised gas outflows detected with high-resolution adaptive optics (AO)-assisted SINFONI observations for a sample of high-z galaxies (see \citealp{newman_sinszc-sinf_2012} and \citealp{Schreiber2014}). The relation of the outflow size with Re can be understood for star forming galaxies in terms of larger star forming region size would produce larger outflows. The origin of a relation is physically less obvious for AGN-driven outflows, where the size must be linked to the AGN power and the geometry and physics of the surrounding ISM and circumgalactic medium (CGM) retaining medium of each individual galaxy. As a consequence, in the case of AGNs we consider the relation approximately valid in a statistical sense, although individual objects may deviate and have a specific size associated with the specific physical properties. {A possible variation of $R_{\text{out}}$, between [$R_{\text{out}}$/2, $R_{\text{out}}\times 2$], is considered inside the +-0.3 dex uncertainty of the mass loading factor in the most massive systems as specified below.}

From these quantities we calculate the mass loading factor as $\eta= \dot M_{\text{out}}/$SFR. Given the assumptions used to calculate each measure the uncertainties in these measurements are assumed to be {0.3 dex of the value.} {{This values takes into account the effect of possible variations of the electron density ($n_e =[200,600]$ cm$^{-3}$ as observed in well studied nearby and high-z galaxies, e.g. \citealp{heckman_nature_1990} for M82 and others strong far-infrared galaxies, \citealp{Arribas2014, Mingozzi2019, schreiber_kmos_2019, Davies_electronDensity2020, Fluetsch2021}) in the mass bins where an estimate was not possible (i.e below $\log(M{\star}/M_{\odot})< 10.6$), as well as variations of $R_{\text{out}}$ in the most massive AGN dominated bin (above $\log(M{\star}/M_{\odot}) \sim 10.6$).}}

v$_{\text{out}}$, $\dot M_{\text{out}}$ and $\eta$ values and errors obtained for all our bins in the H$\alpha$ and [OIII] case are reported on Table \ref{tab:InfoSample2}. {We verified that the obtained values are not sensitive to the variation of the parameters adopted to generate the mock rotating disc emission presented in Section \ref{newmethod}, see Section \ref{perturbations} and Appendix \ref{figurePerturbations} below for more details.}

\subsection{Mass loading factor as a function of stellar mass}\label{mass_loading_factor}
\begin{figure}
   \centering
   \includegraphics[angle=90,width=\hsize]{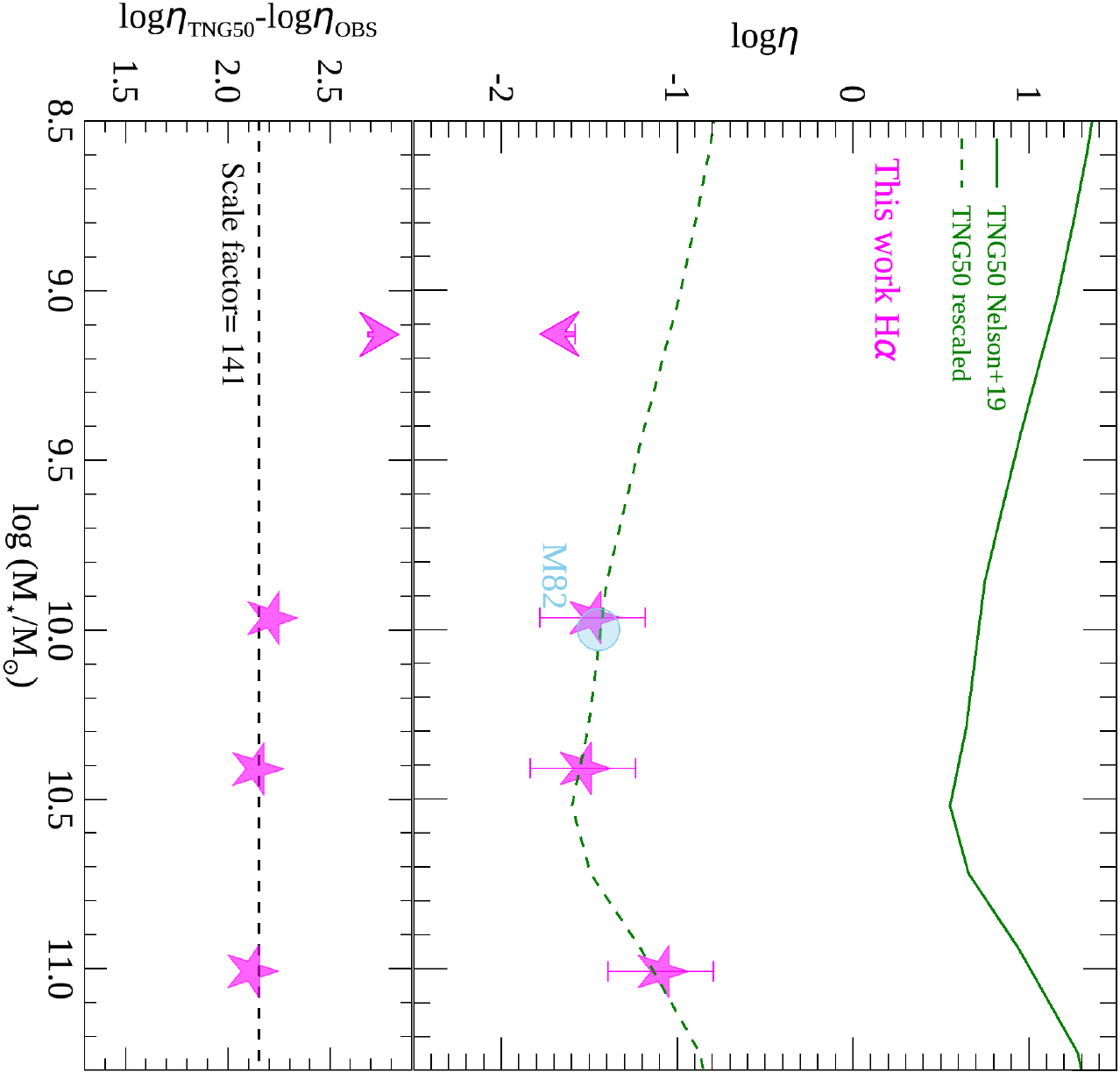}
   \caption[caption.]
  {Mass loading factor, $\eta$, as a function of the stellar mass. KLEVER values, magenta symbols, are compared to the prediction of TNG50 {for gas with $v>0$ km s$^{-1}$ and R$=10$ kpc}  (\citealp{nelson_first_2019}), green solid line. 
  The derived values are significantly lower than what is typically assumed in cosmological simulations,{ $\eta \sim \frac{1}{141} \; \eta_{\text{TNG50}}$.} This result suggests that either a significant fraction of the outflowing mass must be in other gas phases or the theoretical expectation has to be revised. {Once the theoretical curve is re-scaled (dashed green line), the observed mass loading factors scale with stellar mass as expected from TNG50 above $\log(M{\star}/M_{\odot})= 10$, as it can be fully appreciated by the bottom panel where the residuals are shown.  Surprisingly, the discrepancy between theory and observations increases at low $M{\star}$, suggesting that the ionised gas outflows in dwarf galaxies might play a negligible role.} KLEVER results are fully consistent with the ionised mass loading factor measure on M82, cyan circle (see \citealp{heckman_galactic_2019} and references therein). 
  }
              \label{loadingFactor}%
   \end{figure}

Figure \ref{loadingFactor} shows the variation of $\eta$ as a function of stellar mass obtained for the KLEVER stacked spectra (magenta symbols). As previously reported in Section \ref{results}, in the lowest mass bin, $\log(M{\star}/M_{\odot})< 9.6$, we do not detect any clear evidence (above $3\sigma$) of non circular-motions even in the brightest emission lines of our interest ([OIII] and H$\alpha$). We use the values obtained in the fit as an upper limit on the flux associated with non-circular motions providing an upper limit on the maximal mass loading factor for those dwarf galaxies. 

The mass loading factor is approximately constant while M$_\star$ increases from dwarfs to median mass galaxies, $\log(M{\star}/M_{\odot})=10.5$, with a median value of $\sim 0.03$. At the highest masses, $10.8< \log(M{\star}/M_{\odot})< 11.4$, the mass loading factor reaches a maximum of about $\eta= 0.08$. This result is not surprising given the fact that, as already reported in Section \ref{results}, the most massive bin in our sample is characterised by the presence of intense AGN activity. 
\cite{schreiber_kmos_2019} found a similar trend with M${\star}$ and $\eta$ at $0.6<z<2.7$ in the KMOS$^{\text{3D}}$ survey. In particular they found a roughly constant mass loading factor for star formation driven outflows in the stellar mass range of $\log(M{\star}/M_{\odot})= 9.7-11.05$, and higher values of $\eta$ for more massive systems dominated by AGN activity. 

We note that our mass loading factors are relatively low, with $\eta \sim 0.03-0.08$ over the stellar mass range probe by KLEVER. These values are lower than previous observations of ionised gas outflows in galaxies at similar redshifts. In particular, the values found by \cite{schreiber_kmos_2019} are an order of magnitude higher ($\eta \sim 0.1-0.2$ for SF-driven outflows and $0.1-0.5$ for AGNs). Even higher values, 0.3-0.5, are obtained by \cite{davies_kiloparsec_2019} exploiting a sample of 28 star-forming galaxies at $z\sim2.3$ from the SINS/zC-SINS AO survey. Likewise, \cite{swinbank_energetics_2019} found similar values, 0.1-0.4, analysing the averaged H$\alpha$ emission line of $\sim 500$ main-sequence galaxies at $z\sim 1$. We note that all those measurements are obtained by applying the "velocity-subtracted" method to IFU data. As already pointed out in Section \ref{vel_sub_method}, Appendix \ref{vel_sub_residuals} and \ref{comparisonKMOS3D_KROSS}, this technique is quite sensitive to the beam-smearing effect and could have some limitations in the case of unresolved and/or undetermined velocity gradients (e.g. in the case of edge-on galaxies, presence of a bulge). The result is the appearance of spurious broad flux in the proximity of the line wings (see the example shown in Fig. \ref{mock_HR_LR_KROSS_KMOS3Dlike} and \ref{mock_KMOS3D_method},  \ref{mock_stacked_KMOS3D}, \ref{mock_stacked_KROSS} in the Appendix), and a consequent increase of the {broad flux associated with the outflow (as shown in Section 5.2.1 and Appendix D)} and mass loading factor. 

Higher mass loading factors, $\eta \sim0.64-1.4 $, are also detected by \cite{freeman_mosdef_2019} analysing a sample of 127 star-forming galaxies at $1.4<z<2.6$ as a part of the MOSDEF (MOSFIRE Deep Evolution Field) survey \citep{kriek_massive_2016}. As already pointed out by \cite{davies_kiloparsec_2019}, these values are obtained assuming a very low electron density for the outflowing material ($n_e \sim 50$ cm$^{-3}$). If $n_e$ is assumed to be $380$ cm$^{-3}$ the mass loading factor reported by \cite{freeman_mosdef_2019} would decrease to $\eta =0.08-0.2$ (see \citealp{davies_kiloparsec_2019}), but they are still higher than our values. Also in this case, the apparent discrepancy can be explained by the different method used to separate the outflowing flux from the rotation in the global emission line. In particular, \cite{freeman_mosdef_2019} used the galaxy-integrated spectra to search for broad emission, decomposing the line using two Gaussian lines (narrow and broad component) and, finally, they interpreted the broad flux as evidence of galactic outflows. As we already pointed out in Section 3, the large scale rotation velocity and several observational effects (as the spectral response of the instrument and beam smearing effect) may alter the shape of the emission lines generating artificial broad flux at high velocities even without the presence of outflowing material. 

In Figure \ref{loadingFactor} we also compare our findings with the mass loading factor detected in the brightest and best-studied example of local starburst-driven outflow, M82 (cyan circle, \citealp{Lynds_Sandage1963,O_Connell1978}). It is well known that M82 is characterised by intense star-formation activity, with SFR$=7$ M$_{\odot}$yr$^{-1}$ (assuming a Chabrier IMF or SFR$=10$ M$_{\odot}$yr$^{-1}$ if we assume a Kroupa IMF, see \citealp{heckman_galactic_2019}) and stellar mass of about $ \log(M{\star}/M_{\odot})=10$ (\citealp{mayya_star_2006}). As reported by \cite{heckman_galactic_2019}, the mass outflow rate calculated from the warm ionised gas traced by H$\alpha$ and [NII] emission, is about 0.2-0.3 M$_{\odot}$yr$^{-1}$ (\citealp{heckman_nature_1990,Shopbell1998}). The resulting mass loading factor is then $\eta_{\text{M82}} \sim  0.036$ (cyan circle in the figure) which perfectly agrees with the KLEVER value obtained for galaxies with same SFR and M$\star$. This result is astonishing given the different data-sets involved and the different methods applied and it represents a further confirmation of the validity of our procedure and assumptions.

\subsection{Comparison with theoretical expectations}\label{obsVStheory}
Finally, we find that the mass loading factors derived in this paper are in tension with the theoretical predictions of cosmological simulation 
that typically require higher values, $\eta > 1$, to reproduce the low baryon fractions expected at low and high mass halos (e.g. \citealp{dave_galaxy_2011,Hopkins2012,Vogelsberger_2014, muratov_gusty_2015,nelson_first_2019}). In Figure \ref{loadingFactor}, we show, as an example, the comparison between our findings and the theoretical prediction of the TNG50 cosmological simulation \citep{nelson_first_2019,pillepich_first_2019}, the highest-resolution run of the  IllustrisTNG project \citep{Nelson2018,Naiman2018,Marinacci2018,pillepich_simulating_2018,Springel2018}. In particular, we focus on the $\eta_{\text{TNG50}}$ obtained for outflowing gas observed at 10 kpc from the galaxy and with a radial velocity higher than $0$ km s$^{-1}$ (green solid line). Note that considering smaller distances would provide higher $\eta$ from the theoretical side (see Figure 5 of \citealp{nelson_first_2019}). {Above $\log(M{\star}/M_{\odot})= 9.6$ we find a similar trend with the stellar mass: fairly constant values at $\log(M{\star}/M_{\odot})= 10-10.5$ followed by an increase above $\log(M{\star}/M_{\odot})= 10.5$.} Despite the consistent trend, we find that the observed KLEVER mass loading factors are significantly lower compare to theoretical values, accounting for less then $2\%$ of the total mass loading factor expected in TNG50 as represented by the re-scaled (dashed) line in the figure. { The theoretical curve needs to be re-scaled by a factor of $\sim$140 to match the data. Surprisingly, the discrepancy between the theoretical and observed values increases at low stellar masses (below $\log(M{\star}/M_{\odot})= 9.6$), as it can be fully appreciated by the increase of the residuals (theoretical expectations-observations) shown in the bottom panel of Fig. \ref{loadingFactor}. This result suggests that the ionised gas outflows in dwarfs galaxies might play a negligible role even during the peak of the cosmic star-formation history. 
Yet, even neglecting the low mass case, although there is a qualitative agreements of the trends, it remains true that there is a large discrepancy in terms of absolute values of the loading factor.} Therefore, the question arises: Why do the observed values differ from the theoretical expectation? Where is the large amount of outflowing mass that is expected from simulations?

As already pointed out by \cite{schreiber_kmos_2019} and reiterated in \cite{davies_kiloparsec_2019}, from the observational point of view, 
the normalisation of the observed mass loading factor is strongly connected with the uncertainties on the extension of the outflow ($R_{\text{OUT}}$), the electron density ($n_e$) and, the outflow velocity (v$_{\text{OUT}})$. 
However, we note that to reconcile the discrepancy between observations and theory emerged in this paper, we should have or very small values of $R_{\text{OUT}}$ ({150 times} lower than $R_e$) or very low densities ($n_e \sim 10$ cm$^{-3}$) or extremely high velocities (v$_{\text{OUT}}\sim 10^{4}$ km s $^{-1}$) all of which are at odds with the current observations of well studied nearby galaxies, such as M82 (see \citealp{heckman_galactic_2019}) and recent observations of high-z galaxies (e.g. \citealp{schreiber_kmos_2019}). 

Another much more likely explanation for the tension between observations and simulations regards the multi-phase nature of the outflows. As already pointed out, in this paper we are tracing only the ionised gas in the outflow traced by the H$\alpha$ and [OIII] lines. 
Recent studies based on multi-phase observations of local AGNs \citep{morganti_fast_2015,Oosterloo2017} as well as ULIRGs \citep{fluetsch_cold_2018,Fluetsch2021}, local starbursts including M82 (see \citealp{heckman_galactic_2019} and references therein) and high-z galaxies \citep[e.g.]{maiolino_evidence_2012, bischetti_widespread_2019, herrera-camus_molecular_2019}, suggest that the ionised gas only represents a small portion of the global outflowing material, with the majority of the outflowing mass being embedded in the molecular phase \citep[see review in ][]{veilleux_cool_2020}. Despite the enormous progress made in this direction and the rapid growth of outflows simultaneously detected in the ionised and molecular phase (see \citealp{Fluetsch2021}), several uncertainties remain. Firstly, the CO-to-H2 conversion factor ($\alpha_{\text{CO}}$) required to translate the molecular luminosity into outflow mass is still poorly constrained, making the estimate of the molecular outflowing mass really challenging (e.g. \citealp{bolatto_co--h2_2013} for a review on $\alpha_{\text{CO}}$ and \citealp{leroy_multi-phase_2015} for an estimate of $\alpha_{\text{CO}}$ in the outflow of M82). Secondly, most of the current works are based on poorly resolved data which are insensitive to the relative contribution of the two gas phases at different distance of the galactic disc. 

{In the case of M82, 
the molecular gas phase contributes to a large fraction of the outflowing mass in the proximity of the galaxy disc (below 1.5 kpc) but strongly declines with increasing distance, becoming negligible compared to the hot phase already at 4 kpc (\citealp{leroy_multi-phase_2015}). As reported by \cite{leroy_multi-phase_2015}, the cold gas of M82 does not make it far from the disc, as it leaves the outflow and falls back onto the disc, therefore supporting the scenario of a cold fountain. This suggests that, in M82, the contribution of the effective cold gas mass moving out from the disc (R>4 kpc) will be negligible compared to the warm ionised phase. It is clear that if the results of M82 are generalised to the ionised outflows detected in this work, the discrepancy between the predicted and observed mass loading factors cannot be resolved by accounting for the mass contained in the molecular gas phase.} 
In this case the theoretical expectation has to be revised with strong implication 
in the current scenario of galaxy evolution.
Direct probes of the molecular outflows are then fundamental to fully quantify the total mass outflow rate in high-z galaxies and unveil the real strength of the ejective feedback at Cosmic Noon.

{\subsection{Stability of the method}}\label{perturbations}
{As already mentioned in Section 4, our new line decomposition method is based on scaling relations and empirical results used to build the mock rotating disk models (e.g., TF relation, velocity dispersion as a function of z, etc.).
In this Section we test the stability of the method against the variation of the main assumptions used to generate the mock emission lines. In particular, we re-build the mocks by replacing the model's parameters ( n, R$_{e,\star}$, R$_{e,\text{gas}}$, V and $\sigma_{\text{gas}}$) described in Section \ref{mock} with a random value taken from within a factor 1.5 of the original value. We then repeat the steps presented in Section \ref{newmethod} to generate the mock emission lines, fit each observed stacked spectrum, build the stacked mock rotating disc and compare them with the averaged spectrum of each mass bin. Finally, we adopt the prescriptions reported in Section \ref{outflowsProp} to estimate the outflow mass and mass loading factors for the perturbed models and we compare them with the original values presented in the previous Section. As shown in Fig. \ref{loadingFactor_pertubations}, for the H$\alpha$ case, the new perturbed $\eta$ values (magenta symbols) are fully consistent with the original values (grey symbols), confirming the stability of our method against random perturbations.}
\\
\\

\section{Conclusions}
In this paper we investigate the demographics of ionised gas outflows in a representative sample of 141 star-forming main sequence galaxies observed at 1.2<z<2.6. Thanks to the unique multiband (YJ, H and K) coverage offered by the KLEVER survey we have searched for evidence of outflowing gas through telltale high velocity emission underlying the strongest optical rest frame  lines: H$\beta$, [OIII], H$\alpha$, [NII] and [SII]. 
We combine low mass, gravitationally lensed galaxies with more massive systems exploring the kinematics of the ionised gas in an exceptionally wide range of stellar masses, $8.1< \log(M_\star/M_{\odot}) <11.3$, pushing outflow studies to the dwarf regime, which has never been probed before at these redshifts. The single galaxy-integrated spectra are averaged together in four bins of stellar mass providing very high signal-to-noise spectra which are essential to uncover the typical faint flux associated with galactic outflows. We adopt a novel, robust strategy to detect galactic outflows that relies on the direct comparison between the emission of the ionised gas and the expectation of a rotating disc (disc+bulge for  more massive galaxies) model, which was used as a reference of virial motions. Significant deviations from the model parameterized through a broad Gaussian component are interpreted as signature of non-circular motions (as galactic outflows).

Our main results can be summarised as follows:
\begin{itemize}

  \item Surprisingly, we do not find any evidence (above $3 \sigma$ level) of perturbed kinematics (e.g. outflows) in the ionised gas of dwarf galaxies, with $\log(M_\star/M_{\odot})<9.6$: the [OIII], H$\alpha$ and [NII] line profiles are consistent with a simple rotating disc model, suggesting that 
  dwarf galaxies are unexpectedly inefficient at launching massive ionised outflows even during the peak of the cosmic star formation history.
  
 \item Clear signature (above $3\sigma$) of non-circular motions are observed above $\log(M_\star/M_{\odot})>9.6$. The flux associated with perturbed kinematic, F$_{\text{Broad}}$, increases as a function of the stellar mass, from less than $10\%$ of the total line emission to a maximum of F$_{\text{Broad}}= 37 \%$ in the most massive bin at $10.8< \log(M_\star/M_{\odot})> 11.3$.

 \item We compare our findings with those reported in literature based on star-forming galaxies observed with KMOS at similar redshift (e.g. \citealp{schreiber_kmos_2019,swinbank_energetics_2019}). We find a good agreement for the broad fluxes observed in the most massive bin and the fluxes reported by \cite{schreiber_kmos_2019} using the H$\alpha$ spectra of the KMOS$^{\text{3D}}$ galaxy sample. At lower masses, below $\log(M_\star/M_{\odot})=10.8$, our broad component fluxes are $7-8$ times lower than the values reported by the KMOS$^{\text{3D}}$ and KROSS team (see \citealp{schreiber_kmos_2019,swinbank_energetics_2019}). However, we noticed that this discrepancy is {primarily} apparent and attributable to the different technique used to isolate the virial motion { and on the boundaries adopted on the double Gaussian fit.} Indeed, using our mock rotating disc we find that the "velocity-subtracted" method used in previous works can be contaminated by artificial flux at high velocities in case of low spatial resolution, unresolved velocity gradients and disc inclination (see also \citealp{genzel_evidence_2014}). 
 
  \item The analysis of the [OIII]/H$\beta$ and [NII]/H$\alpha$ ratio of the global emission lines, disc  and perturbed emission, reveal that the most massive bin (above $\log(M_\star/M_{\odot})>10.8$) is dominated by AGN activity while the less massive systems are characterised by a stellar photoionization. The broad component, when detected, is always shifted towards the AGN region compared to the disc component, suggesting a possible connection between the non-circular motions and the AGN activity. This indication is corroborated by the similar trend with the stellar mass observed for the broad flux and the incidence of AGN activity reported for similar galaxies (KMOS$^{\text{3D}}$ survey) by \cite{schreiber_kmos_2019}.
  
 \item When detected, the broad flux appears to be blue-shifted with respect to the systemic velocity and the disc component, suggesting a possible connection with the presence of massive galactic outflows. 
 
 \item Ones that the broad flux or non-circular motions are interpreted as gaseous ionised outflows, we find a quite low mass outflow rate ($\dot M_{\text{out}}\sim 0.06-2.3$ M$\odot$ yr$^{-1}$) and mass loading factor ($\eta \sim 0.03-0.08$). These values are fully consistent with the mass loading factor detected in the best-studied example of local starburst-driven outflow, M82. As found for the outflowing flux, we also find that in this case the resulting ($\dot M_{\text{out}}$) and $\eta$ are lower than previous findings (e.g. \citealp{schreiber_kmos_2019,swinbank_energetics_2019,freeman_mosdef_2019,davies_kiloparsec_2019}) {probably} due to the artificial broad flux introduced in previously adopted methodologies.
 
\end{itemize}

Our results suggest that ionised gas outflows might play a negligible dynamical role even during the peak of cosmic SF activity. This seems in tension with most of the current theoretical expectations and numerical simulations that typically require higher mass loading factors, $\eta \geq 0.3-1$ at $\log( \text{M}_{\star}/\text{M}\odot \sim 10$), to explain the low star-formation efficiency at low and high masses. In particular, we show (Fig. \ref{loadingFactor}) the comparison between our mass loading factors and the values expected in TNG50 simulation reported by \cite{nelson_first_2019}. We found a consistent trend with the stellar mass {at $\log( \text{M}_{\star}/\text{M}\odot > 9.6$} but significantly lower values, with the observed mass loading factor accounting for less than $\sim 2\%$ of the total $\eta$ expected on TNG50. This result suggests that either a significant fraction of the predicted outflowing mass is embedded in other gas phases (especially the molecular phase), or the theoretical expectation has to be revised with strong implications for the current scenario of galaxy evolution. {Surprisingly, the discrepancy between the theoretical and observed values increases at low stellar masses, below $\log( \text{M}_{\star}/\text{M}\odot )= 9.6$ (Fig. 11), suggesting that the ionised gas outflows in dwarfs galaxies might play a negligible role even during the peak of the cosmic star-formation history. 
Deep spectroscopy to probe the other gas phases (e.g. with ALMA) are clearly needed to fully quantify the mass associated with outflows at cosmic noon,  unveil the real strength of the ejective feedback, and determine a comprehensive picture of the cosmic baryon cycle.} 

\section*{Data Availability}
The raw KLEVER data underlying this article is publicly available
through the ESO science archive facility (http://archive.eso.
org/cms.html). All other data contributing to
this article will be shared on reasonable request to the corresponding
author.

\section*{Acknowledgements}
AC thanks Federico Lelli for precious suggestions and stimulating discussion. AC is grateful to Hannah \"{U}bler, Antonino Marasco, Sergio Martin Alvarez and Giacomo Venturi for useful discussions and Dylan Nelson for providing a machine-readable version of mass loading factors from the TNG50 simulations. The authors acknowledge all the members of the KMOS$^{\text{3D}}$ (\citealp{schreiber_kmos_2019}) team for sharing additional values related to their Gaussian fitting analysis. AC and MC thank Mark Swinbank for constructive discussion.  AC, RM GCJ and MC acknowledge support by the Science and Technology Facilities Council (STFC) and from ERC Advanced Grant 695671 ‘QUENCH’. 
RM also acknowledges funding from a research professorship from the Royal Society.
AC, AM, GC, FM acknowledge financial support from PRIN-MIUR contract 2017PH3WAT, and PRIN MAIN STREAM INAF "Black hole outflows and the baryon cycle". AM acknowledges financial contributions by PRIN-MIUR 2017WSCC32 "Zooming into dark matter and proto-galaxies with massive lensing clusters" (P.I.: P. Rosati), PRIN MAIN STREAM INAF 1.05.01.86.20: "Deep and wide view of galaxy clusters (P.I.: M. Nonino)" and PRIN MAIN STREAM INAF 1.05.01.86.31 "The deepest view of high-redshift galaxies and globular cluster precursors in the early Universe" (P.I.: E. Vanzella). YP acknowledges National Science Foundation of China (NSFC) Grant No. 12125301, 11773001 and 11991052. This work utilises gravitational lensing models produced by PIs Brada\v{c}, Natarajan $\&$ Kneib (CATS), Merten $\&$ Zitrin, Sharon, Williams, Keeton, Bernstein and Diego, and the GLAFIC group. This lens modelling was partially funded by the HST Frontier Fields program conducted by STScI. STScI is operated by the Association of Universities for Research in Astronomy, Inc. under NASA contract NAS 5-26555. The lens models were obtained from the Mikulski Archive for Space Telescopes (MAST)."



\bibliographystyle{mnras}
\bibliography{myLib} 



\appendix

\begin{figure*}
   \centering
   {\includegraphics[angle=90,width=\hsize]{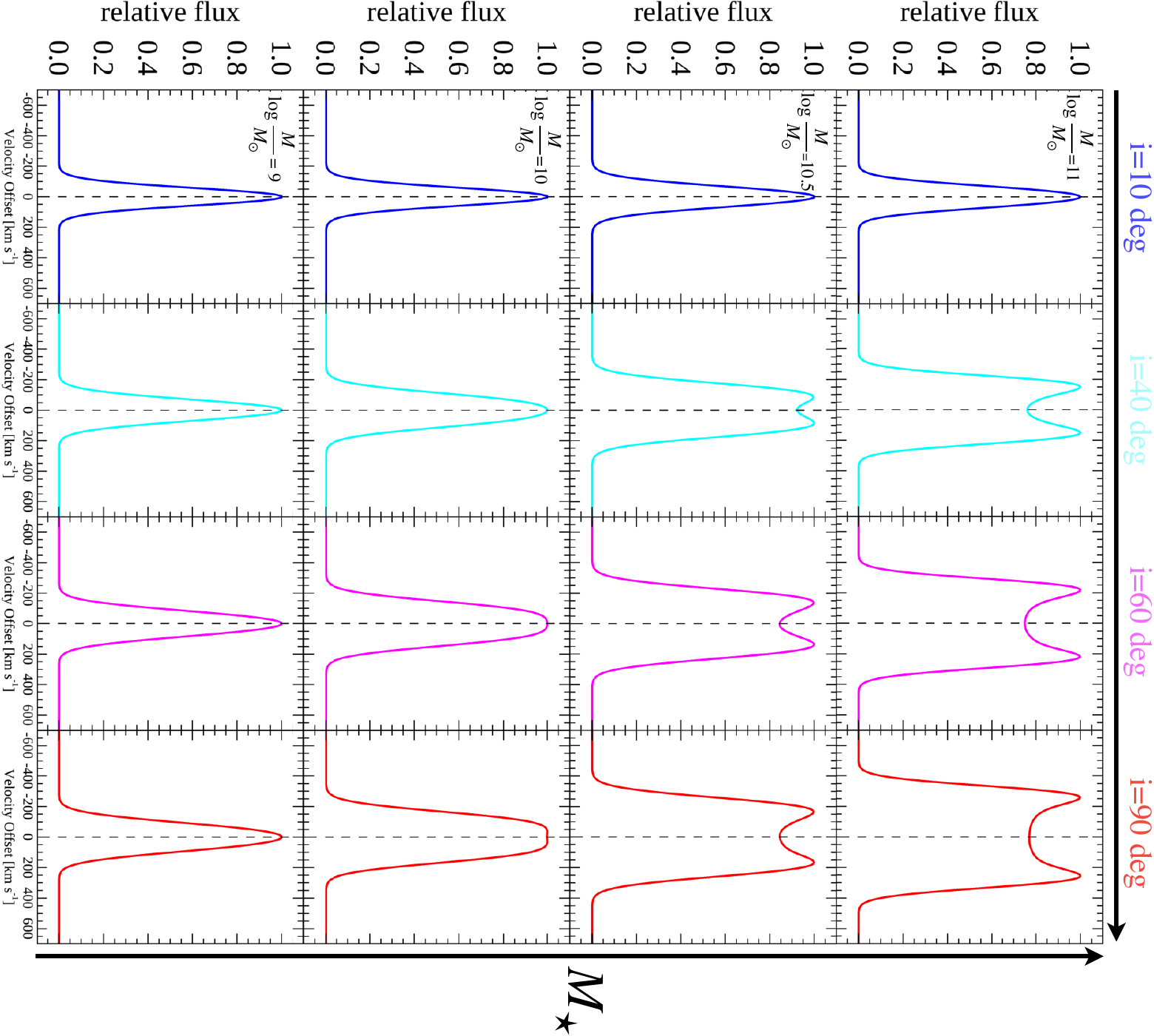}}
   \caption[caption.]{{Variation of the line profiles in the 1D mock spectra as a function of the galactic disc inclination, $i=10, 40, 60$ and $90$ deg from left to right (blue, cyan, magenta and red curves, respectively) for galaxies at different stellar mass, $\log(M_\star/M_{\odot})= 11.0, 10.5, 10.0$ and $9.0$ from top to bottom. The mock spectra are obtained with the \large{K}\small{IN}\large{MS} \normalsize routine (\citealp{davis_atlas3d_2013}) following the procedure presented in Sec. \ref{mock}. Most of the time, the line profile is definitely not Gaussian, especially in massive, $\log(M_\star/M_{\odot})>9$ inclined, i$\geq 40$ deg, galaxies. The line width increases with the increasing of the stellar mass (from bottom to top) and inclination (from left to right). Our model predicts the presence of double peaked emission lines in massive (above $\log(M_\star/M_{\odot})= 10.0$), edge-on (i$\leq 40$) galaxies.}}
                \label{1d_mock_examples}%
   \end{figure*}

{\section{Emission line profile shapes}\label{1D_not_gaussian}
Figure \ref{1d_mock_examples} shows the variation of the emission line profile as a function of the stellar mass (and consequently R$_e$) and galactic disc inclination. The 1D spectra are obtained by using the \large{K}\small{IN}\large{MS} \normalsize code (\citealp{davis_atlas3d_2013}) as discussed in Sec. \ref{mock} for 4 example of galaxies characterised by a stellar mass of $\log(M_\star/M_{\odot})= 11.0, 10.5, 10.0$ and $9.0$ (from top to bottom in the figure) and inclined of i$=10, 40, 60$ and $90$ deg (respectively from left to right). The line profiles are clearly not Gaussian in most of cases and especially above $\log(M_\star/M_{\odot})= 9.0$ and i$>10$ deg. This effect is due to the large scale velocity rotation of the disc which increases with stellar mass accordingly with the Tully-Fisher relation \citep{TullyFisher1977}. As expected for a rotation curve that flattens out at large radii, our model predicts the presence of double peaked emissions for galaxies with stellar mass above $\log(M_\star/M_{\odot})= 10.$ and inclination $\leq 40$ deg. This prediction is confirmed by the line shape of H$\alpha$, [NII] and [OIII] of massive galaxies reported in the next Section \ref{obs_mock_single_EX}.}

\section{Observed and mock galaxy integrated spectra}\label{obs_mock_single_EX}
   \begin{figure*}
   \centering
   {\includegraphics[angle=-90,width=\hsize]{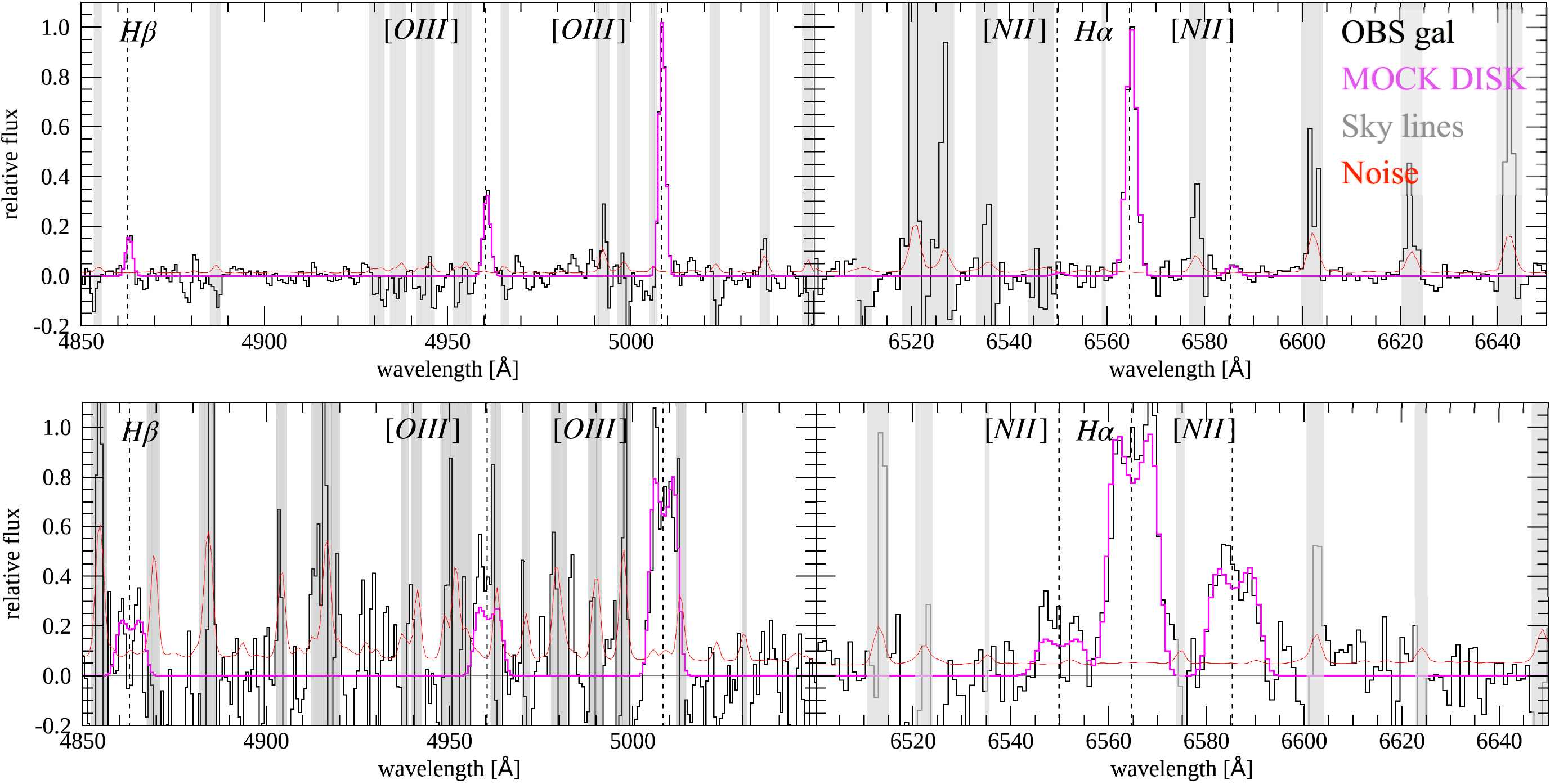}}
   \caption[caption.]{Example of single integrated spectra (black curves) of a low and high mass galaxy (GLASS$\_00333-99-99$ and GS3$\_28464$, top and bottom panel, respectively) in the [OIII] and H$\alpha$ region (left and right panel, respectively) compared with its tailored mock rotating disc model (magenta line) obtained with the \large{K}\small{IN}\large{MS} \normalsize code (see Section. \ref{mock}). The noise spectrum is shown in red, while the grey shaded areas mark the regions affected by the sky lines. Double peaked emission lines are observed in massive systems (e.g. bottom panel) as predicted by our simple disc model (see mock 1D spectra in Fig \ref{1d_mock_examples}). Similar figures are obtained for each galaxy in the KLEVER sample.}
                \label{ha_o3_mock_es}%
   \end{figure*}

In this Section we report some examples of the mock galaxy integrated spectra obtained for two galaxies in our survey. In Fig. \ref{ha_o3_mock_es} the single integrated spectrum of a low mass and massive galaxy (GLASS$\_00333-99-99$ and GS3$\_28464$, top and bottom panel, respectively) in the [OIII] and H$\alpha$ region are compared with the best-fit mock rotating disc emission (magenta lines) obtained following the prescription presented in Section \ref{mock}. Similar figures are obtained for all galaxies in the KLEVER sample. Note that the noise (red line) in the single galaxy spectra is too high to identify the evidence of perturbed flux near the most intense lines. The stacked technique is clearly required to decrease the noise and increase the SNR and enable the disc-outflows decomposition.

\section{Effect of a central bulge in velocity-subtracted spectra}\label{vel_sub_residuals}
\begin{figure} 
\centering
\subfloat{\includegraphics[angle=0,width=\hsize]{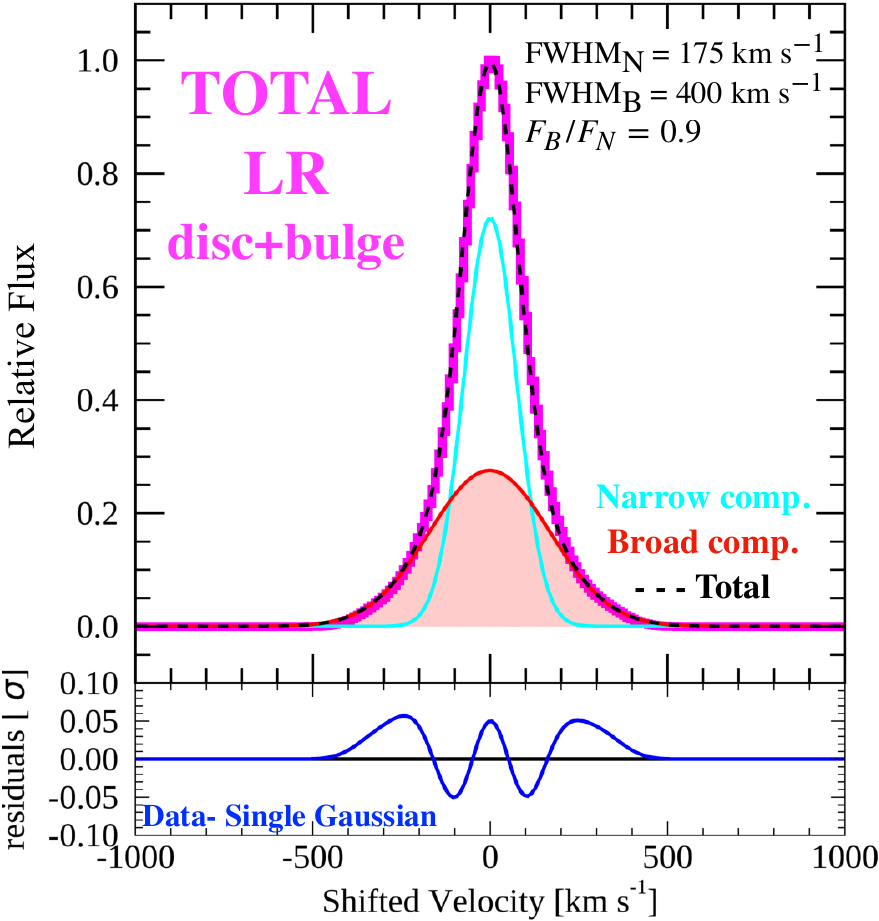}}
   \caption[caption.]{{Same description as right panel in Fig. \ref{mock_HR_LR_KROSS_KMOS3Dlike}. Here we add the bulge component to the mock rotating disc model described in  Fig. \ref{mock_HR_LR_comparison} accordingly with the prescription presented in Section \ref{mock}. {The same priors proposed by \cite{schreiber_kmos_2019} are adopted (FWHM$\geq 400$ km s$^{-1}$).} The inclusion of the bulge in the mock datacube determines an increase of the flux enclosed in the artificial broad component (red area) making the velocity-subtracted method even less accurate on determining the real outflow detection.}}
                \label{mock_KMOS3D_method}%
\end{figure}

{As reported in Section \ref{vel_sub_method} the commonly used velocity-subtracted method (\citealp{shapiro_sins_2009,genzel_sins_2011,genzel_evidence_2014,schreiber_kmos_2019,davies_kiloparsec_2019,avery_incidence_2021,swinbank_energetics_2019}) could be contaminated by an artificial broad flux even in a rotating disc model where no outflows are present. Here we repeat the exercise presented in Section. \ref{vel_sub_method} to show that the strength of this artificial broad component can be even larger if a central bulge is present. Following the prescription presented in Section \ref{mock} we add a bulge to the rotating disc model used in Section \ref{vel_sub_method} and shown in Fig. \ref{mock_HR_LR_comparison}. Note that for galaxies with a stellar mass $\log(M_\star/M_{\odot})\geq 10.5$ at redshift z=1-2 the presence of a central bulge is expected as reported by photometric observations presented by \cite{lang_bulge_2014}. Repeating the same velocity-subtracted analysis presented in Section \ref{vel_sub_method}, and fitting the low resolution KMOS-like velocity-subtracted spectrum with two Gaussian components (see Fig. \ref{mock_KMOS3D_method}), we find that, if the bulge is present, the flux of the spurious broad Gaussian component, F$_{B}$, (red area in the figure), could reach very high values comparable or even larger than the flux of the narrow component, F$_{N}$.} 
{This implies that the velocity-subtracted method could be even less accurate on determining the real outflowing flux in galaxies with a central bulge. }

{The relative variation of the "spurious" broad flux ($F_{\text{Broad}}/F_{\text{Narrow}}$, FWHM$_B$) with the galaxy properties (stellar mass, bulge component, effective radius, etc.), observational effects (inclination, spatial and spectral resolution) and priors adopted in the Gaussian fit, goes beyond the scope of this manuscript but will be further investigated in a forthcoming paper (Concas et al. in prep.). We urge the reader to take into account this "spurious contamination" when the properties of the outflowing gas are estimated using the "velocity-subtracted" method as they could provide an overestimation of the detection rates and mass of the outflowing gas.}

\section{Comparison with previous observations}\label{comparisonKMOS3D_KROSS}

\begin{figure*} 
\centering
\includegraphics[angle=90,width=\hsize]{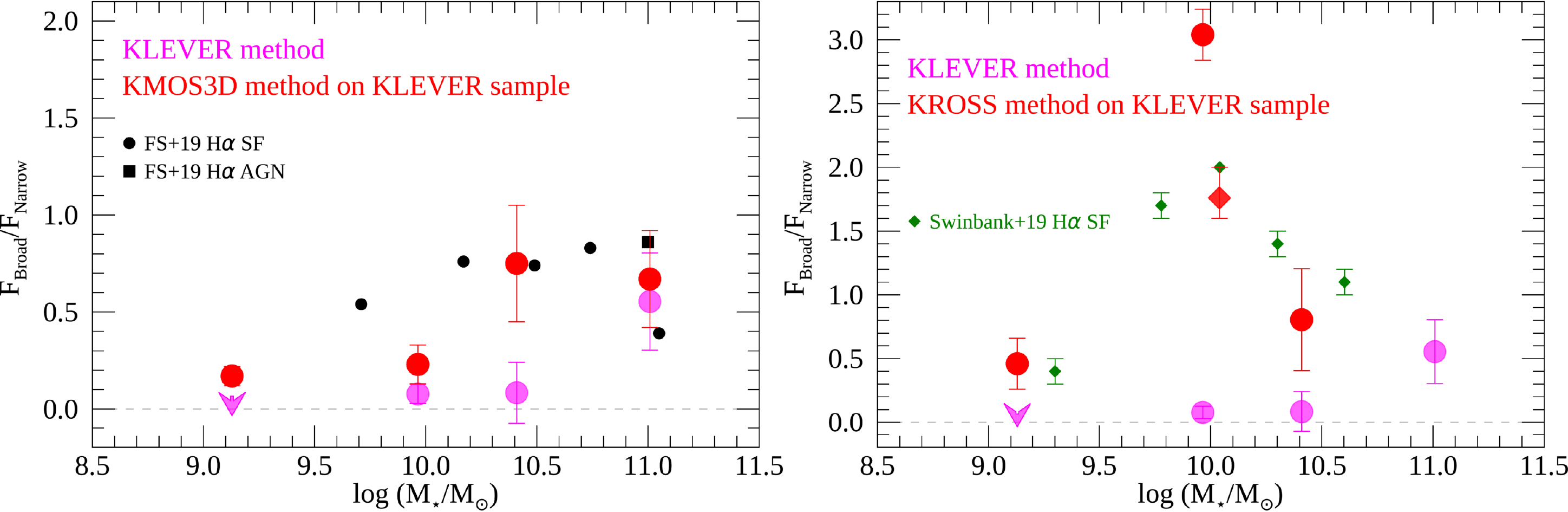}
   \caption[caption.]{{
   Same as Figure \ref{flux_broad} plus the H$\alpha$ broad to narrow flux ratio (red circles) obtained for the KLEVER galaxies using the KMOS$^{\text{3D}}$ method illustrated by \cite{schreiber_kmos_2019} (\textit{Left panel}) and using the KROSS method presented by \cite{swinbank_energetics_2019} (\textit{Right panel}). The big red diamond in the left panel correspond to the flux ratio obtained by considering all KLEVER galaxies except the most massive galaxies where clear AGN contamination is detected.  The discrepancy between the KLEVER flux ratios calculated with the rotating disc decomposition (magenta circles) and the KMOS$^{\text{3D}}$ results (black points) or the KROSS results (green diamonds) is relieved ones the KMOS$^{\text{3D}}$ or the KROSS
 technique is applied to the KLEVER galaxies (red circles), suggesting that the discrepancy is only apparent and mainly driven by the different methodologies. The errors associated with the red points take into account the $F_{\text{Broad}}$/$F_{\text{Narrow}}$ variation obtained weighting the single spectra by (SN)$^2$, (SN) and without any weight in the case of {KROSS} method.}}
                \label{flux_broad_KMOS3D_KROSSmethod}%
\end{figure*}

   \begin{figure*}
   \centering
   \includegraphics[angle=90,width=\hsize]{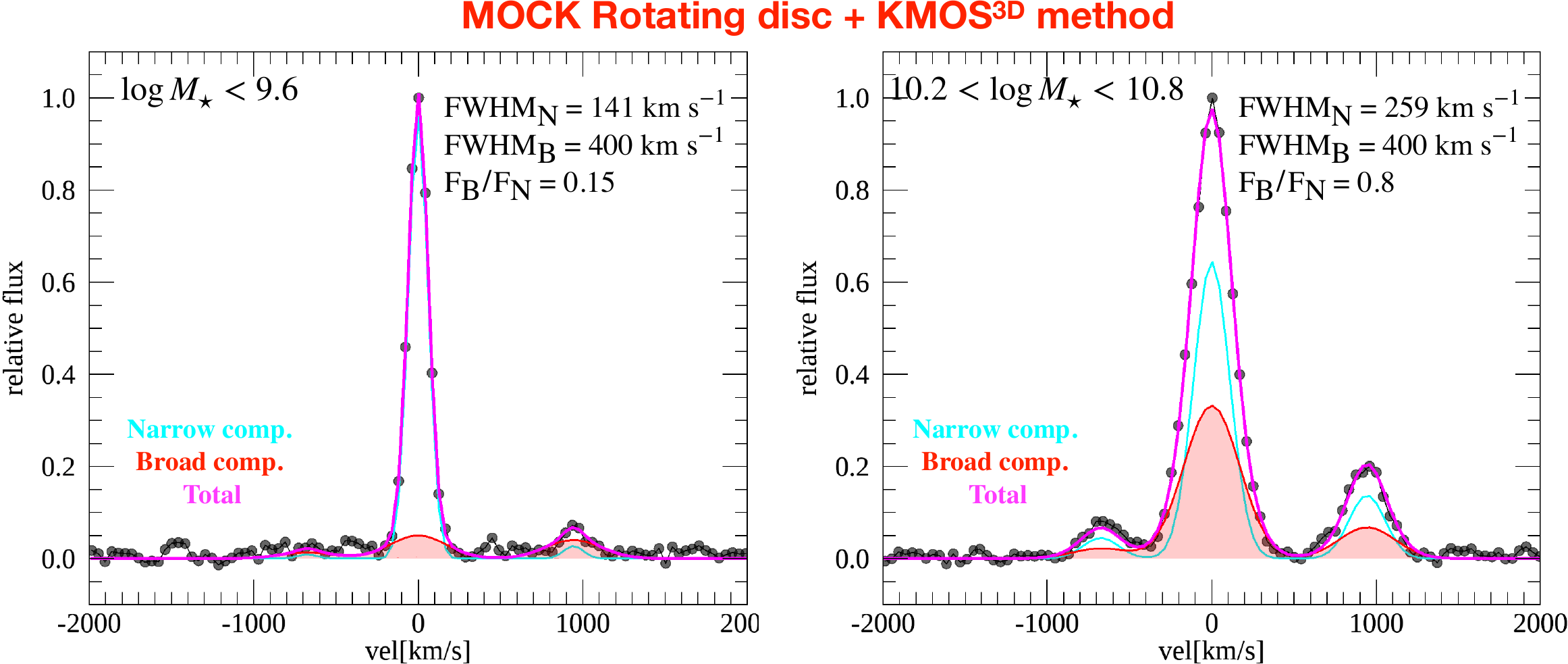}
   \caption[caption.]
  {{Example of KMOS$^{\text{3D}}$ method applied to the averaged mock rotating disk models of dwarf ($\log(M_\star/M_{\odot})<9.6$, left panel) and medium ($10.2< \log(M_\star/M_{\odot}),10.6$, right panel) KLEVER mass bins. The velocity-subtracted spectrum is analysed following the prescriptions presented by \cite{schreiber_kmos_2019}. A substantial artificial broad component (red area) emerges even in these discs models where no outflowing component is present. Line width of the narrow (FWHM$_{\text{N}}$) and broad (FWHM$_{\text{N}}$) Gaussian component as well as their line ratios ($F_{\text{B}}/F_{\text{N}}$) are reported in the figure.}}
              \label{mock_stacked_KMOS3D}%
   \end{figure*}
   \begin{figure*}
   \centering
   \includegraphics[angle=90,width=\hsize]{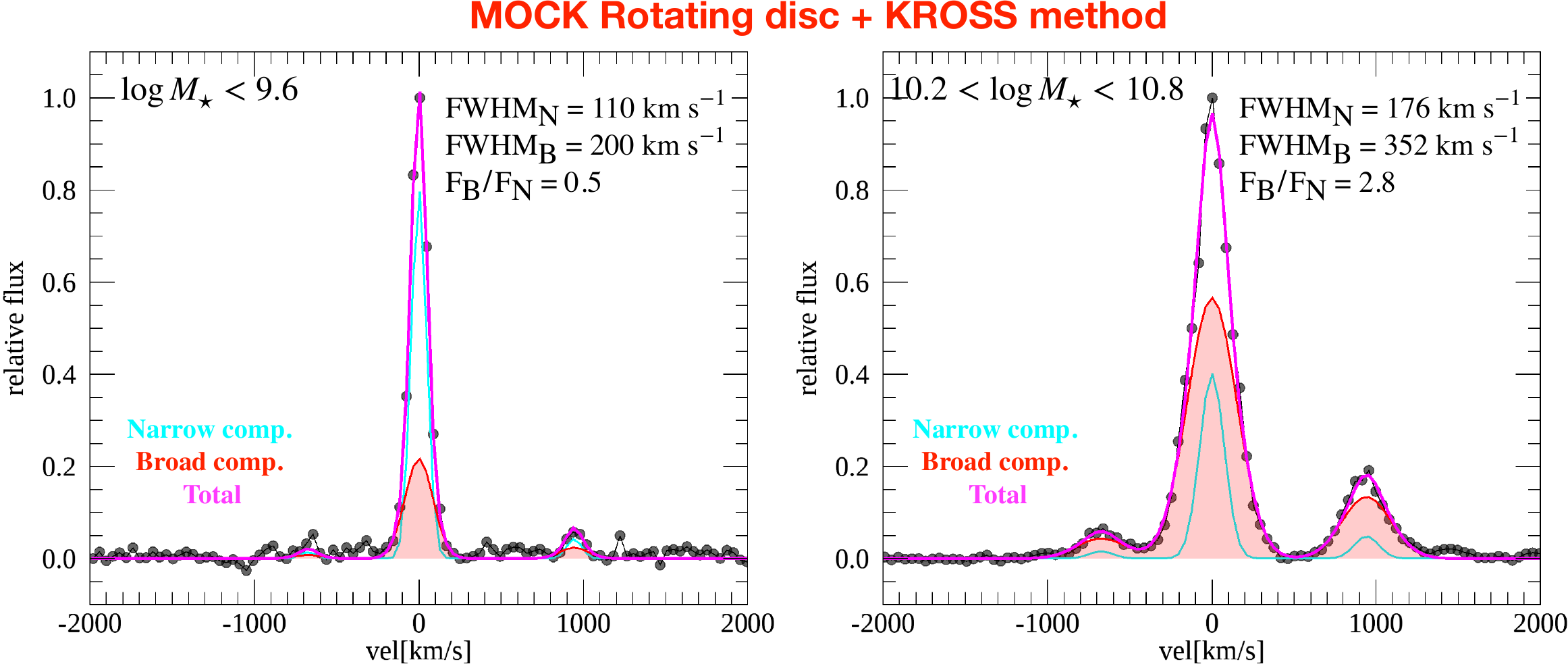}
   \caption[caption.]
  {{Same rotating disc models as Fig. \ref{mock_stacked_KMOS3D} but analysed following the KROSS method discussed in \cite{swinbank_energetics_2019}. Also in this case the spurious broad component (red area) appears without any outflow contribution.}}
              \label{mock_stacked_KROSS}%
   \end{figure*}

Here we investigate the nature of the discrepancy presented in Section \ref{KLEVER_vs_KMOS3D_KROSS} between our results and previous observations of z=1-2 galaxies obtained with KMOS by the KMOS$^{\text{3D}}$ (\citealp{schreiber_kmos_2019}) and KROSS (\citealp{swinbank_energetics_2019}) team. To understand if the discrepancies highlighted in Fig. \ref{flux_broad} are due to differences in the galaxy samples or if they are caused by the methodologies used to analyse the KMOS data, we perform the KMOS$^{\text{3D}}$ and KROSS methods directly on our KLEVER sample following all the fundamental steps described, respectively by \cite{schreiber_kmos_2019} and  \cite{swinbank_energetics_2019}. Starting from the KMOS$^{\text{3D}}$ technique, we 1) $\sigma$-clipped the spectrum before and after the H$\alpha$ region to avoid residual noise, 2) we interpolate the flux over the spectral channels if a sky line is very close to the H$\alpha$ and NII emission, 3) we spatially smooth the cubes with a Gaussian kernel of FWHM=3 pixels, 4) we fit the spectrum of each pixel with a Gaussian line to extract the velocity map, 5) we apply in reverse the moment maps to obtain the velocity-subtracted data-cube, 6) we extract the galaxy integrated spectrum withing an aperture of 0.65 arcseconds of radius, 7) we normalise the final galaxy spectra to the peak amplitude of H$\alpha$ and stacked the spectra weighting by (S/N)$^2$ and, finally, we fit the averaged spectra with a narrow and broad component by assuming FWHM$_B \geq 400$ km s $^{-1}$ and FWHM$_N < 400$ km s $^{-1}$. 
{The results are reported in the left panel of Fig. \ref{flux_broad_KMOS3D_KROSSmethod}.}

{We find that the $F_{\text{Broad}}/F_{\text{Narrow}}$ obtained with the KMOS$^{\text{3D}}$ method using the KLEVER sample (red points) are systematically above the values obtained for the same sample using the disc-decomposition method presented in this paper (magenta points). In particular, $F_{\text{Broad}}/F_{\text{Narrow}}$ derived with the KMOS$^{\text{3D}}$ method are 2.6, 3 and 9 times higher than KLEVER method, respectively in the bins with $\log(M_\star/M_{\odot})<9.6$, $9.6<\log(M_\star/M_{\odot})<10.2$ and $10.2<\log(M_\star/M_{\odot})<10.6$. At high stellar mass, $\log(M_\star/M_{\odot})>10.6$ the two methods give consistent results.}

{An other important point shonw in Fig. \ref{flux_broad_KMOS3D_KROSSmethod} is that the discrepancy between the $F_{\text{Broad}}/F_{\text{Narrow}}$ derived with our method and the values published by \cite{schreiber_kmos_2019} ( black points in the figure)} is relieved ones the KMOS$^{\text{3D}}$ method is applied to our KLEVER sample (red circles). {In particular, the $F_{\text{Broad}}/F_{\text{Narrow}}$ obtained by applying the KMOS$^{\text{3D}}$ method to the KLEVER sample (red points) are consistent with the results  determined by \cite{schreiber_kmos_2019} for the  KMOS$^{\text{3D}}$ sample (black points) with the only exception for the lowest mass bin presented in \cite{schreiber_kmos_2019} ($\log(M_\star/M_{\odot})= 9.71$). We notice that for this particular mass bin the averaged distance from the MS reported by \cite{schreiber_kmos_2019} is 0.41 dex (see their table 1), indicating that the galaxies in this mass bin are on average above the MS and so they are  1) not comparable with our KLEVER data-set (statistically located in the MS as reported in new Section 3.2), 2) they are not representative of the star forming galaxy population at those masses and redshift and, even more important, 3) they could be more contaminated by mergers (e.g. \citealp{Luo2014, Pearson2019}) with resulting disturbed emission lines. Note that a direct comparison for dwarfs galaxies ($\log(M_\star/M_{\odot})< 9.6$) is not possible due to the lack of low mass galaxies in the KMOS$^{\text{3D}}$ sample.}

 {Following the steps described by \cite{swinbank_energetics_2019} we 1) start the analysis by fitting the spectrum of each pixel with a Gaussian line to extract the velocity map, 2) we apply in reverse the moment maps to obtain the velocity-subtracted data-cube, 3) we extract the galaxy integrated spectrum withing an aperture of 0.65 arcseconds of radius, 4) we normalise the final galaxy spectra by its H$\alpha$ luminosity and, 5) we stack the spectra using a median average, 6) we fit the averaged spectra with a single Gaussian component and with a combination of a narrow and broad component without assuming any priors on the line widths and, finally, 6) we use the BIC statistic to evaluate the significance of the broad component by adopting $\Delta$BIC>10.  The results are reported in the right panel of Fig. \ref{flux_broad_KMOS3D_KROSSmethod}, where we can see that, also in this case, the differences disappears once we use the KROSS technique into the KLEVER data. In particular, we notice that, using all KLEVER galaxies except the most massive ones (110 objects) where clear AGN contamination is detected, we find a perfect agreement between the KLEVER data (big red diamond in the figure) and previous results obtained by \cite{swinbank_energetics_2019}.} 
 
  {In conclusion, we find that if the KLEVER data are analysed by following the same methodologies as for the KMOS$^{\text{3D}}$ and/or KROSS data, the $F_{\text{Broad}}/F_{\text{Narrow}}$ are higher that the values obtained with the rotating disc decomposition proposed in this paper. However, both these previous methods are based on the velocity-subtracted technique and,  }
 {as previously shown in Section \ref{vel_sub_method} for a single rotating disk model, this technique could be affected by an artificial broad component in the case of low KMOS-like spatial resolution.} 
 
  {To quantify the average broad flux contamination expected on the stacked spectra arising from the beam smearing effect in the velocity-shifted method, we repeat the KMOS$^{\text{3D}}$ and KROSS analysis directly on our mock rotating discs, where no outflow are present, created for the lower ($\log(M_\star/M_{\odot})<9.6$) and medium massive ($10.2< \log(M_\star/M_{\odot}),10.6$) bins. Before to follow the KMOS$^{\text{3D}}$ and KROSS steps, a realistic noise, including the residual sky lines, is added into the mock cubes. As shown in Fig. \ref{mock_stacked_KMOS3D} and Fig. \ref{mock_stacked_KROSS}, respectively for KMOS$^{\text{3D}}$ and KROSS technique, we find a substantial flux enclosed in the artificial broad component with $F_{\text{Broad}}/F_{\text{Narrow}}$ values comparable or even higher that the ratios obtained in the observations. The relative variation of the "spurious" broad flux ($F_{\text{Broad}}/F_{\text{Narrow}}$) with the galaxy properties, observational effects (inclination, spatial and spectral resolution) and priors adopted in the Gaussian fit, goes beyond the scope of this manuscript but it will be further investigate in a forthcoming paper (Concas et al. in prep.). This simple exercise suggest that the high $F_{\text{Broad}}/F_{\text{Narrow}}$ ratios obtained with the KMOS$^{\text{3D}}$ and KROSS analysis could be strongly contaminated by the artificial broad component. We urge the reader to take into account this "spurious contamination" when the properties of the outflowing gas are estimated using the velocity-subtracted method as they could provide an overestimation of the detection rates and mass of the outflowing gas.}

\section{Mass loading factor from [OIII] line}\label{massLoadingo3}
   \begin{figure}
   \centering
   \includegraphics[angle=0,width=\hsize]{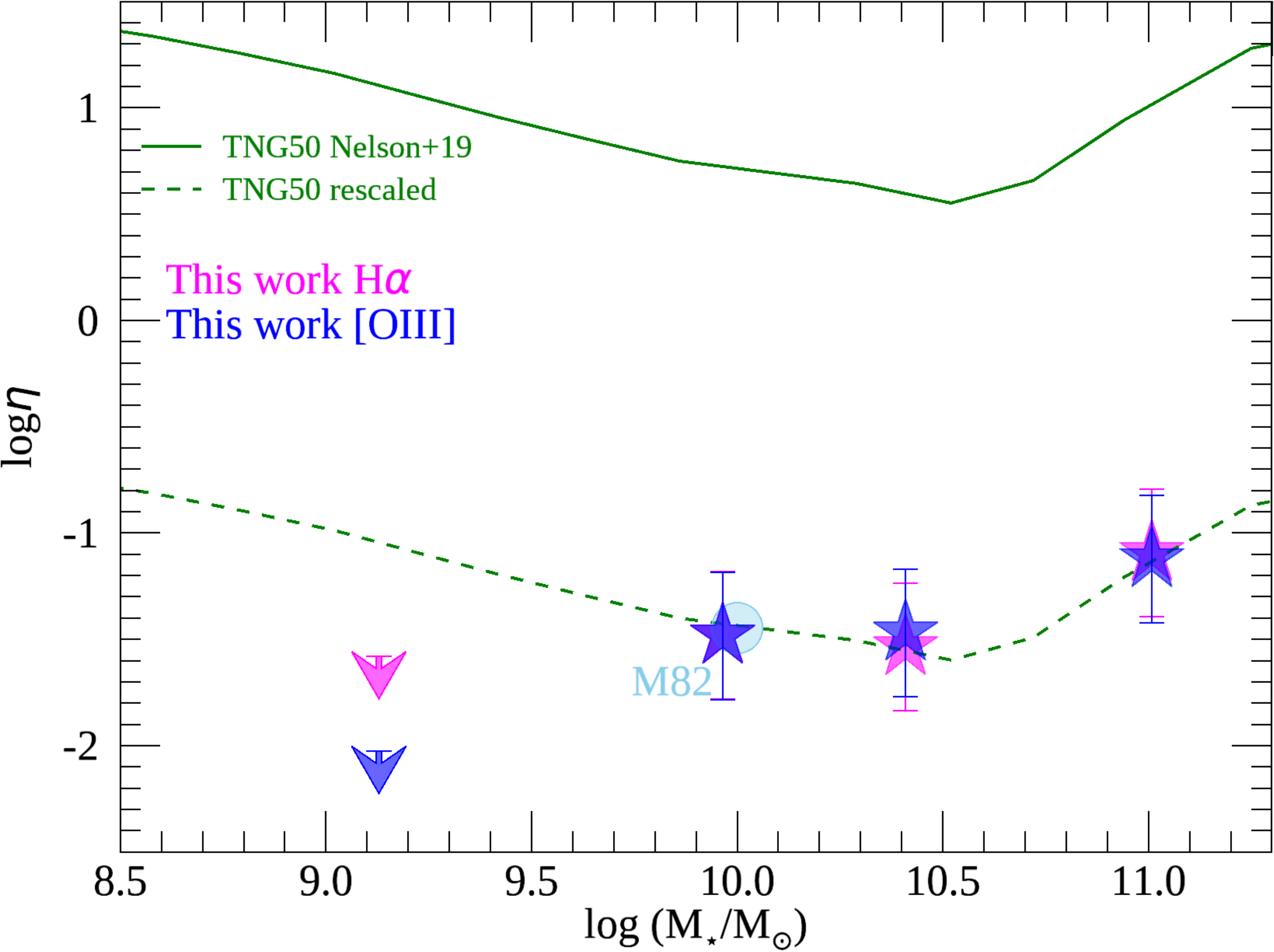}
   \caption[caption.]
  {Same as Figure \ref{loadingFactor} plus the mass loading factors, $\eta$, derived from the [OIII] emission line, blue symbols. $\eta$ obtained with the oxygen emission is consistent with the values obtained with the Hydrogen line once that we assume a sub-solar metallicity {([O/H]$=8.4$ and [O/H]$_{\odot}=8.69$),} accordingly with Equation \ref{massOuto3}. Even assuming the sub-solar metallicity (blue symbols), the observed mass loading factors are lower compared to the theoretical expectation from TNG50 simulation.}
              \label{loadingFactorO3}%
   \end{figure}
In this Section we present how we can measure the mass loading factor using the [OIII] emission line and compare them with the values obtained using the H$\alpha$ line. We start by correcting the [OIII] luminosity of each galaxy for dust attenuation. As done for the H$\alpha$ case (see Section \ref{outflowsProp}), we use the visual extinction obtained from the SED fitting to derive A$_{\text{gas}}$ which is then calculated in the [OIII] region using the Calzetti reddening law assuming that the reddening in the [OIII] line is similar to that of the near H$\beta$ line, A$_{\text{[OIII]}}=\text{A}_{\text{H}\beta}=1.47\text{A}_{\text{H}\alpha}$. Also in this case, we estimate the total weighted L$_{\text{stack}}$ by using Eq. \ref{stackingEq} in the individual [OIII] luminosities. L$_{\text{stack}}$ is then decomposed into disc and broad (L$^{\text{[OIII]}}_{\text{B}}$) component using the flux percentage presented in Section \ref{results}. Following \cite{CanoDiaz2012} and \cite{marasco_galaxy-scale_2020} the outflowing gas mass can be computed from L$^{\text{[OIII]}}_{\text{B}}$ as follows:
\begin{equation}\label{massOuto3}
       \text{M}^{\text{[OIII]}}_{\text{out}}=5.33\times 10^{4} \left ( \frac{ \text{L}^{\text{[OIII]}}_{\text{B}} }{10^{40} \text{ erg s}^{-1}} \right)  \left ( \frac{100 \text{ cm}^{-3}}{n_e}\right) \frac{1}{10^{\text{[O/H]}}} \text{M}_{\odot}
\end{equation}
where $10^{\text{[O/H]}}$ is the oxygen abundance in Solar units. Following the assumptions made in Section \ref{outflowsProp}, we finally calculate the mass outflow rate ($\dot M_{\text{out}}$ using Eq. \ref{Mdot}), and the mass loading factor, $\eta_{\text{[OIII]}}$. As already found in different works, if we assume a solar metallicity, the mass, mass outflow rates and mass loading factor, derived with the [OIII] emission, are lower compared to the same quantities obtained with the hydrogen lines (H$\beta$ and or H$\alpha$, e.g.\citealp{carniani_ionised_2015,Fluetsch2021,marasco_galaxy-scale_2020}). Interestingly, the discrepancy between the two quantities disappears once {we assume the average metallicity observed in KLEVER ([O/H]$=8.4$ and [O/H]$_{\odot}=8.69$, see \citealp{curti_klever_2020}) as shown by the mass loading factors reported in Fig. \ref{loadingFactorO3}.} Also in this case, the observed mass loading factors are lower compared to the theoretical expectations but perfectly in agreement with the mass loading factors observed in the ionised gas phase of M82. See the discussion in Section \ref{obsVStheory}.


\section{Test on the stability of the method}\label{figurePerturbations}
\begin{figure}
   \centering
   \includegraphics[angle=90,width=\hsize]{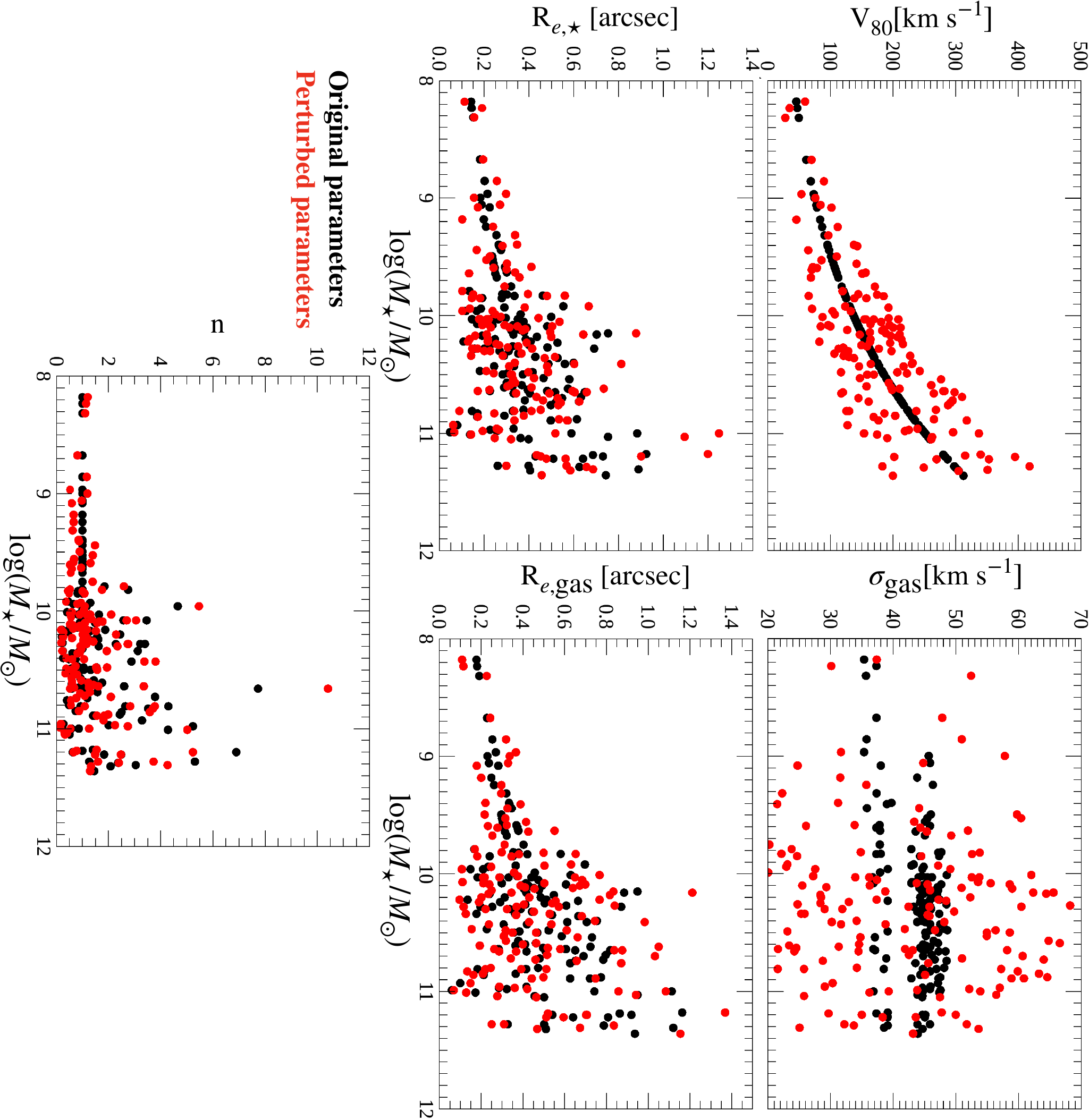}
   \caption[caption.]
  {{Distribution of the original (black) and perturbed (red) model's parameters (n, R$_{e,\star}$, R$_{e,\text{gas}}$, V$_{80}$ and $\sigma_{\text{gas}}$) as a function of the stellar mass. Perturbed values are used to test the stability of our method against random variations on the physical parameters used to define the mock rotating discs. Their effect on the mass loading factor is presented in Fig. \ref{loadingFactor_pertubations}}}
              \label{par_pertubations}%
   \end{figure}

\begin{figure}
   \centering
   \includegraphics[angle=0,width=\hsize]{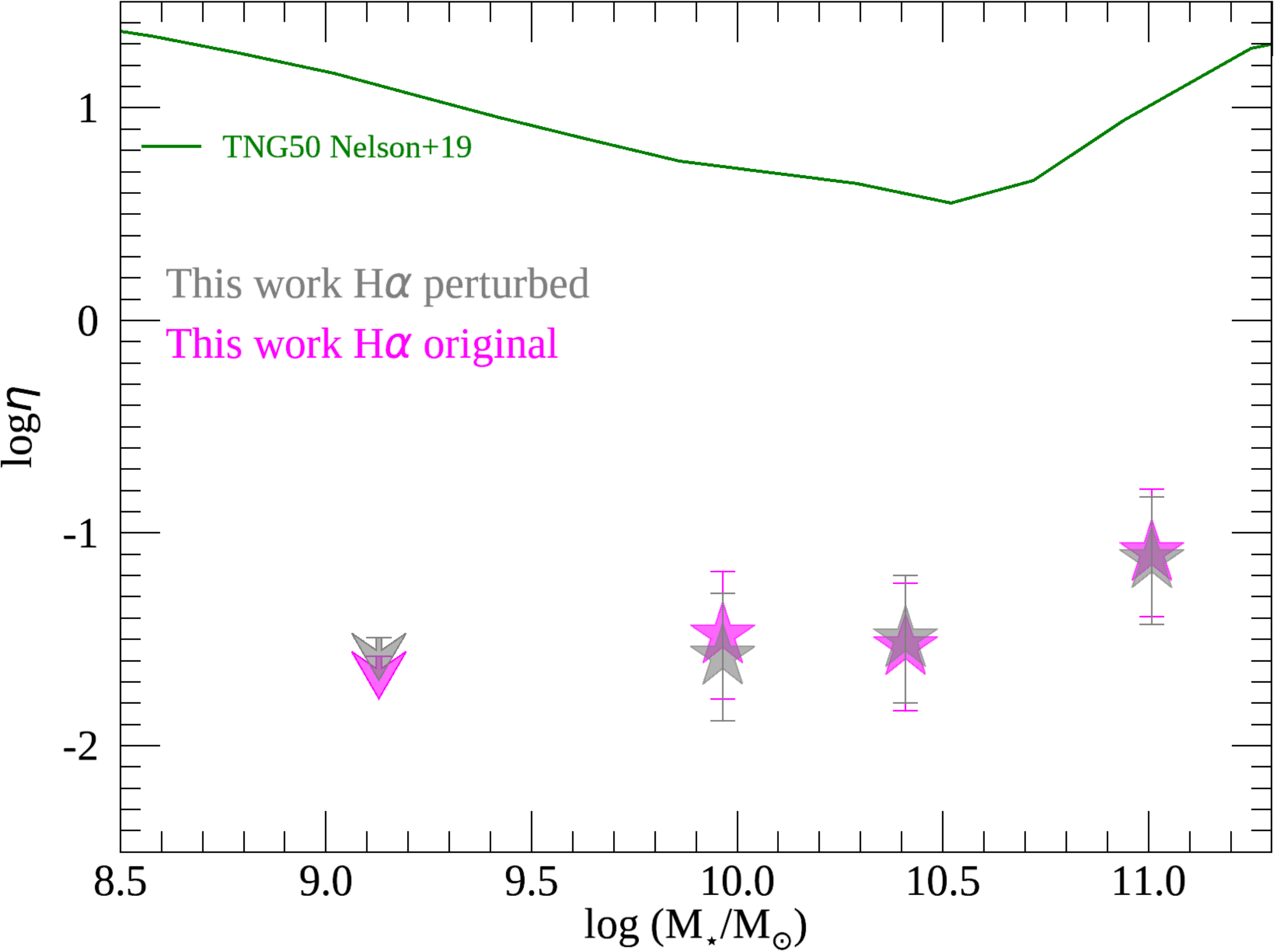}
   \caption[caption.]
  {{Effect of the perturbed parameters (n, R$_{e,\star}$, R$_{e,\text{gas}}$, V$_{80}$ and $\sigma_{\text{gas}}$, {shown in Fig. \ref{par_pertubations})} into the mass loading factor ($\eta$) obtained using the H$\alpha$ line emission. The values obtained with the perturbed mocks (grey symbols) are fully consistent with the original values adopted in the manuscript (magenta symbols), confirming the stability of our method against random perturbations of the adopted physical parameters. The same theoretical curve (green line) as in Fig. \ref{loadingFactor} is shown.}}
              \label{loadingFactor_pertubations}%
   \end{figure}

{As reported in Section \ref{perturbations} we test the stability of our method against random variations on the mock rotating disc parameters. Fig. \ref{par_pertubations} shows the distribution of the original (black points) and perturbed (red points) n, R$\star$, R$_{e,\text{gas}}$, V and $\sigma_{\text{gas}}$ as a function of the stellar mass. Perturbed values are obtained by randomly perturbing the original values within a factor 1.5. Using these perturbed parameters, we find that the derived mass loading factors (grey symbols in Fig. \ref{loadingFactor_pertubations}) are fully consistent with the values obtained with the original parameters (magenta symbols), assuring the stability of our results against random perturbations.}

\bsp	
\label{lastpage}
\end{document}